\documentclass[
preprint,
longbibliography]{revtex4-1}
\usepackage{amssymb}
\usepackage{amsmath}
\usepackage{graphicx}
\usepackage{dcolumn}
\usepackage{bm}

\usepackage{setspace} 
\usepackage{caption}
\usepackage{subcaption}
\usepackage[usenames, dvipsnames]{color}
\usepackage{xcolor}
\usepackage[framemethod=tikz]{mdframed}
\usepackage{soul}
\definecolor{mycolor}{rgb}{0.95, 0.95, 0.95}

\newmdenv[innerlinewidth=0.5pt, roundcorner=2pt,linecolor=black,innerleftmargin=6pt,
innerrightmargin=6pt,innertopmargin=6pt,innerbottommargin=6pt,backgroundcolor=mycolor]{mybox}

\begin{document}

\title{Robust experimental data assimilation for the Spalart-Allmaras turbulence model}


\author{Deepinder Jot Singh Aulakh}
 \email{16djsa@queensu.ca}
\affiliation{%
 Department of Mechanical and Materials Engineering, Queen's University, 130, Stuart Street, Kingston, K7L 2V9, Ontario, Canada
}%
 \altaffiliation[Also at ]{Argonne Leadership Computing Facility, 240, Argonne National Laboratory, Lemont, 60439, Illinois, USA}
\author{Xiang Yang}
\affiliation{%
 Department of Mechanical Engineering,  Pennsylvania State University, E327 Westgate, University Park, 16801, Pennsylvania, USA
}%
\author{Romit Maulik}%
\affiliation{%
Information Sciences and Technology, Pennsylvania State University, E327 Westgate, University Park, 16801, Pennsylvania, USA
}%
\date{\today}
\begin{abstract}
This study presents a methodology focusing on the use of computational model and experimental data fusion to improve the Spalart-Allmaras (SA) closure model for Reynolds-averaged Navier-Stokes solutions. In particular, our goal is to develop a technique that not only assimilates sparse experimental data to improve turbulence model performance, but also preserves generalization for unseen cases by recovering classical SA behavior. We achieve our goals using data assimilation, namely the Ensemble Kalman filtering approach (EnKF), to calibrate the coefficients of the SA model for separated flows. A holistic calibration strategy is implemented via the parameterization of the production, diffusion, and destruction terms. This calibration relies on the assimilation of experimental data collected in the form of velocity profiles, skin friction, and pressure coefficients. Despite using observational data from a single flow condition around a backward-facing step (BFS), the recalibrated SA model demonstrates generalization to other separated flows, including cases such as the 2D NASA wall mounted hump (2D-WMH) and modified BFS. Significant improvement is observed in the quantities of interest, i.e., skin friction coefficient ($C_f$) and pressure coefficient ($C_p$) for each flow tested. Finally, it is also demonstrated that the newly proposed model recovers SA proficiency for flows, such as a NACA-0012 airfoil and axisymmetric jet (ASJ), and that the individually calibrated terms in the SA model target specific flow-physics wherein the calibrated production term improves the re-circulation zone while destruction improves the recovery zone.
\end{abstract}

\maketitle


\doublespacing
\section{Introduction} 
Reynolds-averaged Navier-Stokes (RANS) based simulations are extensively employed for the analysis of turbulent flows, primarily due to their ability to provide swift engineering insights owing to shorter turnover durations. RANS equations consist of time-averaged descriptions of the classical Navier-Stokes equations and are used for the predictive modeling of steady-state characteristics of turbulent flows. Within the RANS framework, instantaneous quantities are decomposed using the Reynolds decomposition into components representing time-averaged and fluctuating aspects. However, the presence of Reynolds stresses, which result from unclosed fluctuation terms, necessitates additional model specifications to achieve RANS closure. One notable closure model utilized extensively in aerospace applications is the Spalart–Allmaras (SA) model \cite{spalart1992one}. Despite its popularity, RANS solutions (using various closure models, including SA) are susceptible to inaccurate predictions in flow regimes involving separation and adverse pressure gradients. These errors primarily stem from the assumptions inherent in RANS models, which are valid for a limited range of flow scenarios.
\newline Despite increased computational power, the utilization of high-fidelity simulations, such as Direct Numerical Simulation (DNS) and Large Eddy Simulation (LES), remains constrained when addressing real-world problems \cite{yang2021grid,li2022grid}. As a result, enhancing the accuracy of Reynolds-Averaged Navier-Stokes (RANS) models continues to be an active area of research \cite{durbin2018some,bush2019recommendations}. Recently, there has been a surge in the application of machine learning (ML) and data-driven techniques to enhance closure models \cite{duraisamy2019turbulence}. The majority of investigations in this field concentrate on either substituting or enhancing the closure model using ML approaches \cite{tracey2015machine,yin2022iterative,ray2018robust,li2024enhancing,vadrot2023survey}. A recently popular method involves substituting the closure model with a trained ML model. In this context, a trained ML model, derived from either high-fidelity DNS data or RANS simulations, replaces the solution variables \cite{tracey2015machine,zhu2019machine,zhu2021turbulence,sun2022high}. While a model trained exclusively on RANS solutions might not lead to accuracy improvements, it holds implications for improving the convergence of the RANS solver, as observed by Maulik et al. \cite{maulik2021turbulent} and Liu et al. \cite{liu2022analysis}. Using DNS data, Ling et al. \cite{ling2016reynolds} introduced the Tensor Basis Neural Network (TBNN) to enhance the accuracy of the RANS solver. This approach employed high-fidelity DNS data to train the neural network (NN), utilizing a tensor combination technique originally proposed by Pope \cite{pope1975more}. Notably, TBNNs inherently uphold Galilean invariance and possess adaptability for capturing nonlinear relationships, thus adhering to some of the foundational principles proposed by Spalart et al. \cite{spalart2023old}.
Additionally, Wang et al. \cite{wang2017physics} introduced an ML model, aiming to learn the disparities between RANS and DNS data. To further enhance the convergence and stability of the RANS-ML model, techniques involving the decomposition of Reynolds stresses in linear and non-linear terms \cite{wu2018physics,wu2019reynolds} and the imposition of non-negative constraints were incorporated on linear terms \cite{mcconkey2022deep}.
\newline In the other approach, i.e., augmenting the closure models, a prominent approach involves the calibration of existing closure models by experimental or DNS data \cite{shirzadi2020rans,grabe2023data, kato2016optimization, kato2013approach}. Ray et al. \cite{ray2018robust,ray2016bayesian}, focused on calibrating three coefficients—namely, $C_{\mu}$, $C_{\epsilon 1}$, and $C_{\epsilon 2}$—within the $k-\epsilon$ model. Calibration was carried out using experimental data for the interaction of a compressible jet with a cross-flow. Notably, the outcomes of the calibrated RANS model demonstrated significantly closer alignment with experimental data in comparison to those obtained using nominal constants.  
\newline Additionally, the concept of field inversion has been extensively explored for the refinement of closure models \cite{duraisamy2015new,singh2017augmentation,singh2016using,yang2020improving}. Duraisamy et al. \cite{parish2016paradigm} and Chongyang et al. \cite{yan2022data} introduced modifications to the production term within the transport equation by incorporating a spatially variable factor. This approach was complemented by the incorporation of flow features as input for the ML model, thereby enhancing the generalizability of the modified model. 
\newline Bin et al. \cite{bin2022progressive} proposed the concept of progressive machine learning, arguing that machine-learned augmentations must not negatively impact the existing calibrations. Following this line of thinking, Bin et al. \cite{bin2023data} calibrated the SA turbulence model using experimental and DNS data. Their work aimed to achieve a more universally applicable improvement for various flow conditions. It involved replacing the SA model's coefficients with NNs trained through Bayesian optimization. Particularly noteworthy was the finding that the most significant enhancements were attributed to the destruction term within the model. {
Subsequently, Bin et al. {\cite{bin2024constrained}} extended their study to two-equation RANS closures, where they further identified importance of preserving the law of wall, which is likely to be disrupted with unconstrained calibration of ML-based models. This disruption could affect behavior of the calibrated model for cases outside the training dataset.}
\newline Recently, another data-driven method, that is data assimilation using Ensemble Kalman Filtering (EnKF) \cite{zhang2022ensemble,evensen2009data, kato2011integration}, has been explored for improving RANS closures. Zhang et al. \cite{zhang2022ensemble} employed EnKF technique to train TBNN originally introduced by Ling et al. \cite{ling2016reynolds}. The application of EnKF for TBNN training exhibited a performance akin to the original study. However, a noteworthy advantage emerged: the capacity to employ sparse and noisy data to effectively train TBNNs. Moreover, EnKF was effectively employed in an online manner, facilitating real-time training of TBNNs using indirect measurements.
\newline Kato et al. \cite{kato2011integration}, used EnKF to integrate experiments with CFD. Surface pressure data from experiments was used to improve accuracy of the simulations.  Yang and Xiao \cite{yang2020improving}, utilized a regularized EnKF approach to enhance the transition model. Experimental data was used to calibrate transition location within the model, leading to notable improvements. Kato and Obayashi \cite{kato2013approach}, used simulation data from  flow over a flat plate to estimate parameters of SA model. The EnKF approach applied in this case resulted in parameters, whose accuracy was consistent with that of the original parameters. Additionally, Kato et al. \cite{kato2016optimization} used EnKF to improve the modified Menter $k-\omega$ model. In contrast with aforementioned studies, our approach performs an optimization of a single parameter $(a)$, which results in improved accuracy for flows with separation and adverse pressure gradients.
\newline Our research \emph{revisits the calibration of the SA turbulence model while utilizing sparse and noisy experimental data}. To achieve this, EnKF is employed to calibrate the coefficients of the SA model. Furthermore, the current study comprehensively employs calibration, encompassing all elements such as production, diffusion, and destruction terms. The calibration process is framed as an inverse problem, wherein iterative corrections of the SA coefficients are performed within an EnKF-based loop. The focus of this work is to harness sparse and noisy experimental data for the calibration of the SA model in scenarios involving separated flows. Specifically, the coefficients are calibrated using the backward-facing step (henceforth denoted BFS1) configuration \cite{driver1985features}, while the subsequent testing encompasses the 2D-WMH case \cite{seifert2002active} and a modified backward-facing step (denoted BFS2) scenario with altered step height \cite{bin2023data,barri2010dns}. To determine that the calibrated model is not detrimental to attached and unbounded flows, tests were also done for flow around an airfoil and zero pressure gradient boundary layer.

\section{Background} \label{SA_Model}

The SA model was proposed in 1992 by Spalart et al. \cite{spalart1992one} and remains a workhorse for aerospace design using RANS. The model takes into account the convection, diffusion, production, and destruction of the eddy viscosity ($\nu_t = f_{v1} \tilde \nu$) in a single transport equation as follows: 
\begin{equation}
 \frac{D \tilde \nu}{Dt} = \underbrace{C_{b1} \tilde S \tilde \nu}_{Production} - \underbrace{C_{w1}f_{w} \left( \frac{\tilde \nu}{d} \right)^2}_{Destruction} + \underbrace{{1 \over \sigma} \{ \nabla . [(\nu +\tilde \nu) \nabla \tilde \nu] +C_{b_2} |\nabla \tilde \nu|^2 \}}_{Diffusion}, 
       \label{SA_eq}
\end{equation}
\newline where, the coefficients $C_{b1}$, $\sigma$, and $C_{b2}$ take the values 0.1355, 0.666, and 0.622, respectively.  $C_{w1}$ and $f_w$ are given by eqs. \ref{cw_1} and \ref{fw_1}, respectively as follows.
\begin{equation}
C_{w1} = \frac{C_{b1}}{\kappa^2} +\frac{1+C_{b2}}{\sigma},
       \label{cw_1}
\end{equation}
where, $\kappa = 0.41$ is von Karman constant.
\begin{equation}
f_{w} = g\left(\frac{1+C_{w3}^6}{g^6 + C_{w3}^6}\right), \quad g = r + C_{w2}(r^6-r),
       \label{fw_1}
\end{equation}
where, $C_{w3} = 2 $ and $C_{w2} = 0.3$. $\tilde S $ is modified strain rate ($S$) of the mean velocity field:
\begin{equation}
\tilde S = S + \tilde \nu \frac{f_{v2}}{\kappa^2 d^2}
       \label{s_tilde}
\end{equation}
where, $d$ is the distance from nearest wall and $f_{v2}$ is dependent on $\chi = \tilde \nu / \nu$, $C_{v1} = 7.1$, and $f_{v1}$ as follows:  
\begin{equation}
f_{v2} = 1 - \frac{\chi}{1+ \chi f_{v1}}, \quad f_{v1} = \frac{\chi^3}{C_{v1}^3+\chi^3}.
       \label{fv2}
\end{equation}
The SA model provides high accuracy for equilibrium flows, but fails to accurately capture separation and recovery in non-equilibrium wall-bounded separating flows. Our study aims to address this drawback by using a data-assimilation-based calibration of the SA model for separating flows. Moreover, we use sparse experimental data to enable this calibration.
\newline The constraints for the calibration in this study are summarised as follows:
\begin{enumerate}
\item Our methodology must utilize noisy and sparse experimental data for calibration.
\item Our calibrated model must generalize, i.e., the data from one type of separating flow at a single Reynolds number ($Re$), boundary condition, and geometry should be enough to improve the accuracy of the model in other separating flows. 
\item The calibration should not distort the model's original behavior in equilibrium flows.
\end{enumerate}
\section{Methodology}
\subsection{Ensemble Kalman Filtering for calibration} \label{enkf} Ensemble Kalman Filtering (EnKF) is commonly used in data assimilation to aid in the estimation of system states, such as velocity and pressure in a flow field, given sparse observational data \cite{zhang2022ensemble}. In the current study, we use EnKF for calibrating the SA model. Before going into further details about calibration, we first introduce the observation matrix $(H)$ free implementation of EnKF used in the current study \cite{mandel2006efficient}:
\begin{subequations}
\begin{equation}
X^p = X + \frac{1}{N-1}A(HA)^TP^{-1}(D-HX),
       \label{Enkf_1}
\end{equation}
where:
\begin{equation}
P = \frac{1}{N-1}(HA)(HA)^T +R,
       \label{Enkf_2}
\end{equation}
\begin{equation}
A = X-E(X), \quad HA = HX-E(HX)
       \label{Enkf_3}
\end{equation}
\end{subequations}
Here, matrix $X$ is denoted the prior ensemble and $X^p$ represents the posterior ensemble. Both matrices have dimensions $n \times N$. The value $n$ pertains to the number of coefficients selected for calibration from eq. \ref{SA_eq} and $N$ refers to the number of members in the ensemble. To exemplify, if we consider the instance of selecting $C_{b1}$, $C_{b2}$, and $\sigma$, then $n$ would be equal to 3. $D$ is an $m \times N$ matrix containing experimental data, where $m$ signifies the number of probe points. Additionally, $R$ represents the covariance matrix of random noise in $D$. The central objective revolves around employing the coefficients derived from the SA model as elements of matrix $X$ while utilizing the data matrix $D$ to refine and calibrate these coefficients. The outcome is the matrix $X^p$, which encapsulates the calibrated coefficients. This process aligns with the broader aspiration of refining understanding and enhancing the accuracy of the SA model through the integration of both theoretical insights and empirical observations. Notably, for covariance reduction an identity matrix is added to the eq. \ref{Enkf_2}. This covariance reduction considerably reduces the required $N$, hence reducing the computation cost. The trade-off here is an increase in the stochastic nature of the optimization. This was deemed to be acceptable for the current application.  
\newline The selection of $HX$ is contingent on the available experimental data. For instance, if experimental data for velocity is available, the $HX$ will be the velocity obtained after solving the RANS equations by using $X$ as the SA coefficients. It should be noted that the $HX$  will be formulated by only extracting the locations where the experimental data is available. This is illustrated in figure \ref{Enkf_loop}c, where the values are extracted along the lower wall of the BFS.
\subsection{Calibration loop} \label{cal_loop_secion}
\begin{figure}
\hspace{-5pt}
\includegraphics[width=15cm]{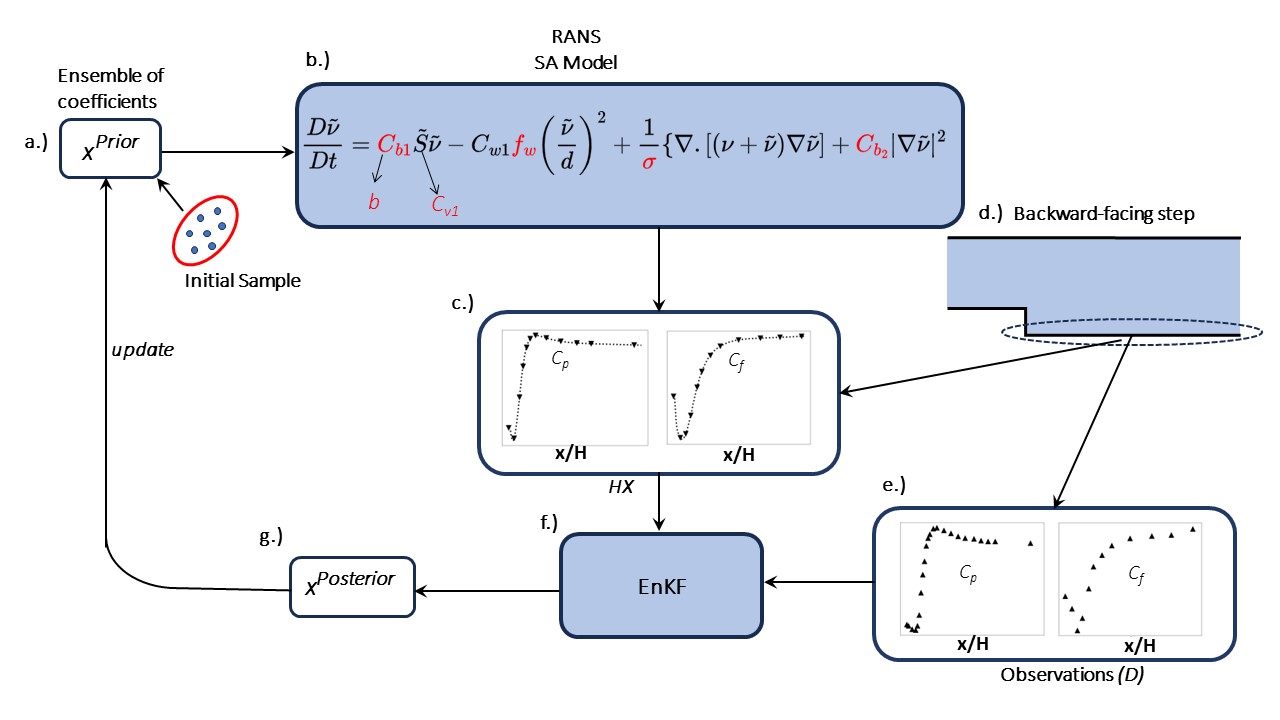}
\caption{EnKF calibration loop, a.) An ensemble ($X$) of the SA coefficients - for the first iteration $X$ is sampled from an initial distribution determined by a parametric analysis in \ref{senstivity}. b.) The RANS solver and SA model calculate the flow variables based on the SA coefficients in $X$, the coefficients used in the ensemble are highlighted in red, and the coefficient $b$ is defined later. c.) The extracted quantities of interest (QOIs) from {\tt OpenFOAM } at specified locations serve as the $HX$ for the EnKF. d.) The QOI in the current study are $C_f$ and $C_p$ along the bottom wall. e.) $D_{m\times N}$ matrix for the available data for QOI, $m$ is number of probe points $N$ is the ensemble size picked from the Gaussian distribution. g.) Updated $X^p$ based on the EnKF in f.}
\label{Enkf_loop}
\end{figure}

The iterative EnKF calibration loop used in the current study is outlined as follows:
\begin{enumerate}
\item Sample SA coefficients $X$ based on an initial prior distribution (figure \ref{Enkf_loop}a). The initial distribution's upper and lower bounds are determined after undertaking a parametric analysis (\ref{senstivity}) of the SA model for a flow over the BFS case.
\item The $X$, i.e. the sampled coefficients, are used to obtain evaluations for $HX$. The observation matrix $H$ encompasses a RANS simulation and the extraction of the quantity of interest (QOI) at a given location in the domain (figure \ref{Enkf_loop} b and \ref{Enkf_loop}c). The locations are dictated by the available experimental data for QOI. In the current case, the QOI are the friction ($C_f$) and pressure ($C_p$) coefficients, available at the lower wall downstream of the step in BFS flow (figure \ref{Enkf_loop}d).
\item The $HX$ is further substituted into eq. \ref{Enkf_1} to obtain posterior ensemble $X^p$. The $X^p$ serves as an ensemble distribution for the next iteration.  
\end{enumerate}
EnKF possesses desirable attributes such as the ability to accommodate noisy data, quantification of uncertainty, and enabling gradient-free optimization. Notably, the gradient-free optimization not only streamlines the EnKF's implementation process but also endows it with heightened adaptability for integration with diverse computational fluid dynamics solvers, thereby reducing the necessity for intrusive modifications of its source code.
\newline In addition, the $HX$ can be easily modified to use any QOI for which the calibration data is available. This customization process requires the (usually straightforward) extraction of the requisite QOI directly from the solution field.
\newline Moreover, the EnKF's proficiency in effectively managing noisy data aligns well with experimental data that has inherent measurement variabilities. This particular attribute underscores the EnKF's suitability for assimilating real-world experimental observations into the calibration process, thereby enhancing its efficacy in bridging theoretical models with empirical data.

\subsection{Calibration data (D)} \label{calibration_data} 
\begin{figure}
\centering
\includegraphics[width=12cm]{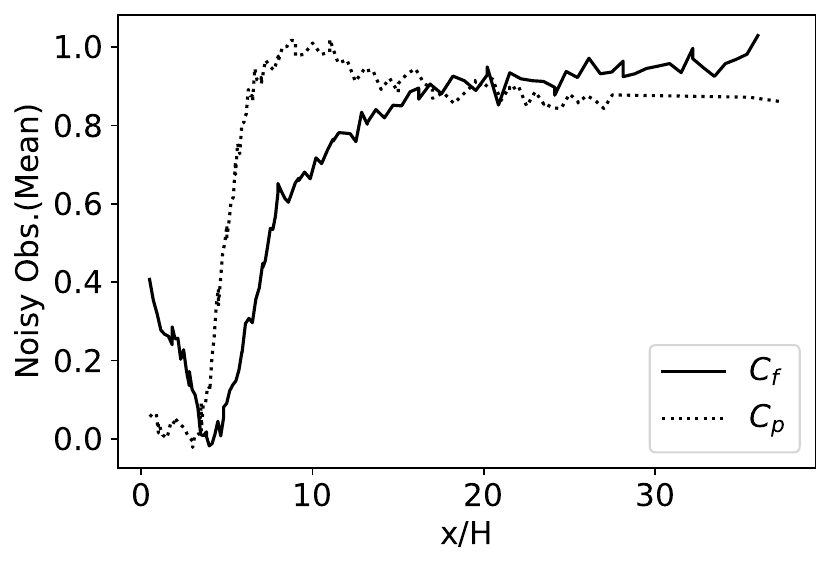}
\caption{Mean of the calibration data $D$, used for current study. The values of $C_f$ and $C_p$ are plotted along the bottom wall of BFS.  The baseline data was obtained from the NASA turbulence repository \cite{nasa} and a noise $\epsilon \sim  \mathcal{N}(0, \sigma_D = 0.05)$ is added to formulate the $D$ matrix.}
\label{noisy_obs}
\end{figure}

We begin this section by noting that the data-driven calibration proposed in this research relied solely on experimental data. This is in contrast to similar studies that relied on DNS data for improving RANS models \cite{bin2023data}. The requirement of only using experimental data presented challenges since the available data covered only a limited portion of the domain compared to DNS. This aligns with common data acquisition practices, as data is typically gathered predominantly along the walls, making it sparse, and it naturally contains some level of noise in the readings. Consequently, the calibration process proposed necessitated a robust handling of sparsity and noise inherent in the experimental data.
\newline The experimental data for the current study is from Driver and Seegmiller \cite{driver1985features} and is retrieved the from NASA turbulence repository \cite{driver1985features,nasa}. The data consists of $C_f$ and $C_p$  measurements along the bottom wall of the BFS downstream of the step. The data was interpolated to 112 locations along the bottom of the wall, i.e. $m = 112$. Notably, the magnitude of $C_p$ is approximately 3 orders of magnitude higher than the $C_f$. This difference in magnitude can cause the calibration to be weighted more towards $C_p$. Hence, the values of both $C_p$ and $C_f$ are separately scaled between $(0,1)$. Furthermore, a normally distributed noise $\epsilon \sim  \mathcal{N}(0, \sigma_D = 0.05)$, is added to the experimental data, and $N = 5$ data vectors are sampled to formulate our $D$ matrix. Figure \ref{noisy_obs} shows the mean value of observations.
\subsection{Ensemble matrix (X)} \label{ensemble_matrix} To reiterate, the $X$ matrix for this study is formulated by coefficients of the SA model. A parametric analysis outlined in \ref{senstivity} was performed to select the SA coefficients for calibration, namely, $C_{b1}, C_{w2}, C_{w3}, \sigma $ and $C_{v1}$ were selected. The parametric space for ensemble members is: $C_{w2} \in [0.75, 1.75]$, $C_{w3} \in [1,2]$, $\sigma \in [0.3,2]$, and $C_{v1} \in [6,9]$.
\newline In the original SA, $C_{b1}$ has a value of 0.1355 for the entire domain. However, after an initial study (\ref{senstivity2}) it was found that implementing a varying $C_{b1}$ yields more flexibility in the model and may aid in improved calibration. Hence, the $C_{b1}$ is further parameterized in terms of non-dimensional field $r$ as follows:
\begin{equation}
C_{b1} = min\left( max\left(r*b,c*r + d\right), 0.1355 \right)
\label{cb_1_eqn}
\end{equation}
where, $ r \equiv \nu_t/Sk^2d^2$, 0.1355 is the original value of $C_{b1}$.  The parameter $b \in [1.5,2.5]$ is also added to the ensemble matrix of the EnKF. To summarize, there are a total of five parameters that are calibrated using EnKF in this study.
\newline It is important to highlight that the initial decision involved parameterizing $C_{b1}$ as $min(r*b, 0.1355)$. While this parameterization yielded enhanced accuracy in the targeted flows of interest, it had the drawback of disabling the SA model when $r\approx 0$, for free shear flows. Consequently, the parameterization was modified for the inclusion of $c * r + d$ in equation \ref{cb_1_eqn}(figure \ref{fw_cb}a) to ensure the retention of $C_{b1}$ values for $r \approx 0$, thereby preserving the model's applicability in free shear flows. Additionally, to mitigate the expansion of the parameter space for EnKF calibration, a decision was made to fix the values of $c$ and $d$ at -50 and 0.2, respectively. These fixed scalars effectively maintain the SA model's behavior by linearly increasing the $C_{b1}$ near $r\approx 0$  while yielding comparable behavior for the targeted flows as achieved by $min(r * b, 0.1355)$. It should be noted that equation \ref{cb_1_eqn} was used for calibration.
\subsection{Calibration metrics} \label{calibration_metrics} 
\begin{figure}
\centering
\includegraphics[width=12cm]{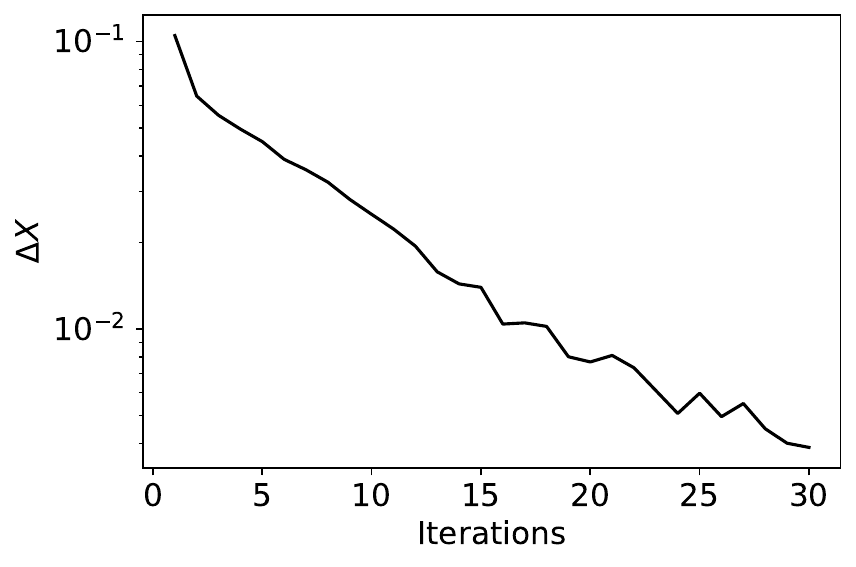}
\caption{$\Delta X$ vs. iterations for the calibration loop.}
\label{residuals}
\end{figure}

\begin{figure}
\hspace{-40pt}
\subfloat[]{\includegraphics[height=6.4cm]{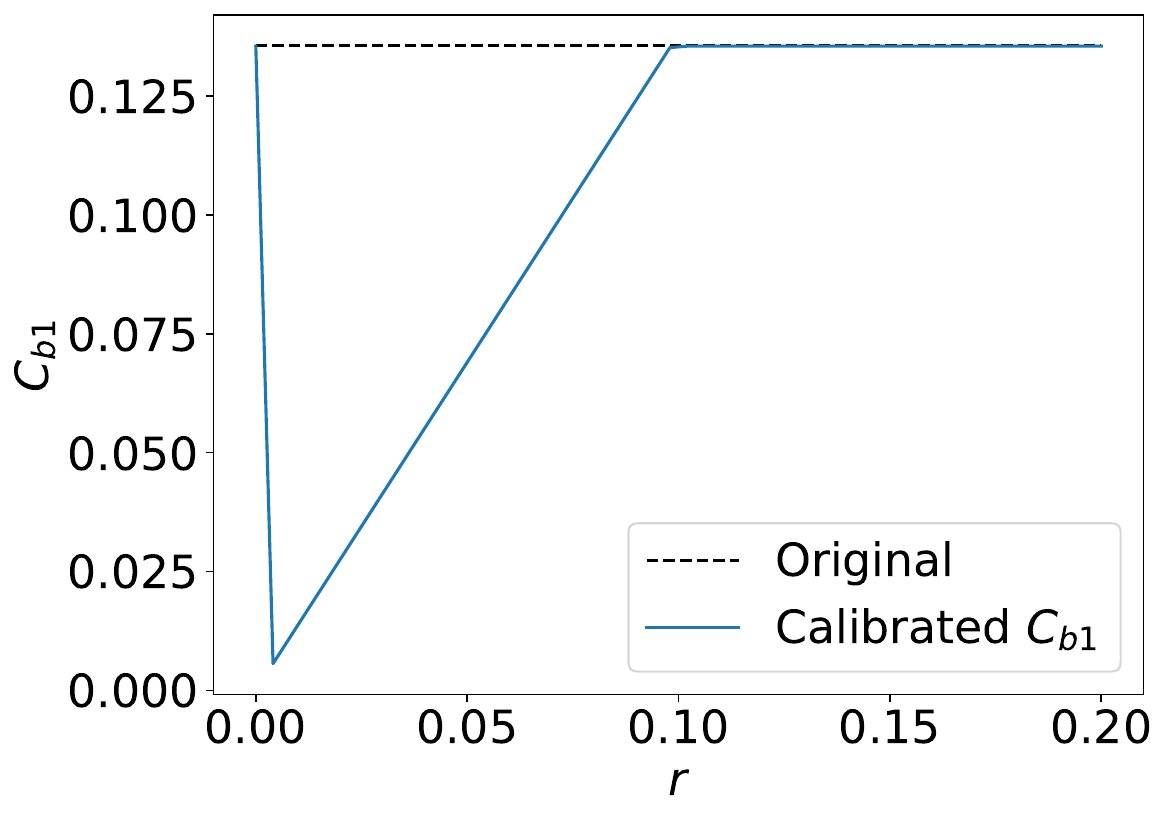}}
\subfloat[]{\includegraphics[height=6.4cm]{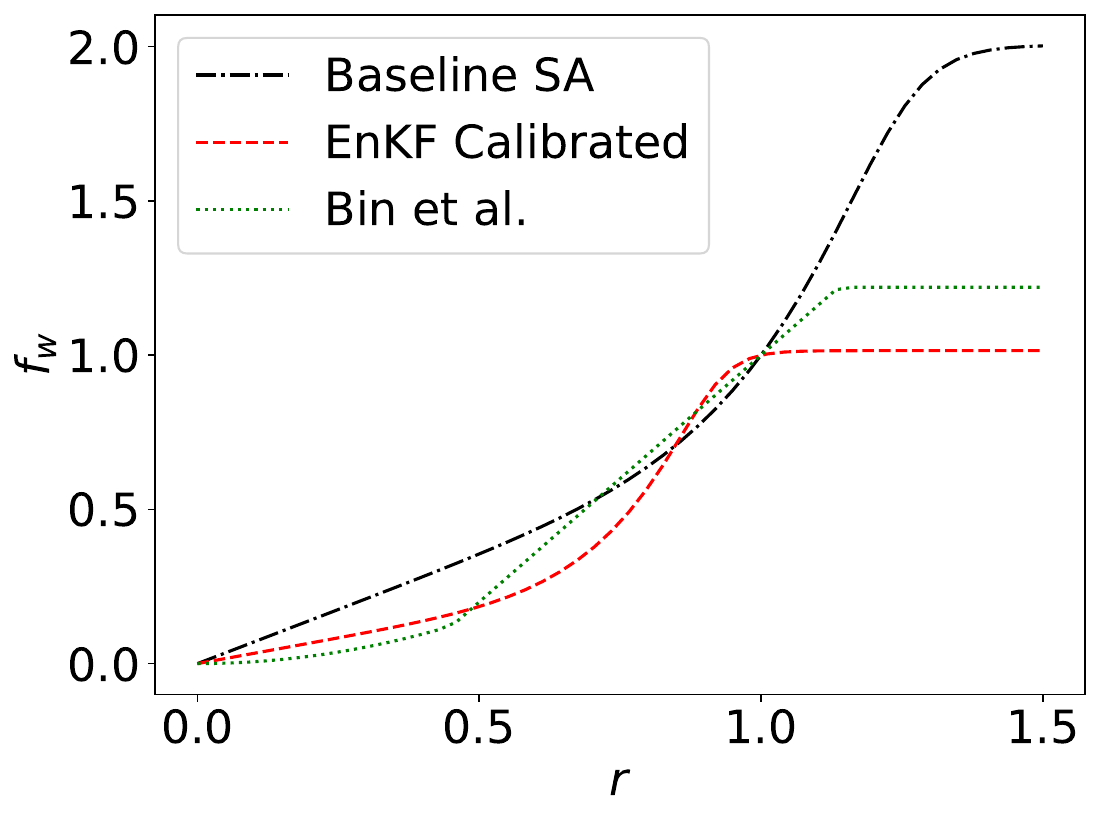}}
\caption{a. $C_{b1}$, b. $f_w$ vs. $r$ obtained from the calibrated coefficients. The original values are also plotted in corresponding plots. For additional comparison, the $f_w$ obtained by Bin et al. \cite{bin2023data} is also plotted. }
\label{fw_cb}
\end{figure}

The $X$ matrix is calibrated in an iterative manner as discussed in section \ref{cal_loop_secion}. The performance of the calibration is monitored using the change in $X$ each iteration, specifically, as $\Delta X^i = mean(|X^i-X^{i-1}|)$ for $i^{th}$ iteration. The EnKF loop is run for 30 iterations. Figure \ref{residuals} shows $\Delta X$ vs. iterations. The $\Delta X^{30} =0.0038$ at the end of the EnKF deployment is considered acceptable for convergence. This is also evident from \ref{senstivity}, where such small changes in $X$ do not result in any variation of the QOI. 
\newline The mean of the members of the posterior $X^p$ is obtained as: $b = 1.39$, $ \sigma = 0.97$, $C_{w2} = 0.78$, $C_{w3} = 0.67$, and $C_{v1} = 8.24$. The plots of $C_{b1}$ and $f_{w}$ using the calibrated coefficients are given in figure \ref{fw_cb}a and \ref{fw_cb}b.  In figure \ref{fw_cb}b, the mapping (NN) of $f_w$ from the study of Bin et al. \cite{bin2023data} is also compared to the current mapping. From here it can be concluded that the $f_w$ in Eq \ref{fw_1} provides enough flexibility to learn new mappings by changing $C_{w2}$ and $C_{w3}$.
\newline In terms of computational costs, each of the 30 iterations utilized 5 different sets of SA coefficients, resulting in 150 CFD runs on a mesh with approximately 13,000 cells for the BFS. The calibration was parallelized by assigning one core to each of the five CFD simulations. Although the parallelization used in this study is relatively straightforward, improved load-balancing methods can be employed to accelerate convergence.
\section{Results} 
In this section, the calibrated model is tested for various flows. The tested flows are classified into three categories: separated, attached, and unbounded flows. The calibrated model is evaluated for improved performance on separating flows while retaining the same good performance in attached and unbounded flows.
\subsection{Separated flows} 
In this section, three separated flows are analyzed namely, BFS \cite{driver1985features}, 2D-WMH \cite{seifert2002active}, BFS2 (changed step height) \cite{barri2010dns,bin2023data}. 
\subsubsection{Flow over a BFS} \label{flow_over_bfs} 
\begin{figure}
\begin{center}
\subfloat[]{\includegraphics[height=6.7cm]{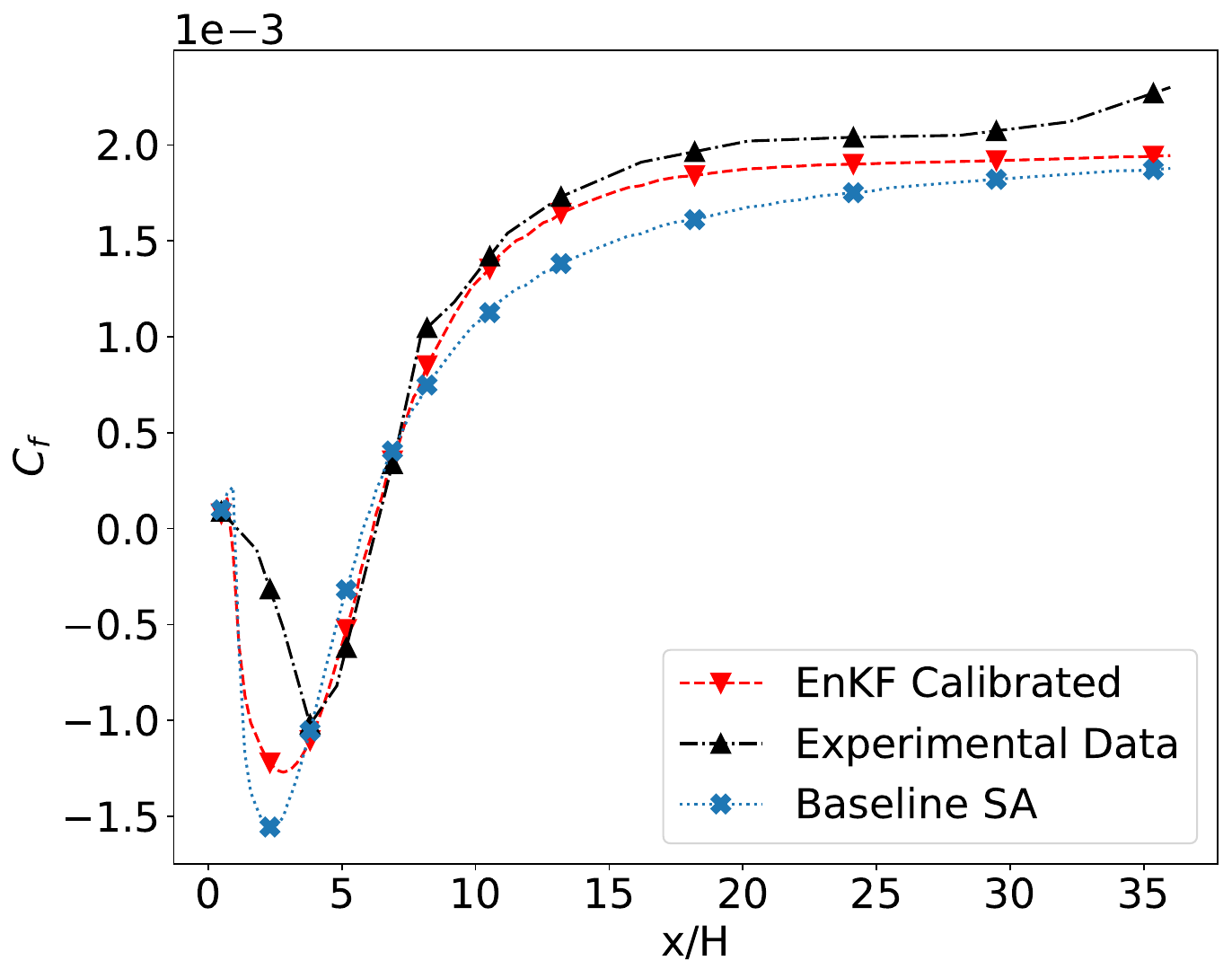}}
\subfloat[]{\includegraphics[height=6.7cm]{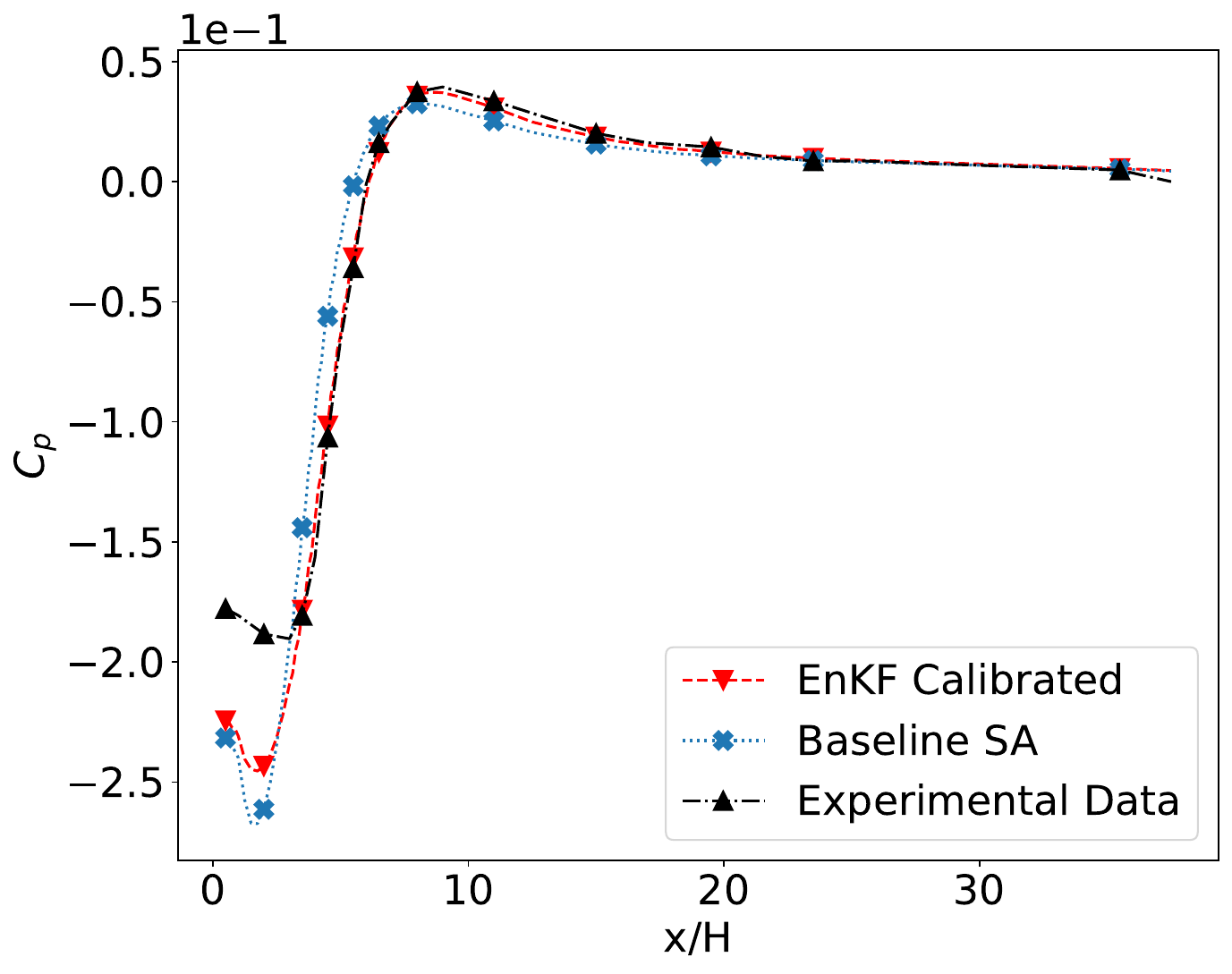}}
\end{center}
\caption{BFS: a.) $C_f$, b.) $C_p$ vs. $x/H$. The calibrated SA shows improvement in both the recovery zone and separation bubble as compared to the baseline model. The improvement in this case is defined as proximity to the experimental results.}
\label{bfs_cp_cb}
\end{figure}

\begin{figure}
\centering
\includegraphics[height=8cm]{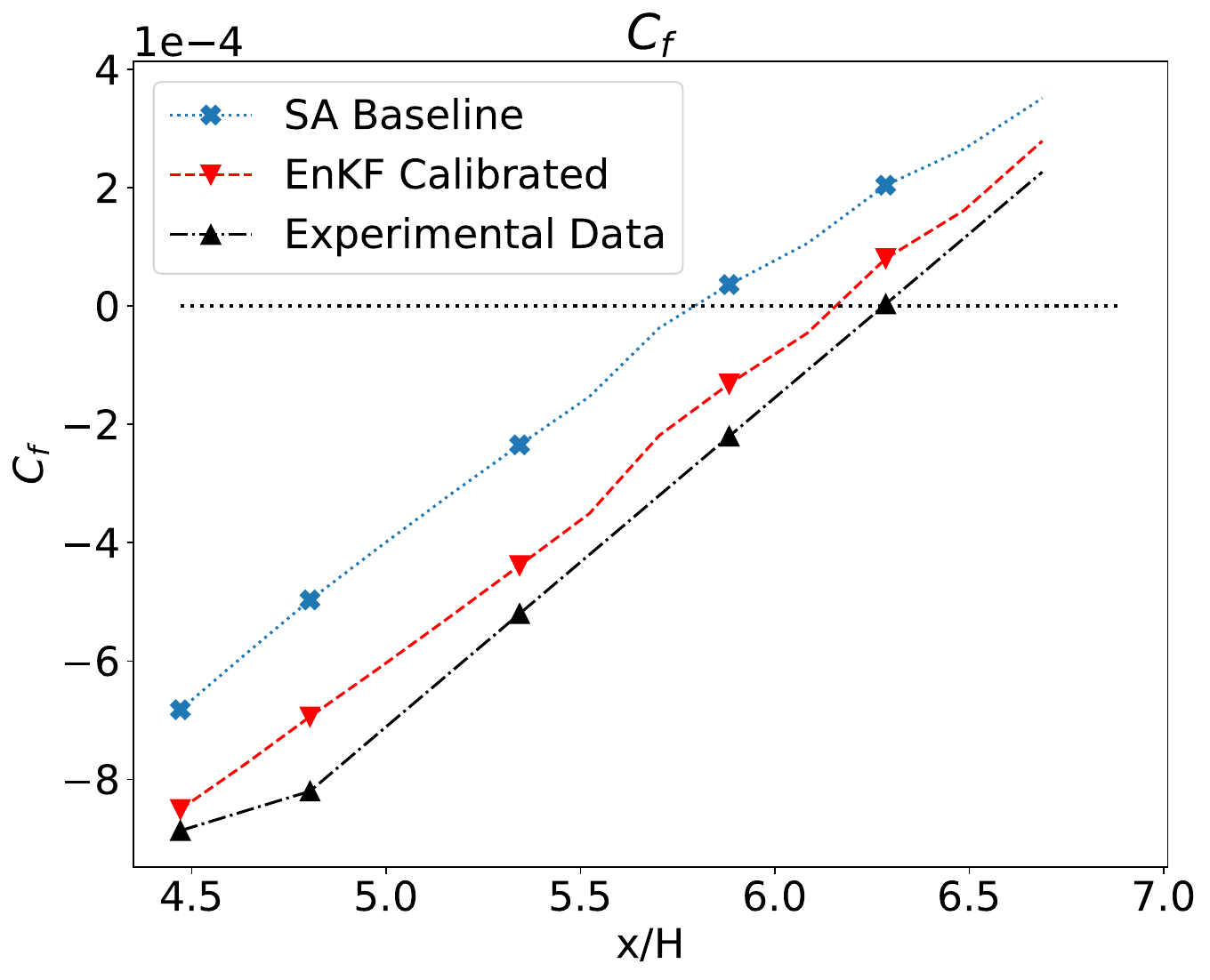}
\caption{BFS: $C_f$ vs. $x/H$ showing the improved reattachment location ($C_f = 0$) for the calibrated model.}
\label{bfs_reattach}
\end{figure}

\begin{figure}
\hspace{-50pt}
\subfloat[]{\includegraphics[height=6.7cm]{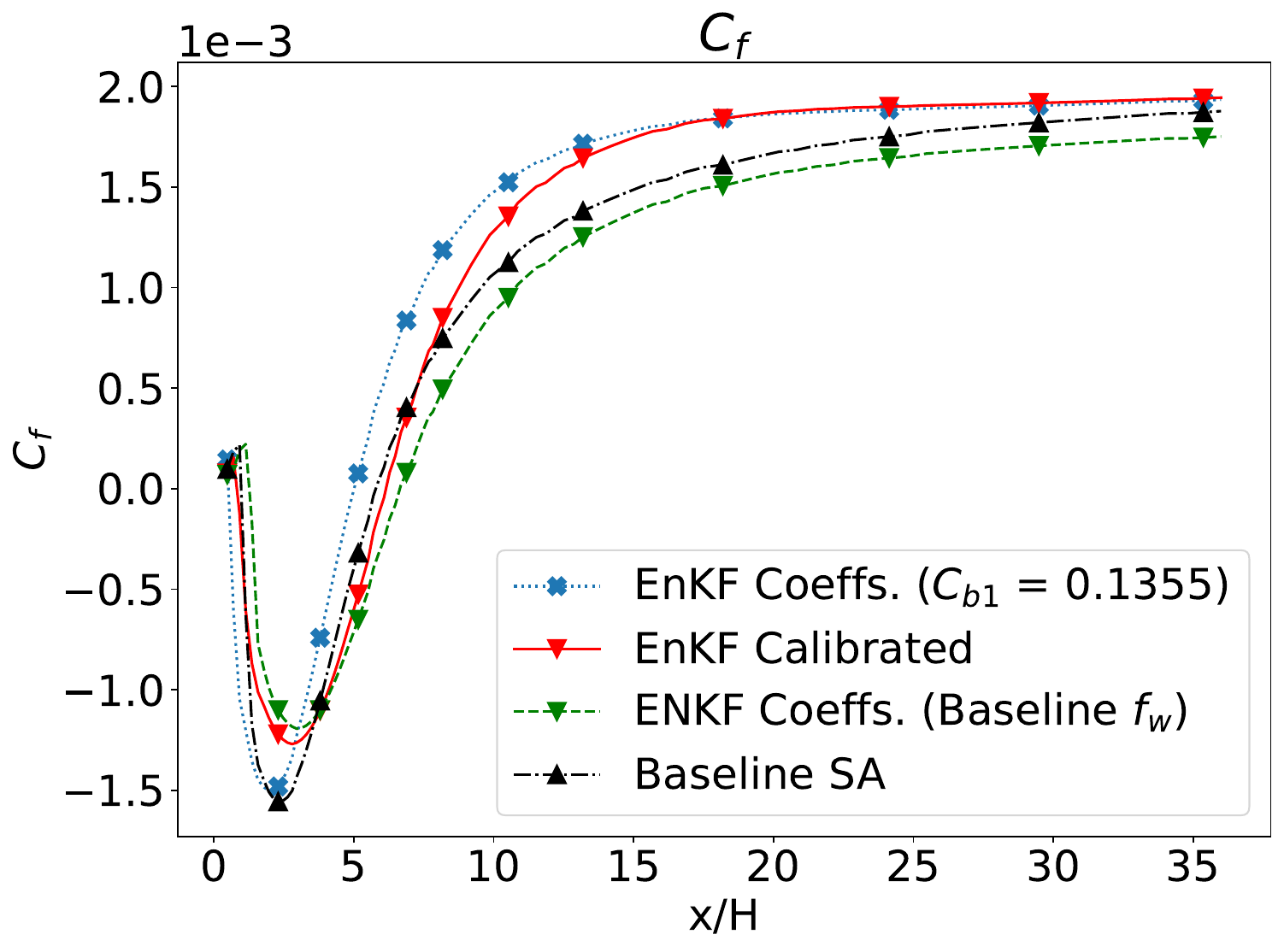}}
\subfloat[]{\includegraphics[height=6.7cm]{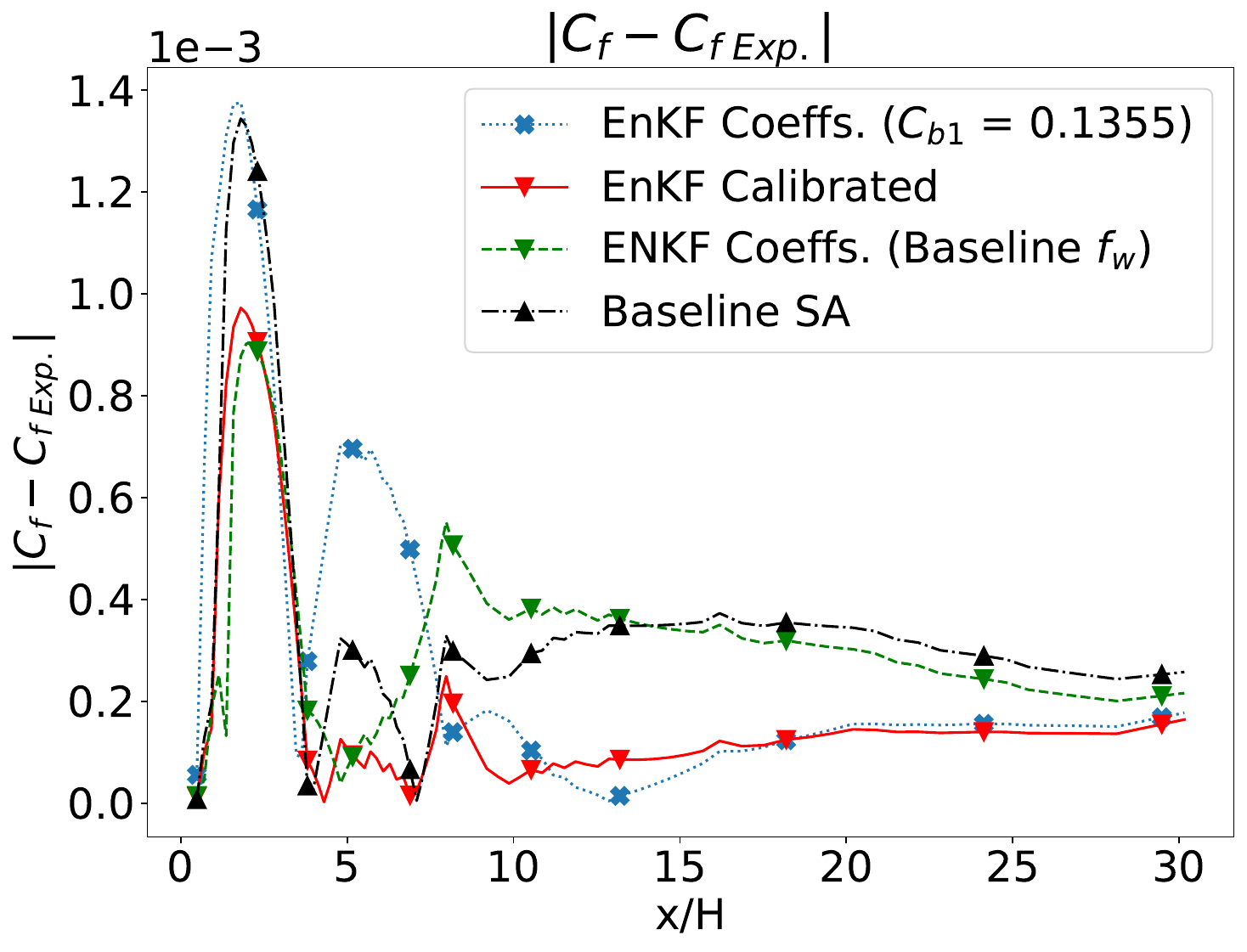}}
\caption{BFS: a.) $C_f$ vs. $x/H$ showing the impact of $C_{b1}$ and $f_w$ individually on the calibration results. If baseline $C_{b1}$ is used along with the rest of the calibrated values (dotted blue), the results only show improvement in the recovery zone, while remaining almost similar in the separation bubble. On the other hand, if baseline $f_w(C_{w2}, C_{w3})$ is used along with the rest of the calibrated coefficients (dashed green) the improvement is mainly observed in the separation zone while the results in recovery zone remain close to the baseline SA. b.) Absolute error w.r.t  the experimental data for each configuration vs. $X/H$.}
\label{bfs_1355}
\end{figure}

\begin{figure}
\begin{center}
\subfloat[$C_{b1}$]{\includegraphics[width=13cm]{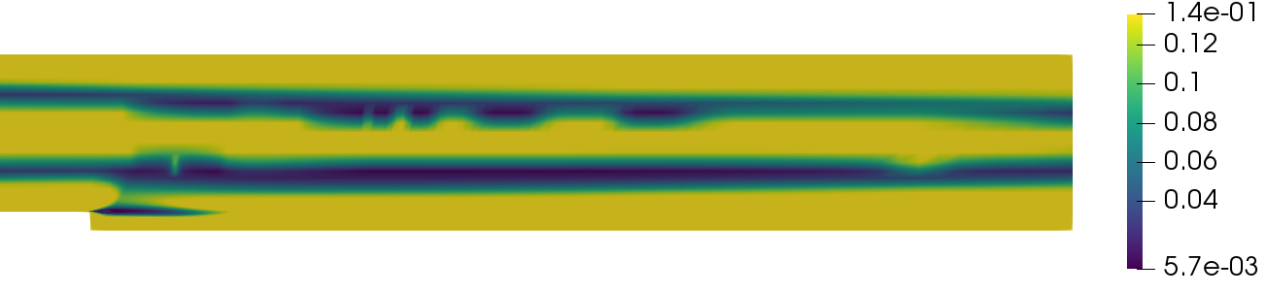}}\\
\subfloat[$f_w$]{\includegraphics[width=13cm]{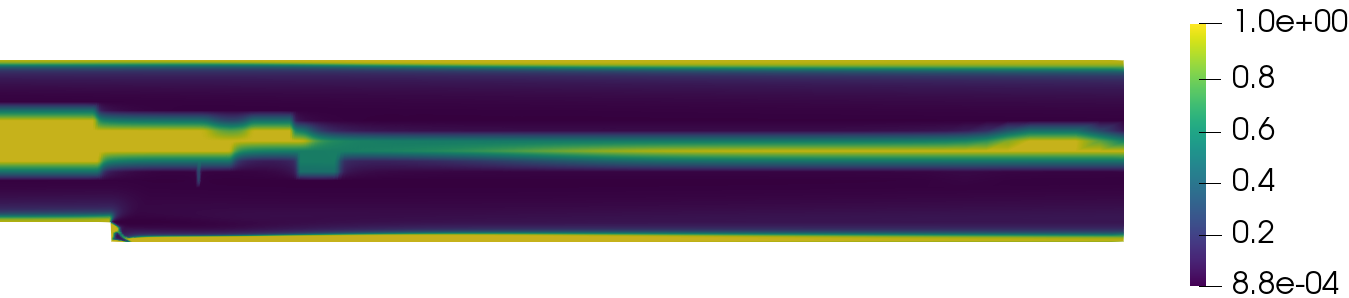}}
\end{center}
\caption{BFS: Variation of $C_{b1}$ and $f_w$ in the domain. Note that the original $C_{b1}$ possesses a constant value throughout the domain.}
\label{cont}
\end{figure}

In this section, the results for the flow over BFS are outlined. This flow was also used to calibrate the SA model using a relatively coarser mesh ($\approx 13000$ cells). The testing was done on multiple finer meshes (Mesh1-4), $\approx 28000, 43000, 47000,$ and $53000$ cells. These test cases also serve as a good indicator for the soft evaluation of the calibration across different meshes. Here, mesh 1 (28000 cells) is tested, while appendix \ref{meshes} covers the rest of the test cases.
\newline Figure \ref{bfs_cp_cb} compares the $C_f$ and $C_p$ plots from the calibrated SA model with that of the original SA; experimental data are also plotted for the purpose of comparison. It can be seen that the calibration significantly improves the results for the SA model, i.e., the results are closer to the experimental data used for calibration. Figure \ref{bfs_cp_cb}a shows the $C_f$ in the recovery zone is more accurately predicted by the calibrated model. In addition, there is a significant improvement in the magnitude of the $C_f$ in the separation bubble. Similar improvements are also observed in $C_p$ figure \ref{bfs_cp_cb}b. The reattachment length in the calibrated model is also closer to the experimental values as shown in figure \ref{bfs_reattach}.  
\newline Further analysis suggested that each coefficient of calibrated SA impacted improvement in results in a very specific manner. Notably, $C_{b1}$ and $f_w(C_{w2},C_{w3})$ influenced separation and recovery zone respectively. The effect is more evident in $C_f$, hence figure \ref{bfs_1355} shows only the results for $C_f$. 
\newline As shown in figure \ref{bfs_1355}, if the baseline value of $C_{b1} = 0.1355$ was used while using the calibrated values for the rest of the coefficients the improvement is mainly observed in the recovery zone, while the separation zone remains similar in magnitude to the baseline SA model. On the contrary, if baseline $f_w(C_{w2}, C_{w3})$ is used in combination with the rest of the calibrated coefficients the separation zone improves while the recovery zone remains closer to the baseline SA. However, the best results are obtained by using all the calibrated coefficients. These results are consistent with the observation of Bin et al. \cite{bin2023data} who used $f_w$ as NN to calibrate the SA model while keeping the rest of the values identical. They also observed an improved recovery zone with a slight reduction of accuracy in the separation zone. 
\newline From figure \ref{bfs_1355}, it is evident that the $C_{b1}$ and $f_{w}$ have a domain-specific effect. In order to gain a further understanding of how these coefficients vary throughout the domain, we provide contour plots in Figure \ref{cont}. The $C_{b1}$ takes particularly low values near the separation zone. It can also be observed that the trend of $C_{b1}$ and $f_w$ is almost opposite to each other throughout the domain. This opposite trend shows correlation between $C_{b1}$ (production) and $f_w$ (destruction). We may also gather that instead of putting the onus of balancing region-based production and destruction solely on the $f_w$ as in the baseline model, the current formulation involves $C_{b1}(r)$ working with $f_w$ to balance these quantities. 
\subsubsection{2D-WMH} 
\begin{figure}[h!]
\subfloat[]{\includegraphics[height=6.8cm]{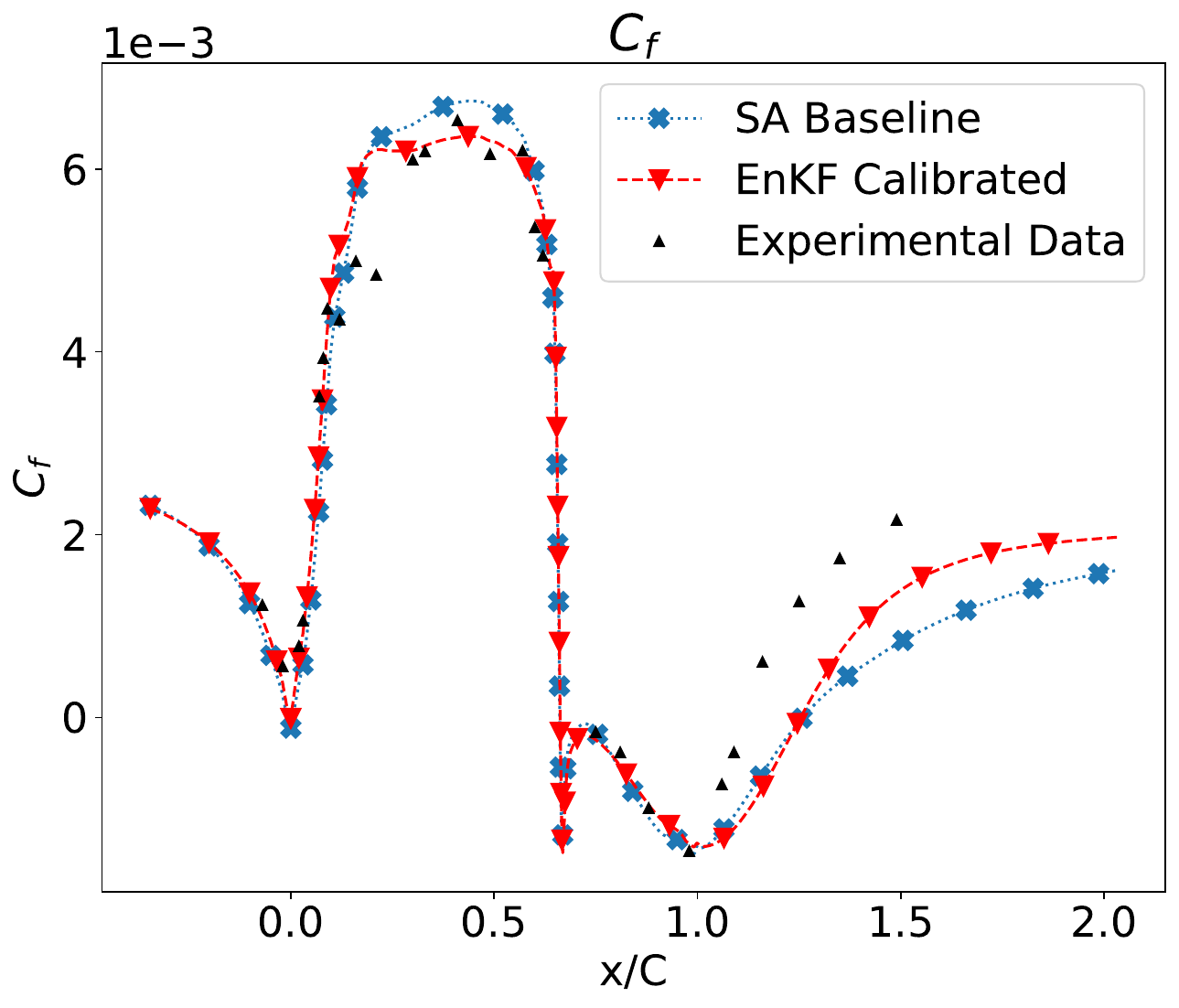}}
\subfloat[]{\includegraphics[height=6.8cm]{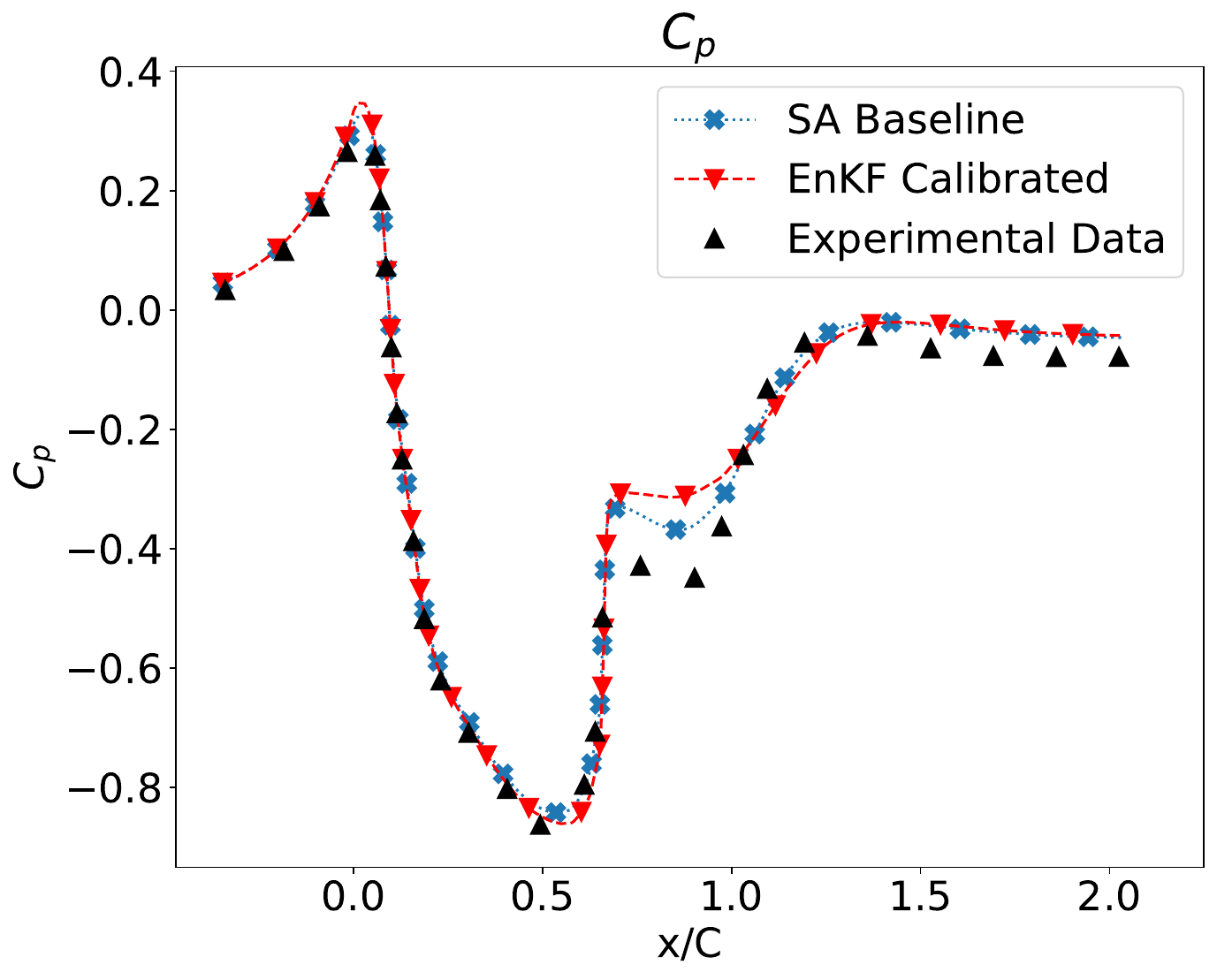}}\\
\subfloat[]{\includegraphics[height=6.8cm]{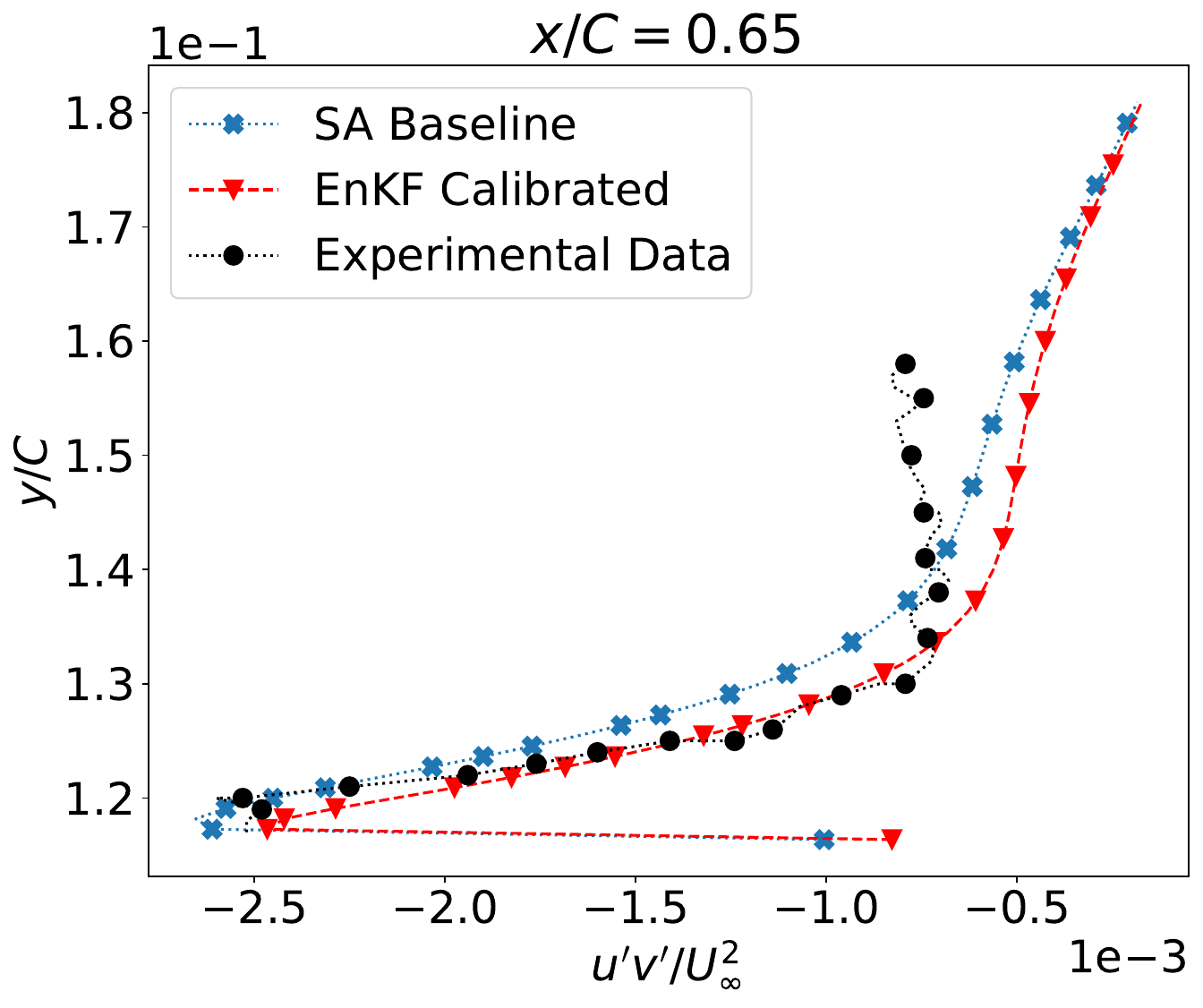}}
\subfloat[]{\includegraphics[height=6.8cm]{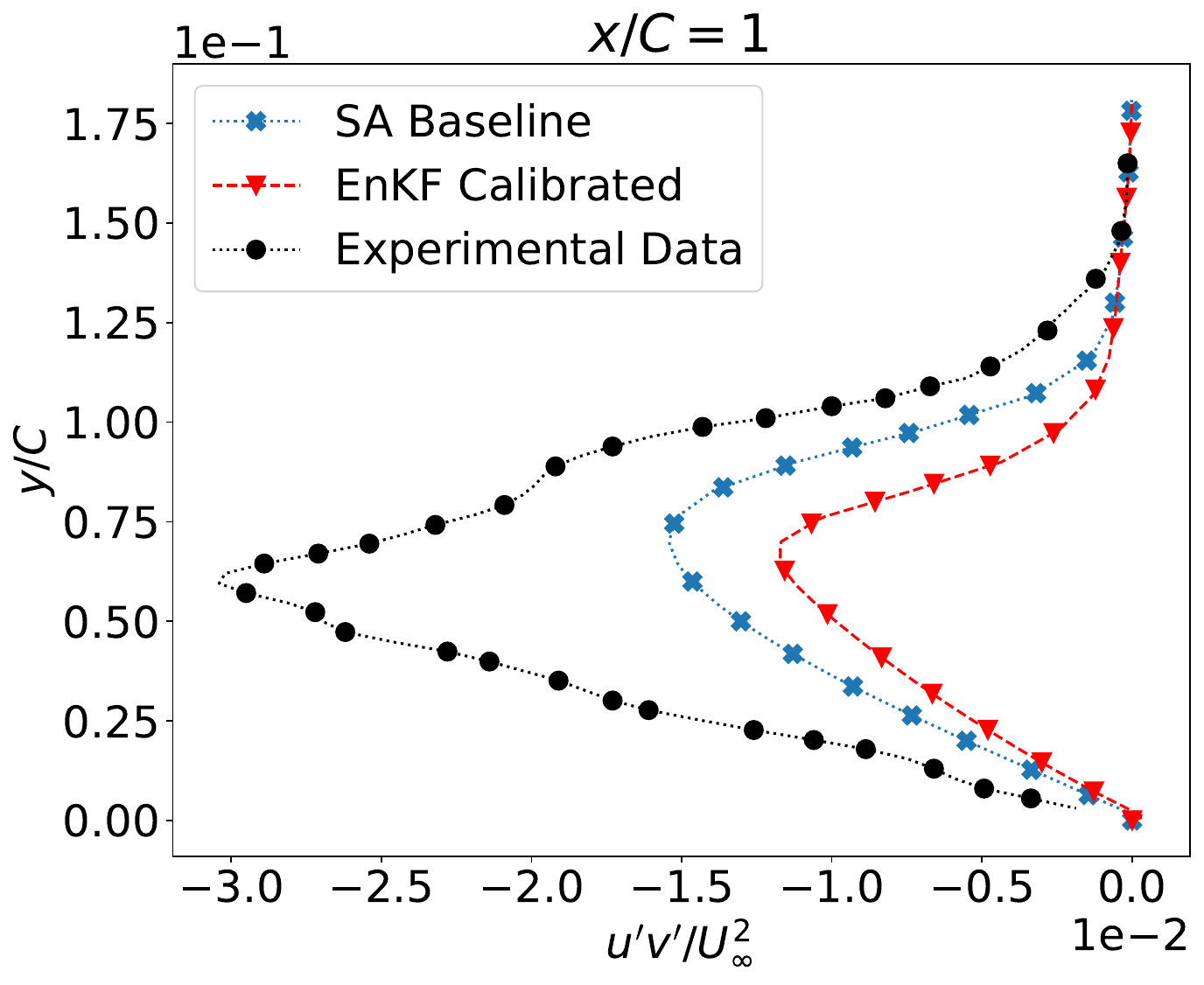}}
\caption{2D-WMH: a.) $C_f$, b.) $C_p$ vs. $x/C$, where $C$ is the chord length of the bump. c.) and d.) show the $u'v'/U_{\infty}^2$ vs. $y/C$ at location $x/C = $ 0.65 and 1, respectively. The $u'v'/U_{\infty}^2$ for calibrated model matches the experimental data for $x/C = 0.65$,i.e., separation. For $x/C = 1$, both models have significant deviation from the experimental data, however the SA model is slightly better. } 
\label{hump_cp_cb}
\end{figure}

The 2D-WMH is a standard flow geometry in NASA turbulence repository \cite{nasa_bump}. The results for the 2D-WMH are plotted in figure \ref{hump_cp_cb}. The flow is attached to the bump up to $x/C=0.655$, after which separation is observed. In figure \ref{hump_cp_cb}a and b, until separation, the SA baseline and calibrated model are in good agreement with each other as well as experimental data \cite{seifert2002active,nasa_bump}. This agreement is encouraging as the calibration did not distort the performance of the data-enhanced model in the attached flows. Furthermore, the calibrated model shows better recovery characteristics for $C_f$ for $x/C >1$. Overall, in figure \ref{hump_cp_cb}b, the $C_p$ data from the calibrated model shows a good agreement with experimental data. There are some deviations observed around $0.7<x/C<1$. However, it can be concluded that the calibrated model $C_f$ predictions are in better agreement with the experimental data as compared to the baseline case. On the other hand, the $C_p$ predictions are slightly worse than baseline model.

\begin{figure}
\centering
\includegraphics[height=8cm]{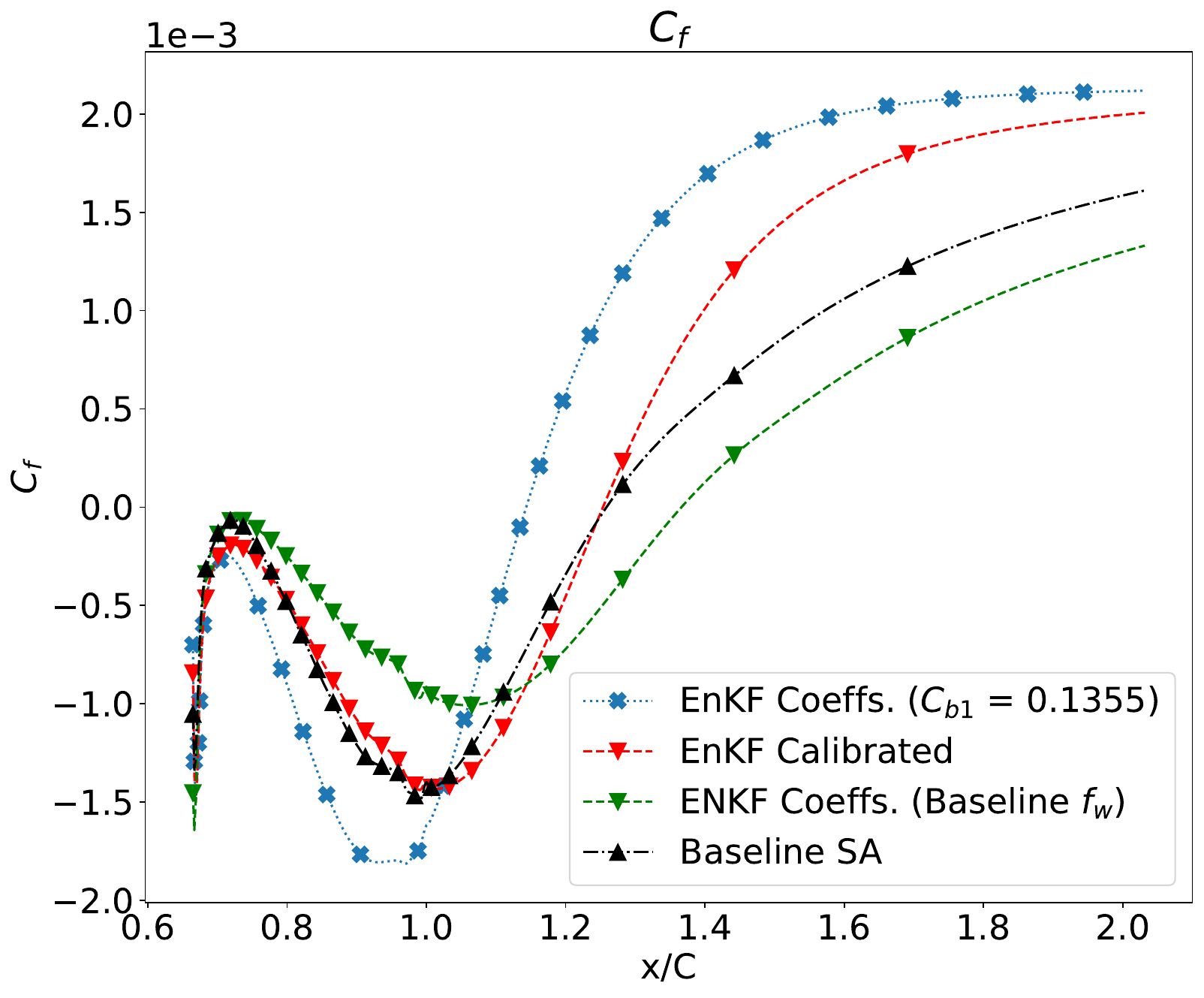}
\caption{2D-WMH: $C_f$ vs. $x/C$, plotted after after the separation point $x/C>0.655$. Parametric analysis of calibrated coefficients by using baseline values in $C_{b1}$ and $f_w$ alternatively. The dotted blue line shows $C_f$ where calibrated coefficients are paired with $C_{b1}$. The dashed green line shows the calibrated coefficients paired with baseline $f_w$. The red and black lines show the results from calibrated and baseline models respectively.}
\label{hump_param}
\end{figure}

\begin{figure}[h!]
\begin{center}
\subfloat[$C_{b1}$]{\includegraphics[width=12cm]{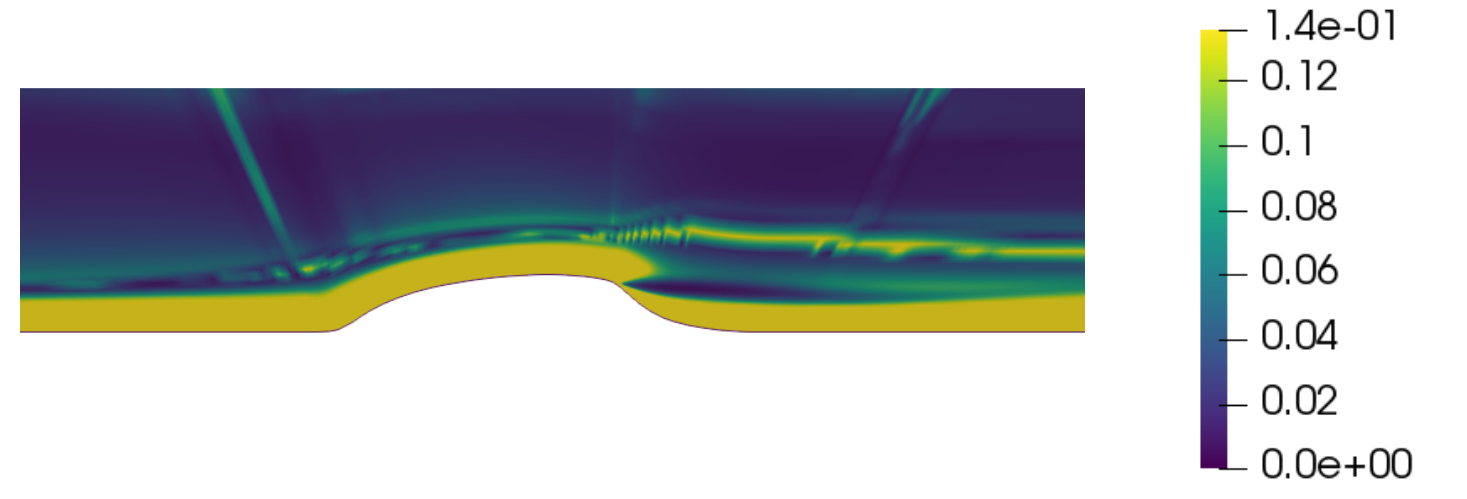}}\\
\subfloat[$f_w$]{\includegraphics[width=12cm]{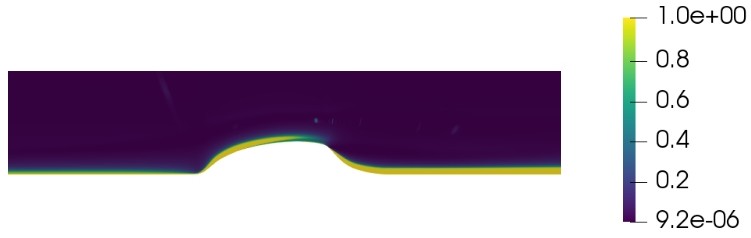}}
\end{center}
\caption{2D-WMH: Variation of $C_{b1}$ and $f_w$ in the domain.}
\label{cont_hump}
\end{figure}

Similar to figure \ref{bfs_1355}, a parametric analysis was also done for the 2D-WMH case as shown in figure \ref{hump_param}. The results show similar trends to figure \ref{bfs_1355}, which is encouraging for consistency and generalization of the calibrated model. As observed previously, the $C_{b1}$ affects the separation zone while $f_w$ affects the recovery zone. In Figure \ref{hump_param}, choosing $C_{b1} = 0.1355$ (original value) while keeping other values from calibration, results in faster recovery of $C_f$, implying the effect of calibrated coefficients ($f_w$) on recovery zone. On the contrary, using a baseline $f_w$ with other coefficients being calibrated shows a very slow recovery.  Figure \ref{cont_hump} shows the contour plots of $C_{b1}$ and $f_w$ for the 2D-WMH. As with the BFS case (figure \ref{cont}), the $C_{b1}$ and $f_w$ also follow an opposite trend to each other. The $C_{b1}$ takes lower values near the separation zone and increases to 0.1355 downstream of the bump.
\newpage
\subsubsection{Flow over backward-facing step (changed height, BFS2)}
\begin{figure}
\centering
\includegraphics[height=7cm]{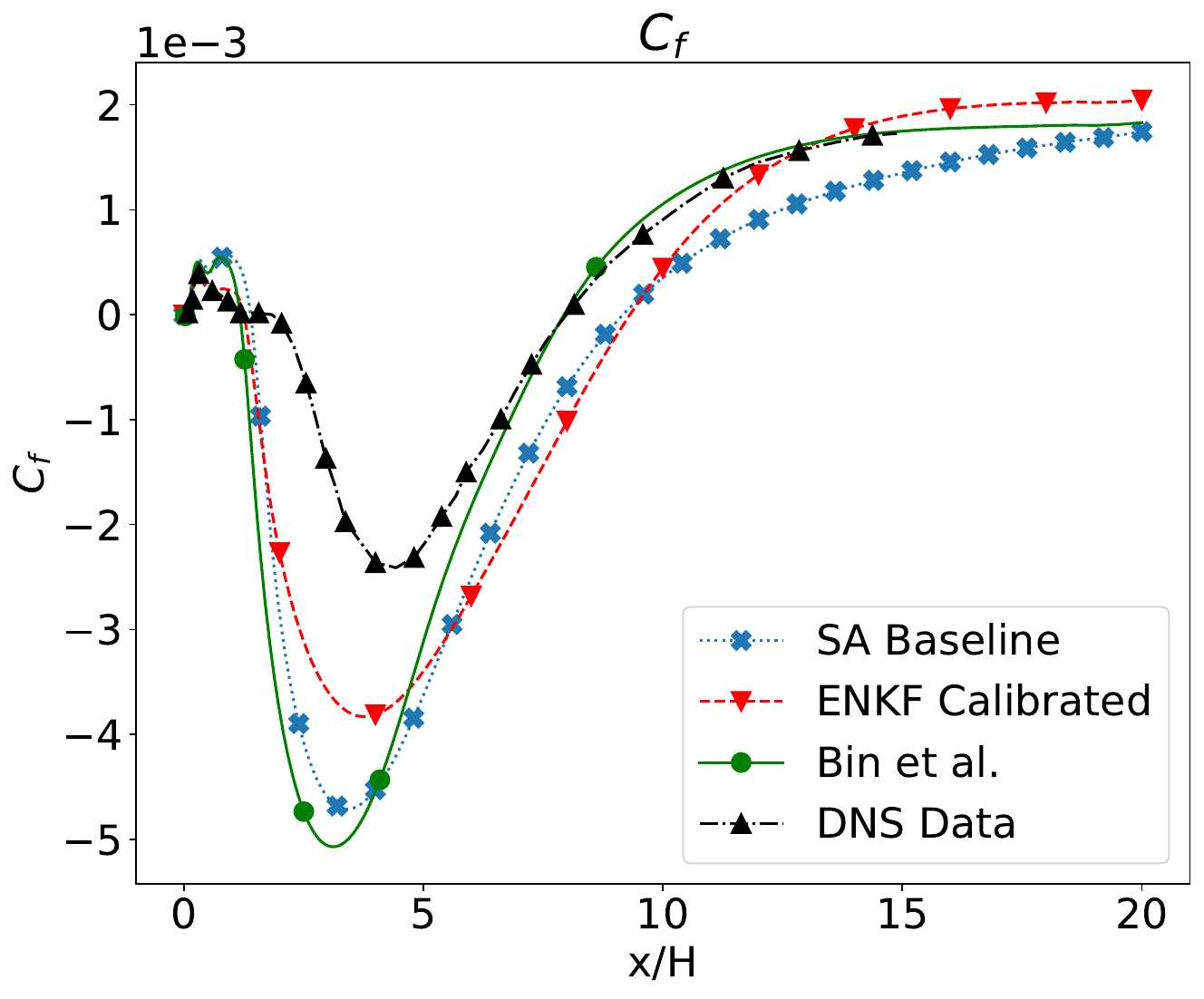}
\caption{BFS2: $C_f$ vs. $x/H$ along bottom wall obtained from the calibrated coefficients is compared with DNS \cite{bin2023data} and baseline SA. Additionally, the data from the calibrated SA model of Bin et al. \cite{bin2023data} is also given.}
\label{bfs2_cf}
\end{figure}

\begin{figure}
\centering
\includegraphics[height=7cm]{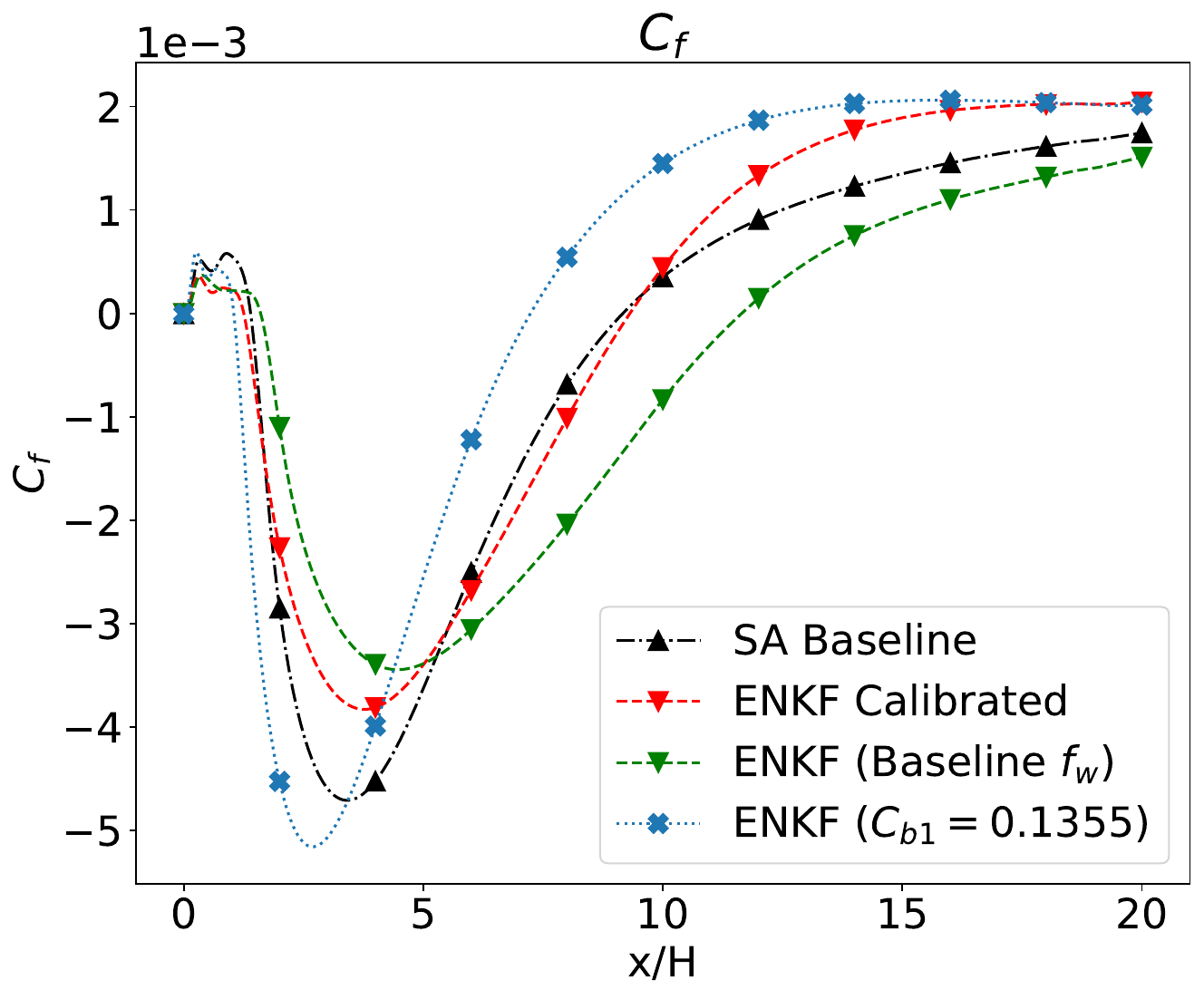}
\caption{BFS2: Parametric analysis plotting $C_f$ vs. $x/H$ along bottom wall post step. The dashed green line displays the baseline $f_w$ combined with the rest of EnKF calibrated coefficients, maintaining the initially observed improvement (figure \ref{bfs2_cf}) within the re-circulation zone but with a reduction in the recovery zone. Conversely, the dotted blue line depicts the baseline ($C_{b1} = 0.1355$) along with the rest of the calibrated coefficients, sustaining the improvement in the recovery zone but a decrease in the re-circulation zone.}
\label{bfs2_param}
\end{figure}

The calibrated model was further tested on a new backward-facing step case with a changed step height and $Re \approx 5600$ \cite{bin2023data,barri2010dns}. The case is derived from the study of Bin et al. \cite{bin2023data}, who also used to test it for their DNS-calibrated SA model. The step height is 2m i.e., half of the domain height and $Re \approx 5600$. Figure \ref{bfs2_cf} compares the agreement of $C_f$ from calibrated and baseline SA with that of direct numerical simulation(DNS) from Bin et al. \cite{bin2023data}. In addition, the results of the Bin et al. model are also presented for comparison. The results of figure \ref{bfs2_cf} show better agreement of $C_f$ between our calibrated model and DNS when compared to the baseline SA model. In addition, the results show a better accuracy of the current calibrated model in the recirculating zone than the Bin et al. \cite{bin2023data} calibrated model. The better results in the re-circulation zone are attributed to the calibration of $C_{b1}$ in addition to other coefficients whereas Bin et al. \cite{bin2023data} only calibrate $f_w$ in their study.
For further establishing the role of $C_{b1}$ and $f_w$ on $C_f$, a similar analysis to figures \ref{bfs_1355} and \ref{hump_param} is performed on figure \ref{bfs2_param}. The results are consistent with the previous results i.e., $C_{b1}$ affects the accuracy in the recirculating zone whereas the $f_w$ effect is predominant in the recovery zone.  
\vspace{-15pt}
\subsection{Unbounded or attached flow} We have established that the EnKF-based calibration improves the SA model's performance in separated flows. However, the change in model parameters may distort the good performance of the model in unbounded or attached flows. In this section, the calibrated model is compared with the original SA model to measure any deviation.
\vspace{-15pt}
\subsubsection{NACA0012} 
\begin{figure}[h!]
\begin{center}
\subfloat[]{\includegraphics[height=6.7cm]{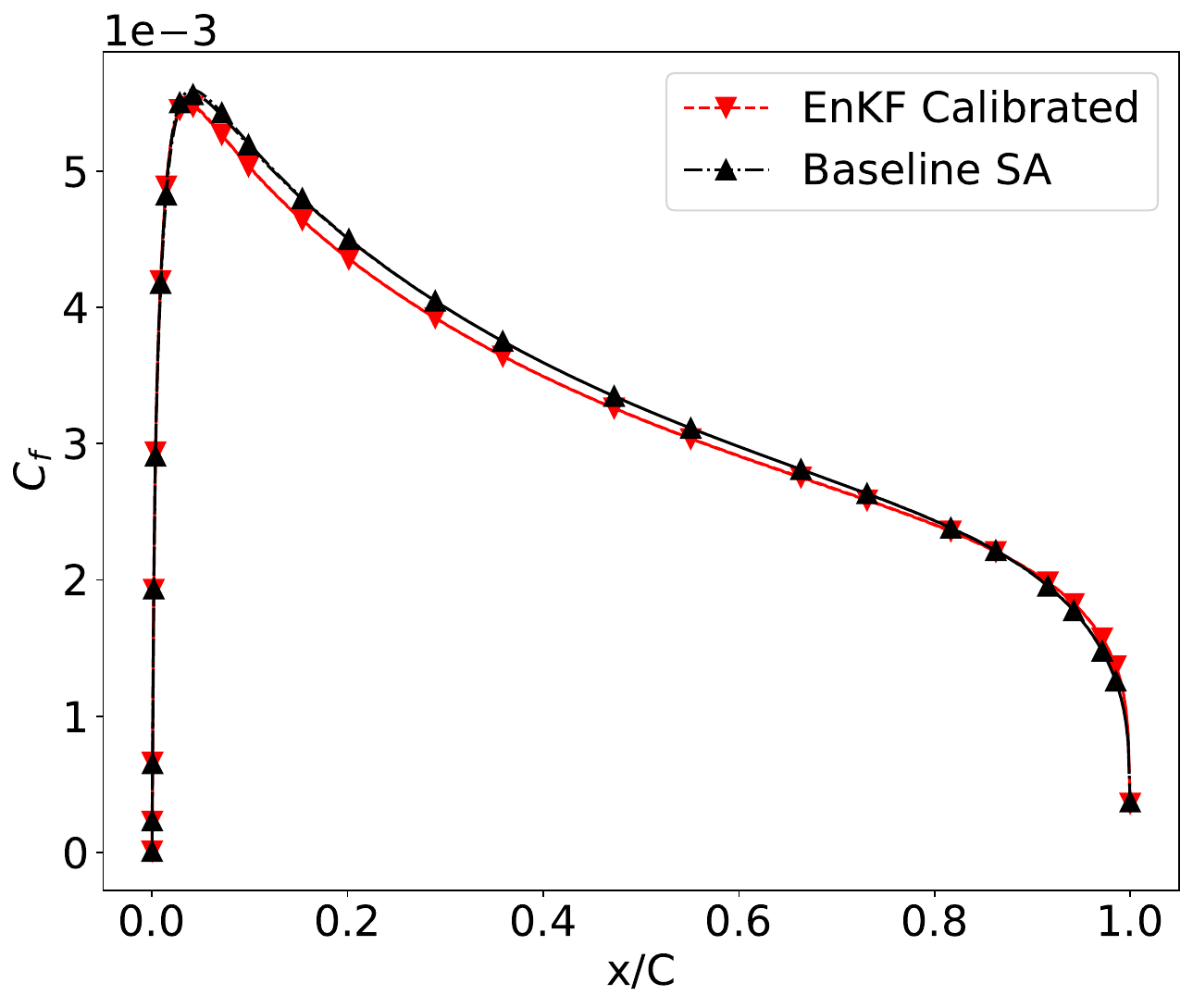}}
\subfloat[]{\includegraphics[height=6.5cm]{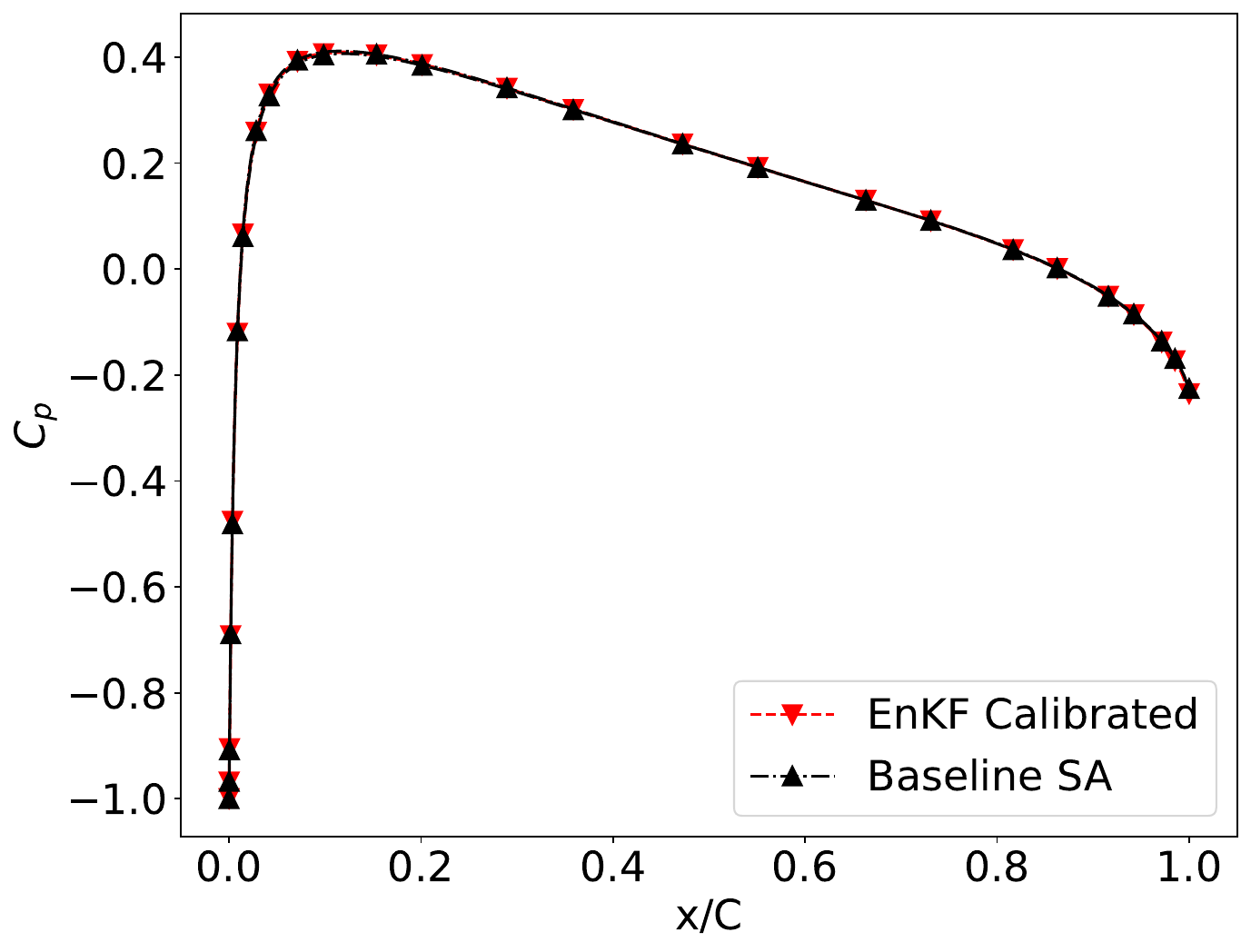}}
\end{center}
\caption{NACA0012 airfoil at angle of attack $0^{\circ}$: a. $C_f$, b. $C_p$ vs. $x/C$. The calibrated model shows good agreement with the original model. Hence, no distortion in good behavior of the SA model is observed for unbounded flows over an airfoil.}
\label{naca0012_cf_cb_0}
\end{figure}

\begin{figure}
\begin{center}
\subfloat[]{\includegraphics[height=6.7cm]{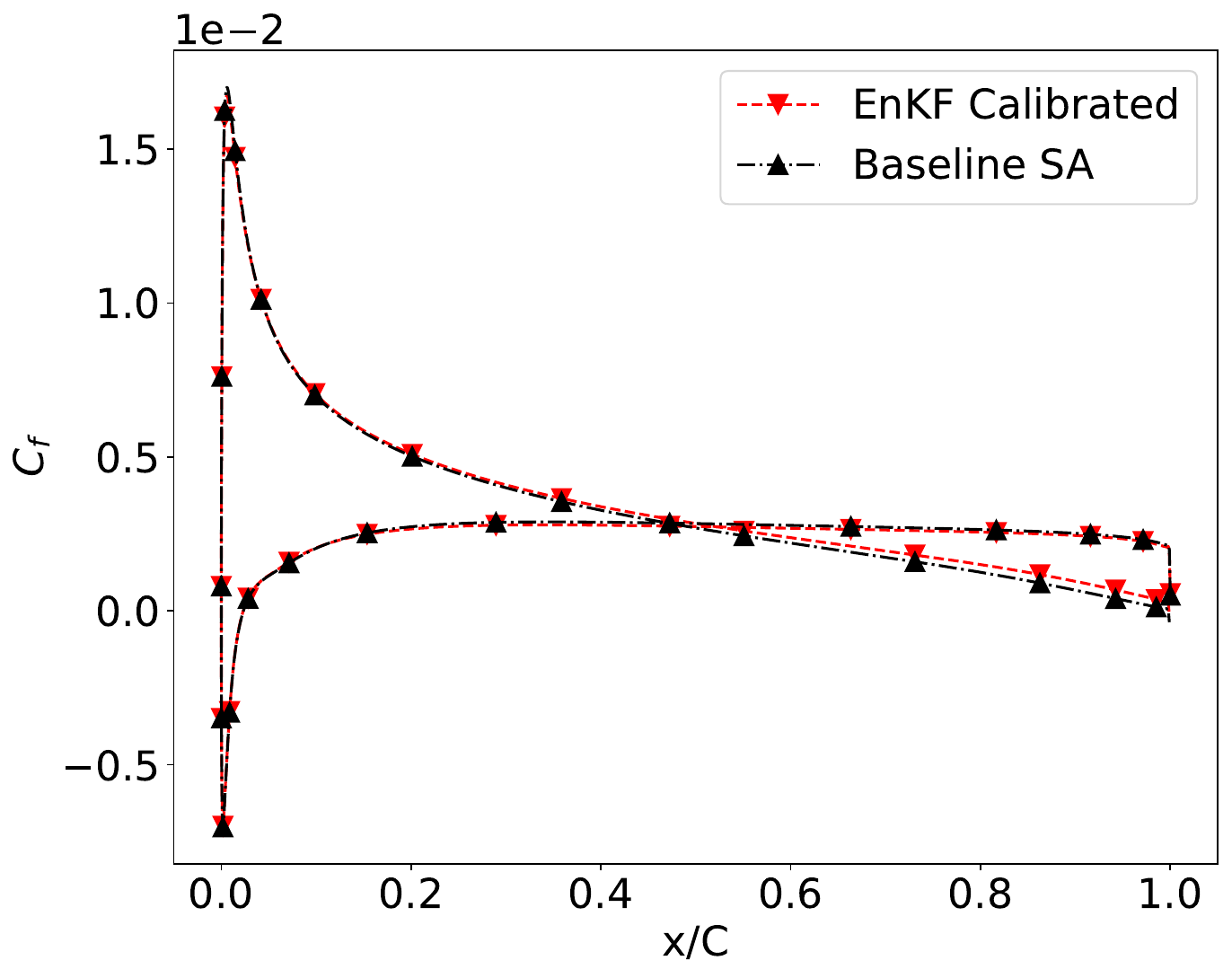}}
\subfloat[]{\includegraphics[height=6.5cm]{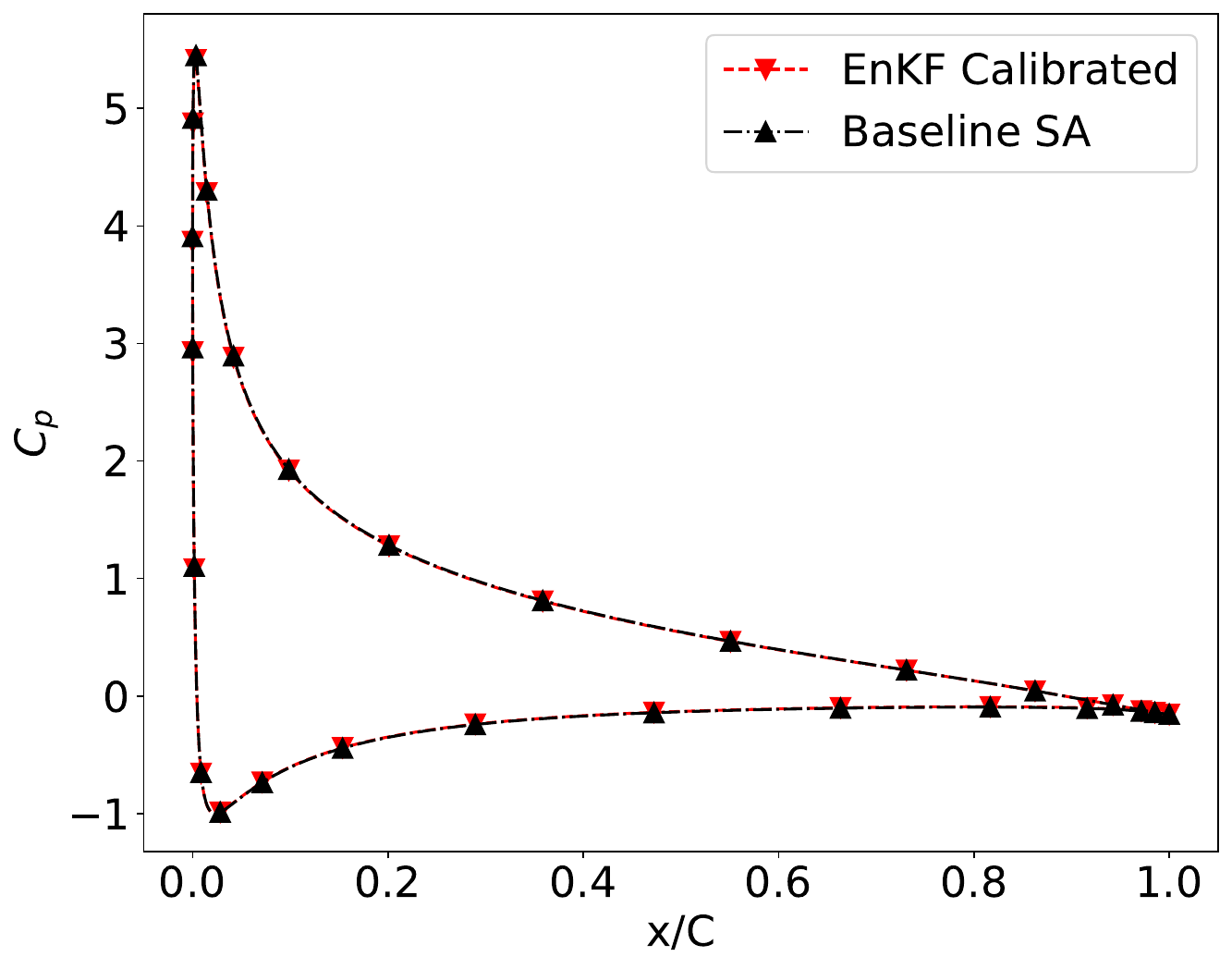}}
\end{center}
\caption{NACA0012 airfoil at angle of attack $10^{\circ}$: a. $C_f$, b. $C_p$ vs. $x/C$.}
\label{naca0012_cf_cb_15}
\end{figure}

The model is first tested for an unbounded flow over a NACA0012 airfoil at the angles of attack $0^{\circ}$ and $10^{\circ}$ in figure \ref{naca0012_cf_cb_0} and \ref{naca0012_cf_cb_15}, respectively \cite{brooks1986airfoil,di2009prediction}. The calibrated model shows a good agreement with the original SA model with minimal distortions -- and we conclude that the performance of the original model is preserved in the calibrated variant. Furthermore, a comparison of the lift ($C_l$) and drag ($C_d$) coefficients was also made, for 10$^\circ$, the $C_l$ was $\approx 1$,  and $C_d$ was $\approx 0.013$ for both models. For 0$^\circ$, the $C_d$ was $\approx 0.008$  and $C_l$ was $\approx 0.001$ for both models. This demonstrated a minimal deviation from base SA performance for the NACA0012 case.
\subsubsection{Flat plate boundary layer} 
\begin{figure}[h!]
\begin{center}
\subfloat[]{\includegraphics[height=6.8cm]{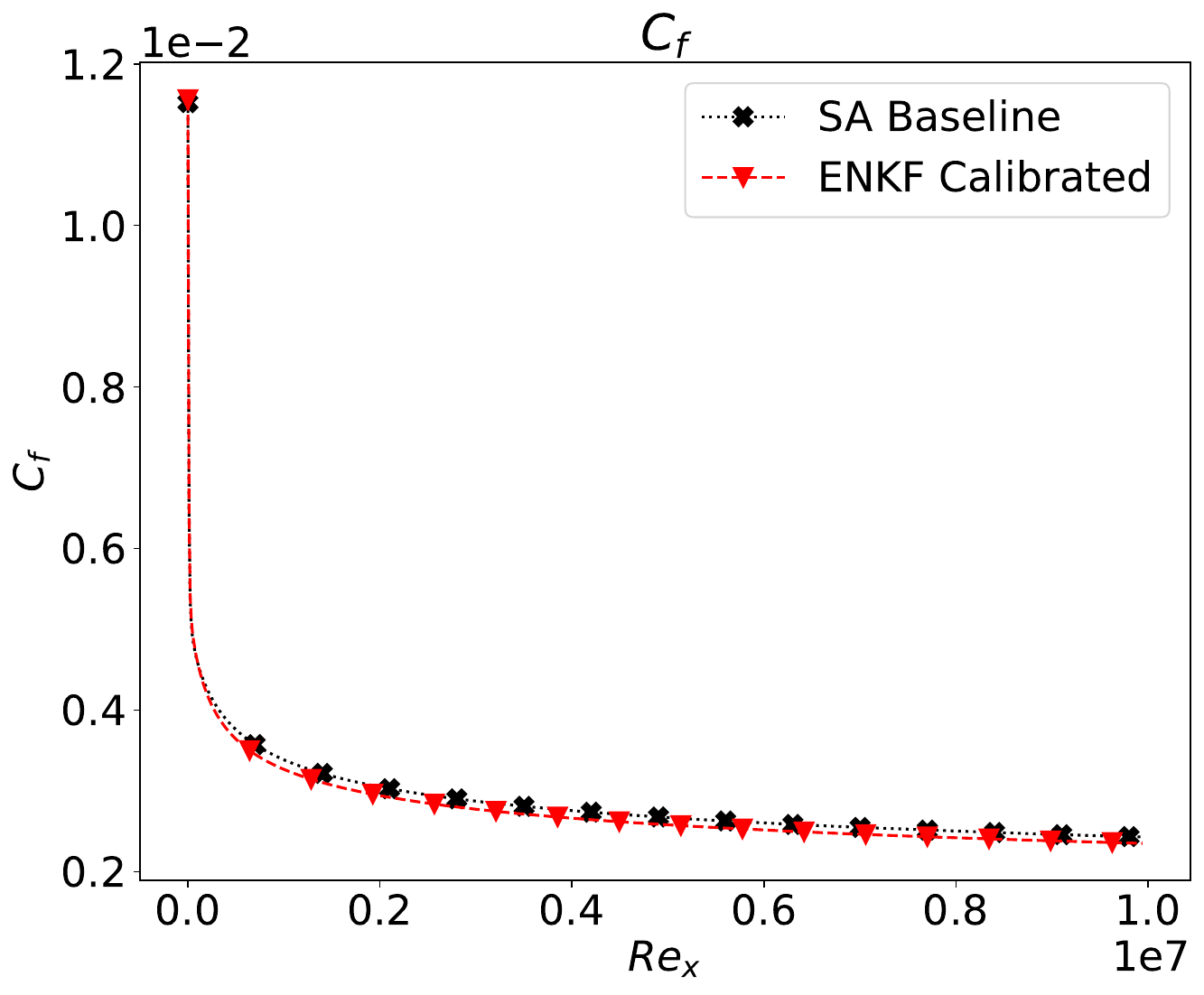}}
\subfloat[]{\includegraphics[height=6.8cm]{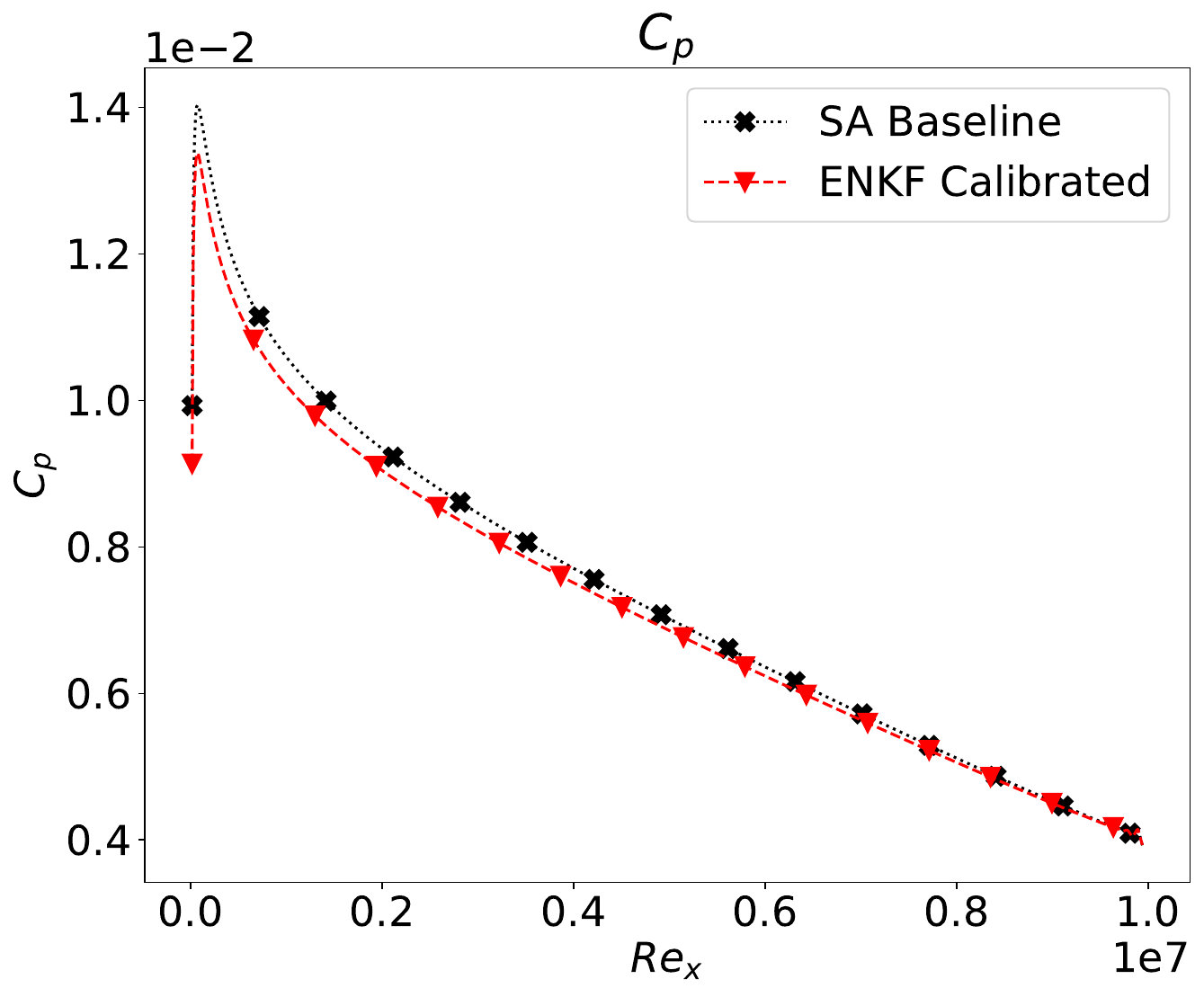}}\\
\subfloat[]{\includegraphics[height=6.8cm]{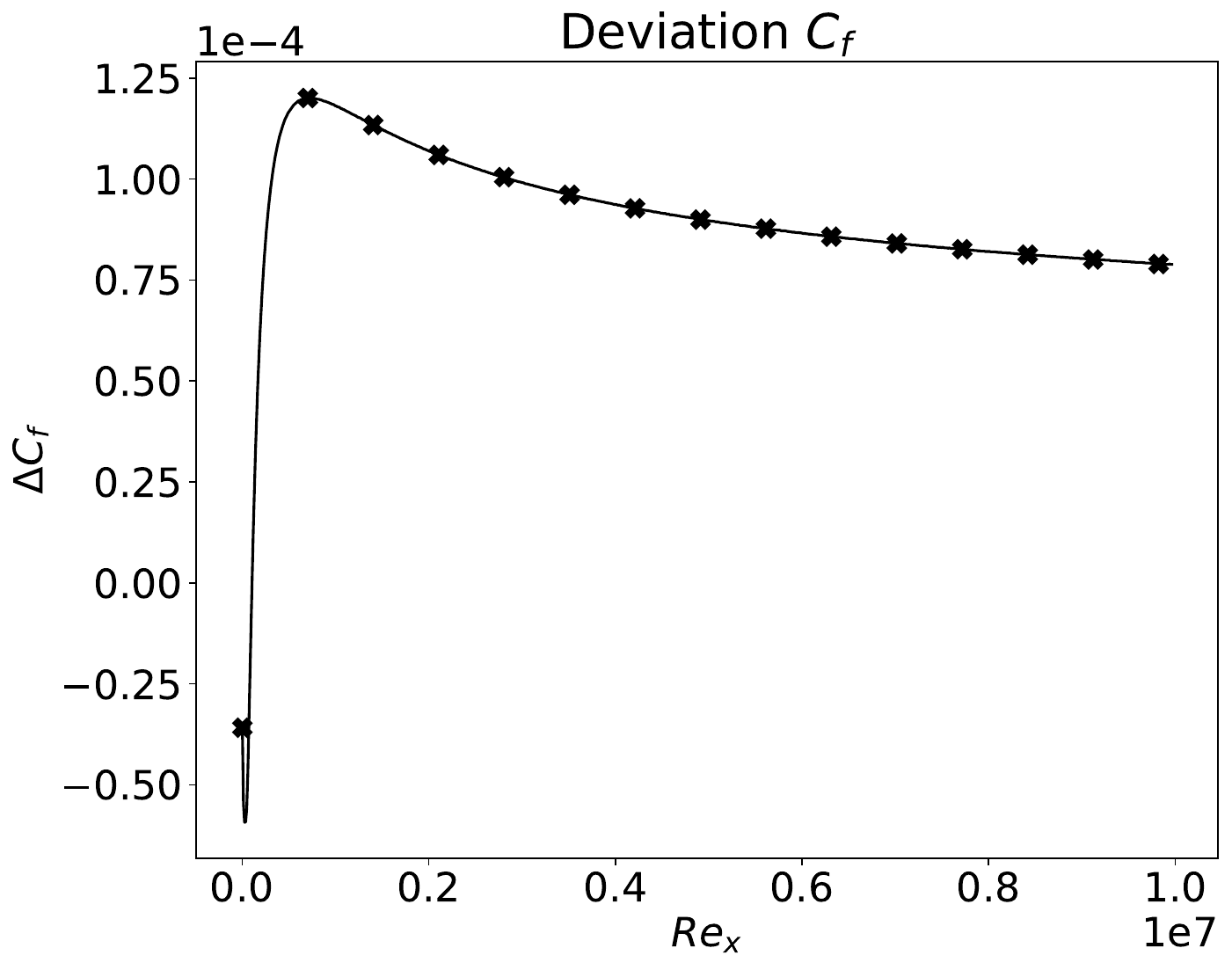}}
\subfloat[]{\includegraphics[height=6.8cm]{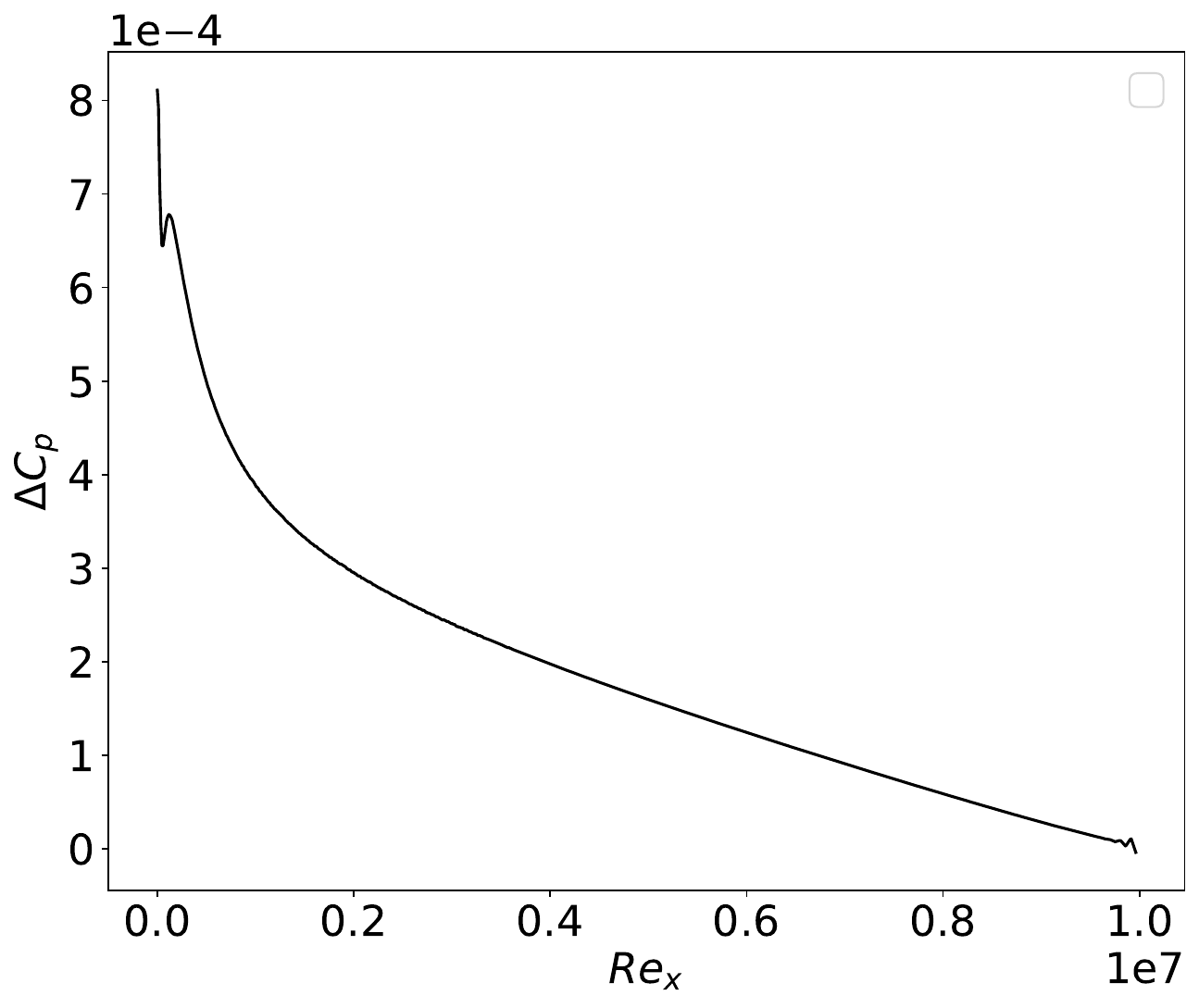}}

\end{center}
\caption{Zero pressure gradient boundary layer: a. $C_f$, b. $C_p$ vs. $Re_x$ for flow over a flat plate showing agreement in the behavior of calibrated model for attached flows. c and d respectively show deviation between the two models for $C_f$ and $C_p$.}
\label{fp_cf_cb}
\end{figure}

\begin{figure}[h!]
\centering
\includegraphics[height=6.7cm]{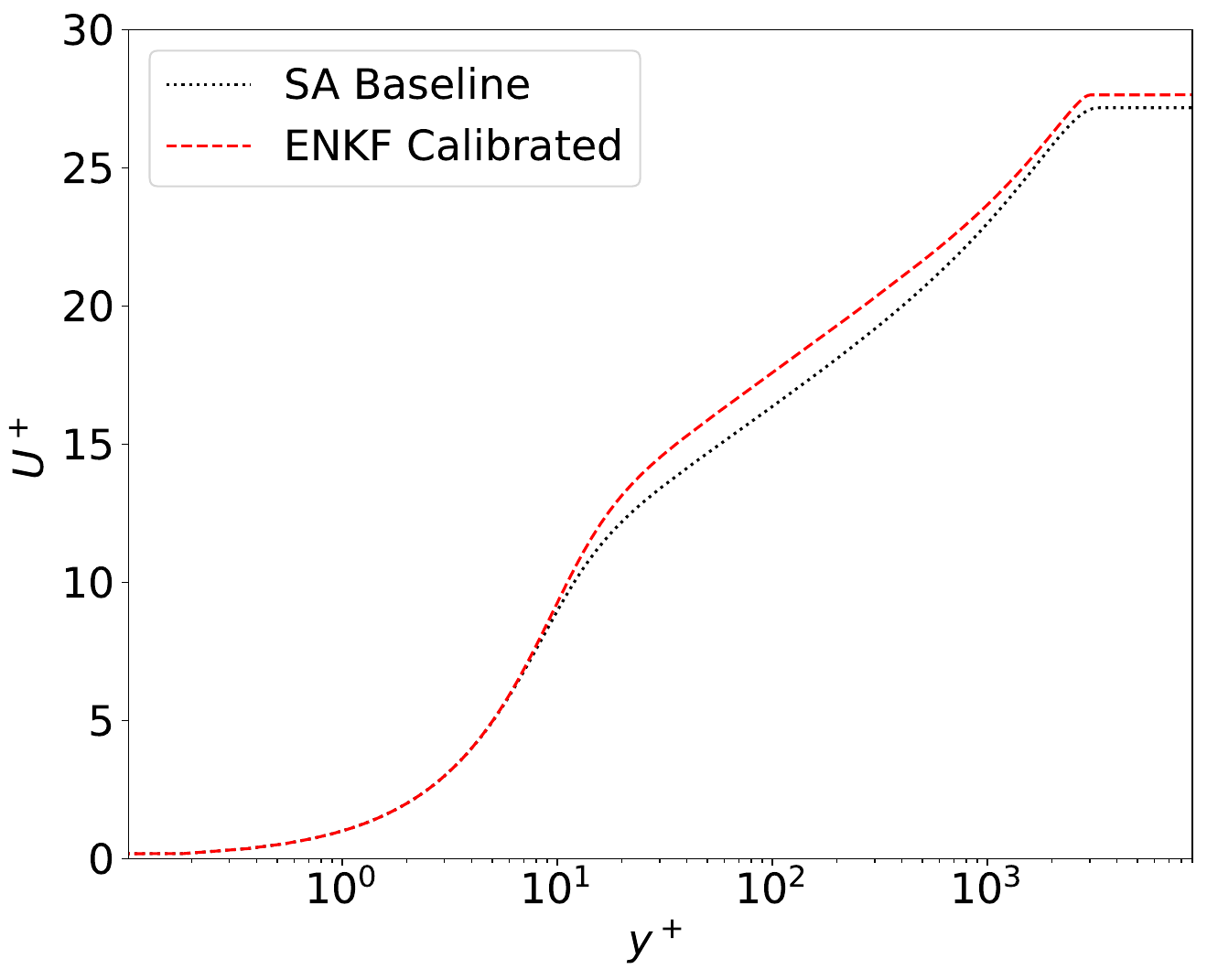}
\caption{Zero pressure gradient boundary layer: $U^+$ vs. $y^+$ for both models.}
\label{u+}
\end{figure}

The improved model was further tested on a zero pressure gradient flat plate boundary layer flow to determine any distortion from the standard SA model in attached flows. Figure \ref{fp_cf_cb}a and b show the $C_f$ and $C_p$ vs. $Re_x$ plots for the calibrated and baseline model. It can be seen the results for baseline and calibrated model are in good agreement, whilst having slight distortions. Figures \ref{fp_cf_cb}c and d show the difference between baseline and calibrated model for $C_f$ and $C_p$. Additionally, figure \ref{u+}, shows $U^+$ vs. $y^+$ plot at $Re_{\theta} \approx 8000$ for both the models. It can be seen both models agree well. From here, it can be inferred that the calibrated SA model does not significantly distort the behavior of the baseline model in attached or unbounded flows.  
\subsubsection{Axisymmetric jet} 
\begin{figure}[h!]
\centering 
\includegraphics[height=8cm]{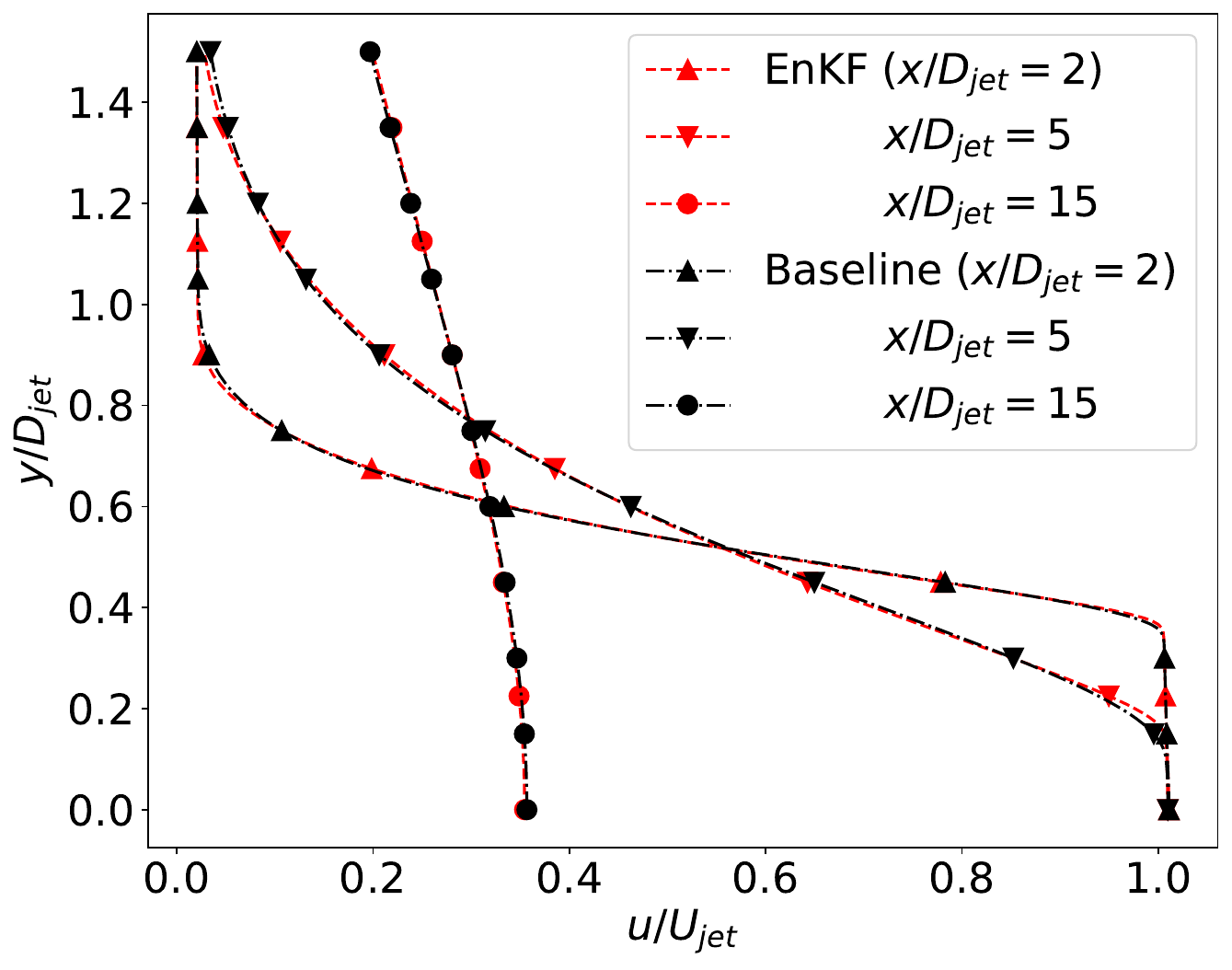}
\caption{ASJ: $u/U_{jet}$ vs. $y/D_{jet}$ along $x/D_{jet} =$ 2, 5 and 15,  obtained from the calibrated coefficients is compared with baseline SA. Here, $D_{jet} = 50.8 mm$, is the diameter of the jet, and $U_{jet}$ is the velocity at the center of the jet exit.}
\label{ASJ_U}
\end{figure}

The calibrated model is further tested on the free shear flow, i.e., axisymmetric jet (ASJ), \cite{bridges2010establishing,bridges2011nasa}.  This test will further provide insights into the calibrated model's behavior for ($r$) close to zero. The primary objective of the ASJ test is to assess whether the calibrated model can accurately capture the behavior of the original SA model for ASJ conditions. In Figure \ref{ASJ_U}, the velocity profile $(u/U_{jet})$ is plotted against radial distance $(y/D_{jet})$ for various axial positions $(x/D_{jet} = 2, 5, \ and \ 15)$. The results from the EnKF model closely align with those from the original SA (baseline) model.

\subsubsection{Mixing Layer (MIXL)} 
\begin{figure}[h!]
\begin{center}
\subfloat[]{\includegraphics[height=6.7cm]{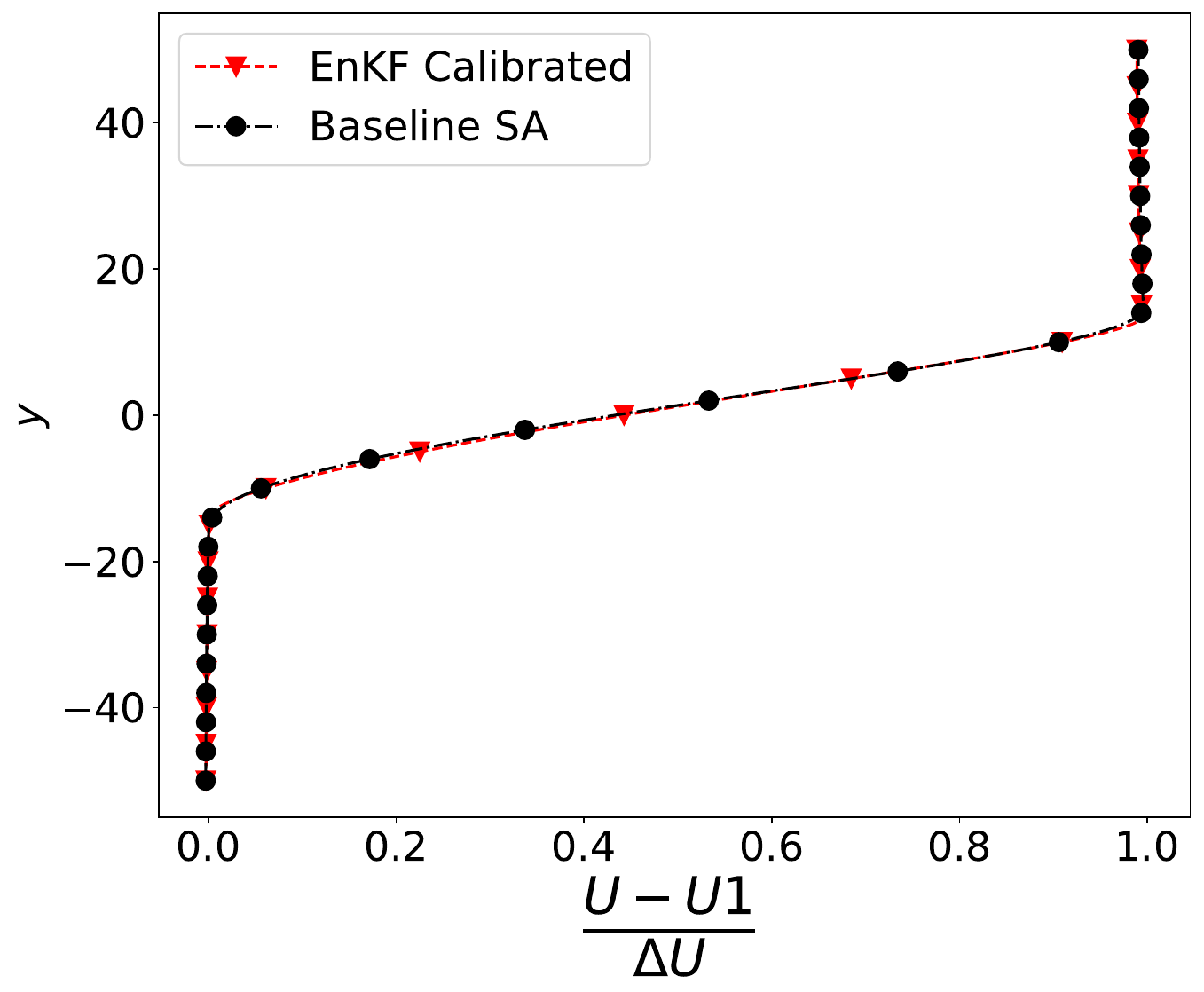}}
\subfloat[]{\includegraphics[height=6.7cm]{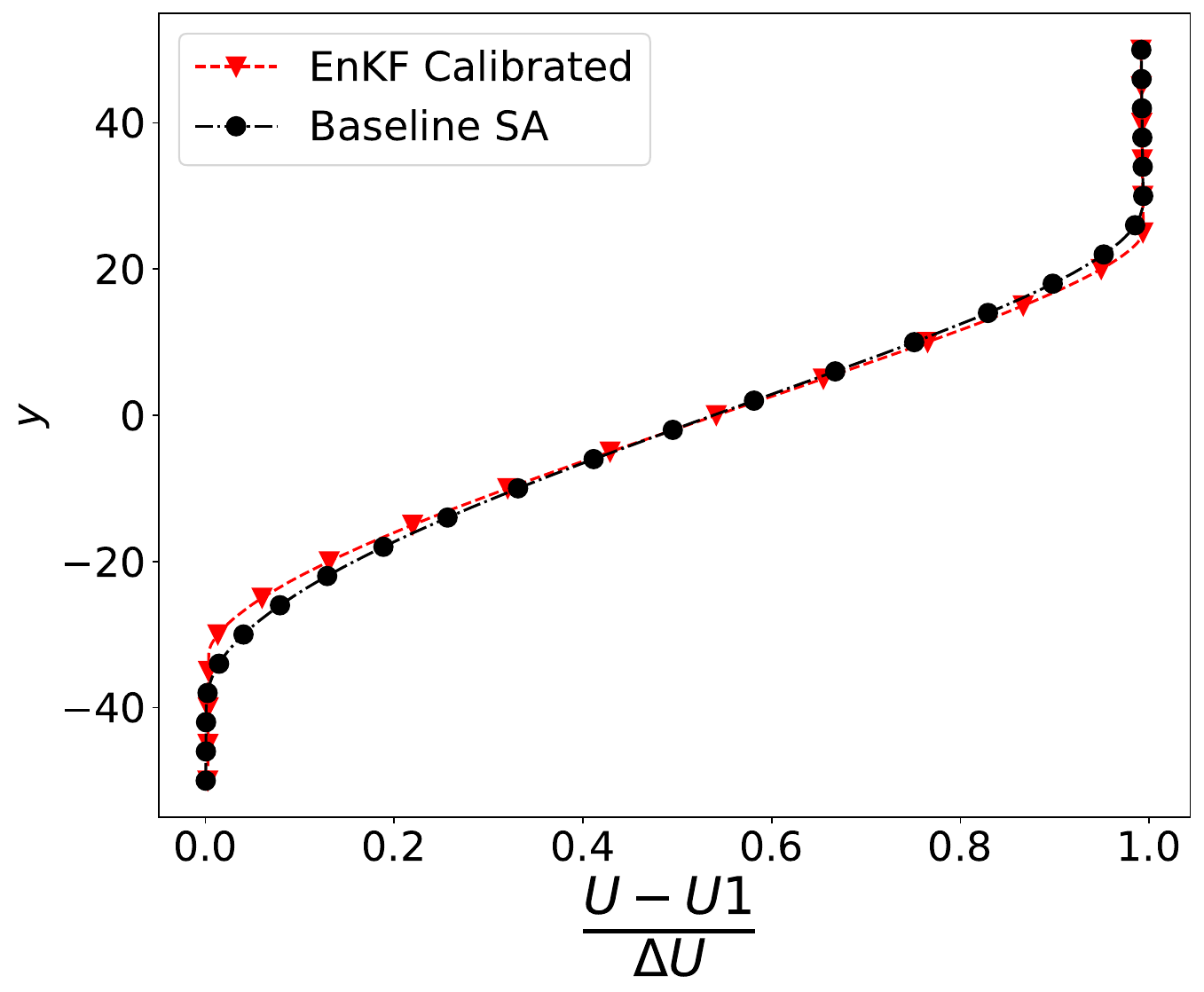}}
\end{center}
\caption{MIXL: $U$ vs. $Y$ for mixing layer showing good agreement between calibrated and baseline SA.}
\label{MIXL_U}
\end{figure}

For further validation the calibrated model was also tested on a classical mixing layer test case. Figure \ref{MIXL_U} shows the results of velocity profile at two $x$ locations. Here, $U$ is the velocity, $U1$ is lower inlet velocity, and $\Delta U$ is the difference between velocity of upper and lower inlet.It can bee seen that the results are in good agreement with some mindor distortion towards the extremes. This further illustrates that the calibrated model recovers SA performance on classical test cases.
\section{Conclusions} An EnKF-based calibration methodology has been introduced for improving RANS closure models with limited and noisy experimental data. This study's focus was on the calibration of SA model for cases where sparse, quantity of interest experimental data may be available. The SA coefficients were effectively adjusted using a simple parameterization, potentially minimizing the need for black-box ML models like NNs which may fail to generalize. The calibrated SA model exhibited improved accuracy in predicting important flow quantities, specifically $C_{p}$ and $C_{f}$, in scenarios involving separated flows. Notably, this improvement was achieved without compromising the SA model's accuracy in predicting behavior in attached and unbounded flows, aligning well with the progressive nature of SA enhancements.
\newline The findings further corroborate the hypothesis put forth by \cite{ray2018robust}, demonstrating that a substantial portion of closure model errors can be rectified by coefficient adjustments. The calibration process utilized only one geometry, namely BFS, at a single $Re$. Owing to the generalization provided by the SA model, this approach successfully extended the calibrated model's applicability to extrapolated cases and scenarios beyond its training range. In contrast, such stringent training criteria could lead to overfitting and extrapolation in the context of deep neural network applications.
\newline The adaptability of the original $f_w$ function (equation \ref{fw_1}) was also evident in this study, as the calibrated $C_{w2}$ and $C_{w3}$ coefficients displayed trends similar to those captured by a trained NN (as seen in figure \ref{fw_cb}b). Another notable advancement involved the treatment of $C_{b1}$ as a function of $r$ instead of a fixed scalar value found in the baseline SA model. This increased flexibility in representing $C_{b1}$ notably enhanced the calibrated model's predictive capabilities within re-circulation zones. The interplay between $C_{b1}$ and $f_w$ was also evident, where the former significantly impacted re-circulation zone prediction while the latter influenced recovery zone predictions.
\newline The stochastic nature of the EnKF necessitated a thoughtful selection of coefficient sampling ranges, a process guided by a parametric analysis (\ref{senstivity}). This range determination intricately hinged on an understanding of the relationships between production, destruction, and diffusion terms. Nonetheless, the possibility of relaxing this selection criterion through calibrations with multiple flow conditions remains a viable avenue for future exploration within the study's scope.

\begin{acknowledgments}
 We gratefully acknowledge the insights provided about the Spalart-Allmaras turbulence model by Dimitrios Fytanidis at Argonne National Laboratory. This material is based upon work supported by the U.S. Department of Energy (DOE), Office of Science, Office of Advanced Scientific Computing Research, under Contract DE-AC02-06CH11357. This research was funded in part and used resources from the Argonne Leadership Computing Facility, which is a DOE Office of Science User Facility supported under Contract DE-AC02-06CH11357. RM acknowledges support from DOE-ASCR-2493 - ``Data-intensive scientific machine learning".
\end{acknowledgments}

\appendix
\section{Parametric analysis for SA model} 
\label{senstivity} 
\begin{figure}
\begin{center}
\subfloat[]{\includegraphics[height=5.5cm]{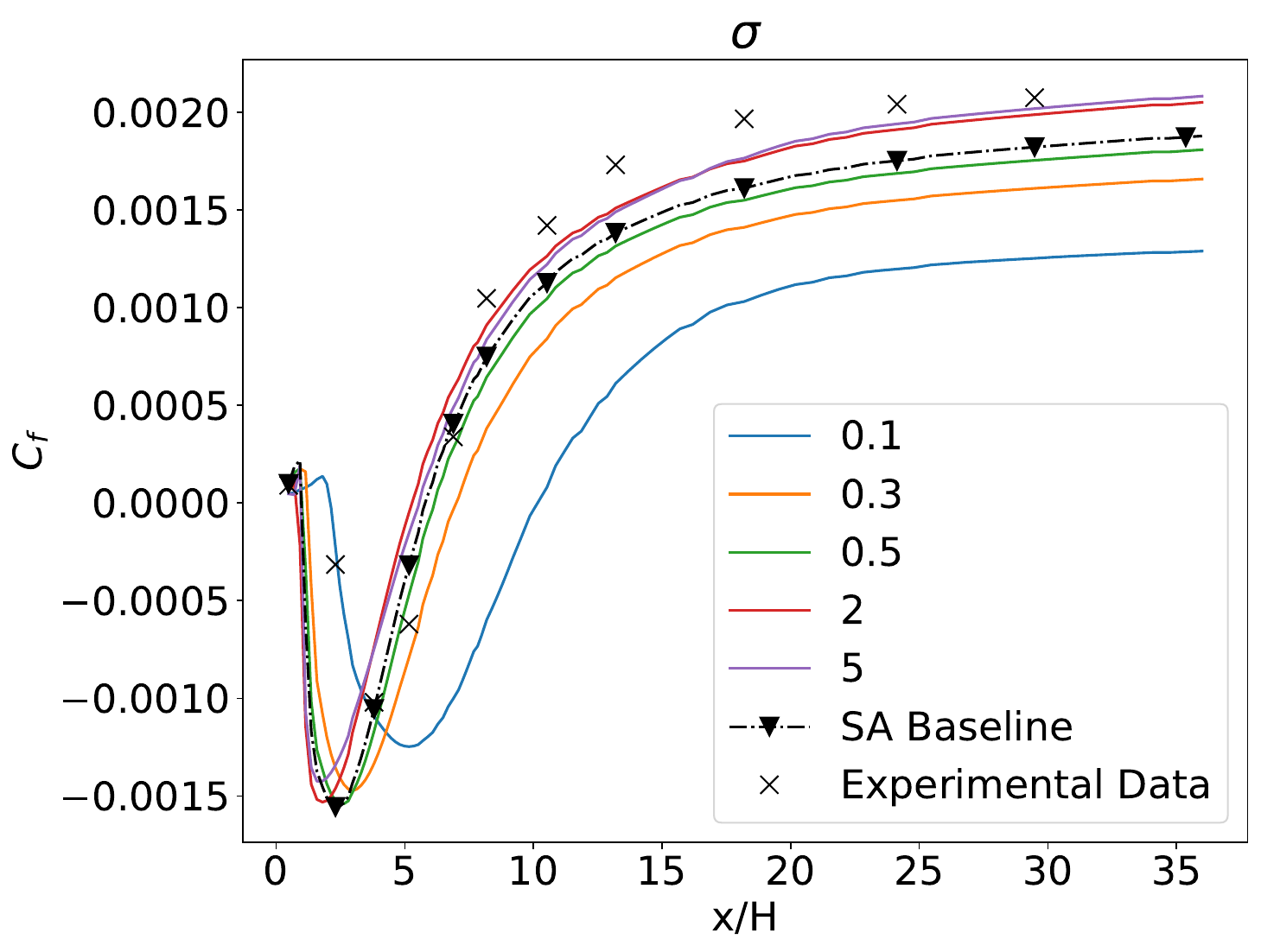}}
\subfloat[]{\includegraphics[height=5.5cm]{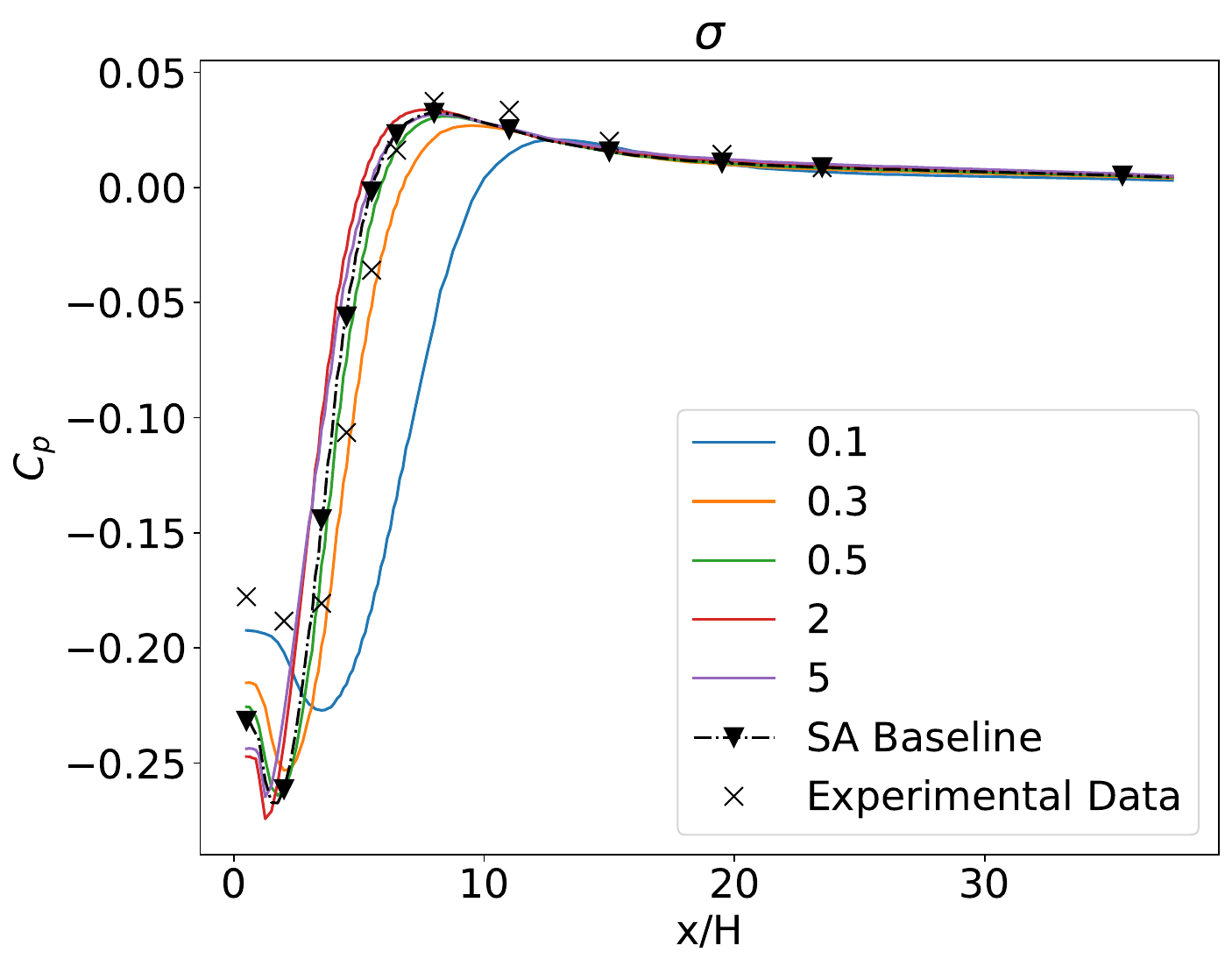}}\\
\subfloat[]{\includegraphics[height=5.5cm]{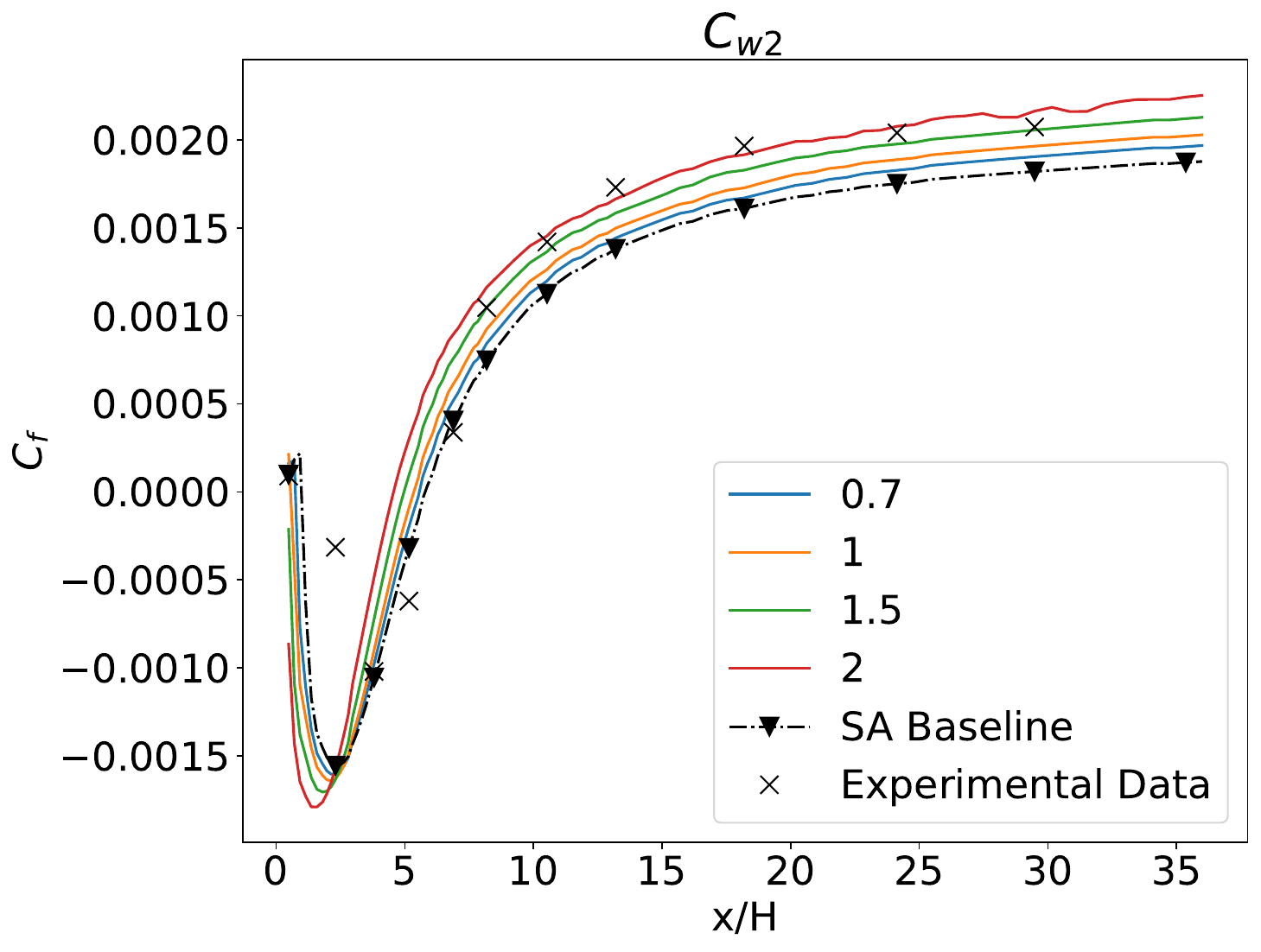}}
\subfloat[]{\includegraphics[height=5.5cm]{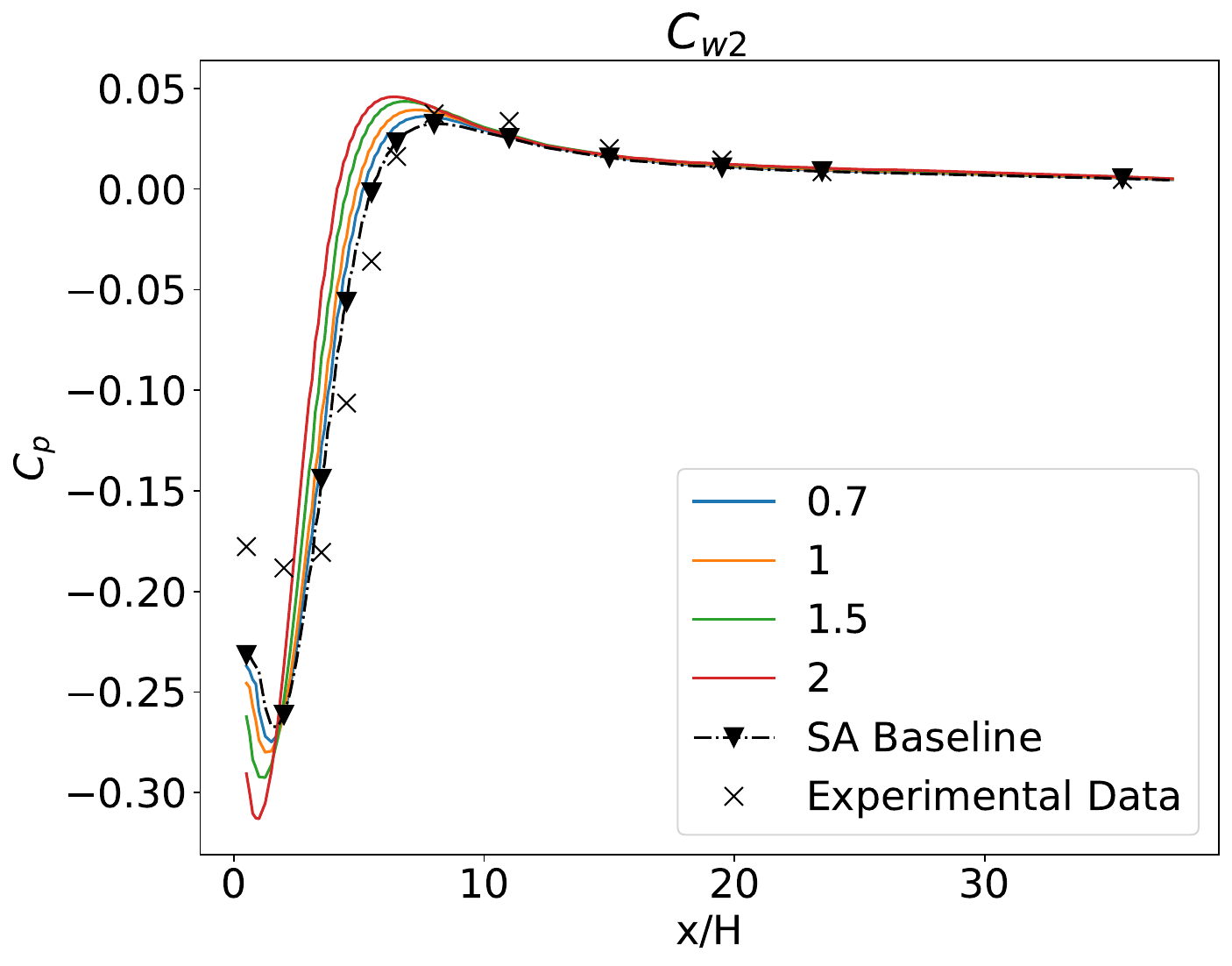}}\\
\subfloat[]{\includegraphics[height=5.5cm]{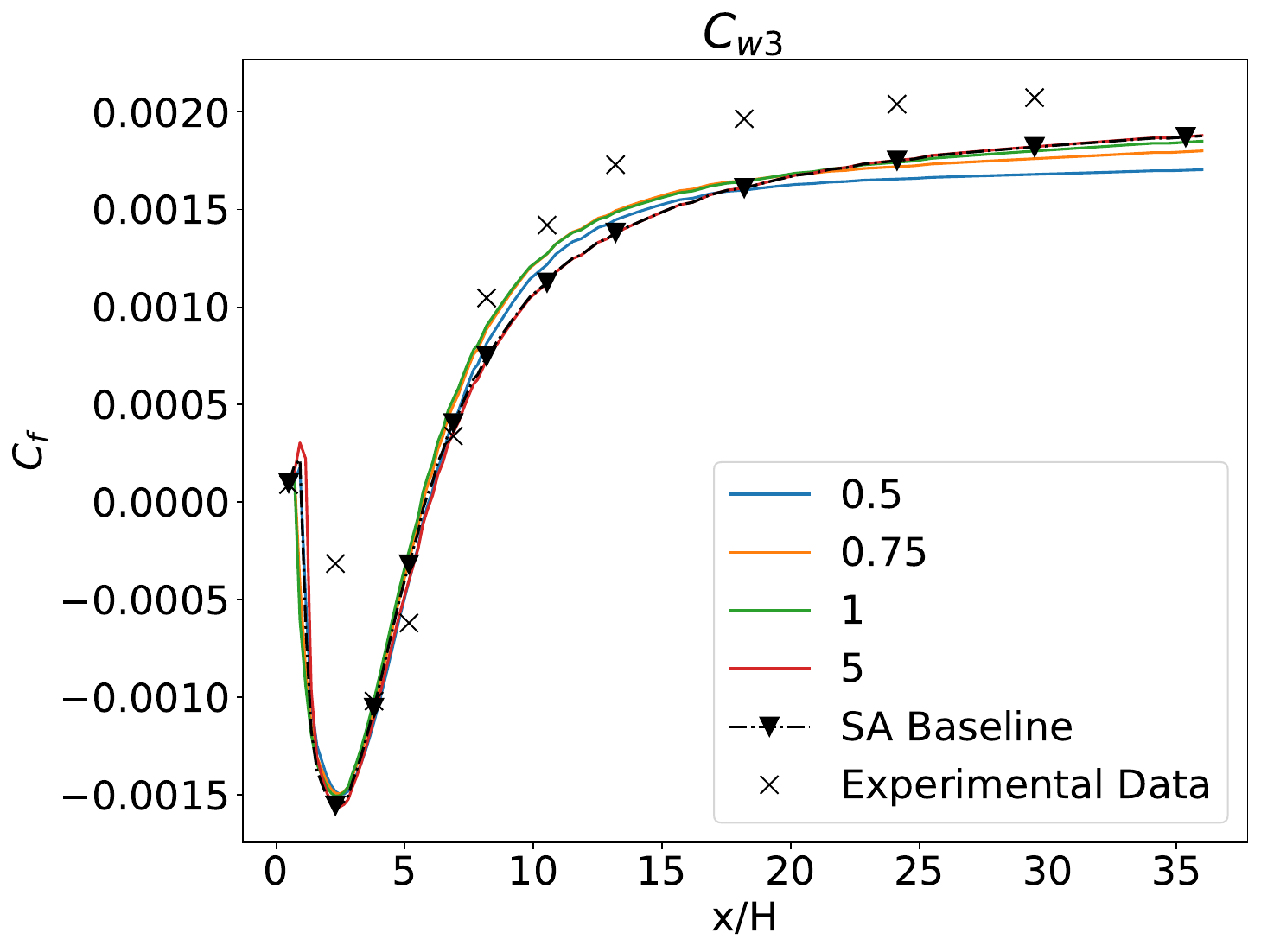}}
\subfloat[]{\includegraphics[height=5.5cm]{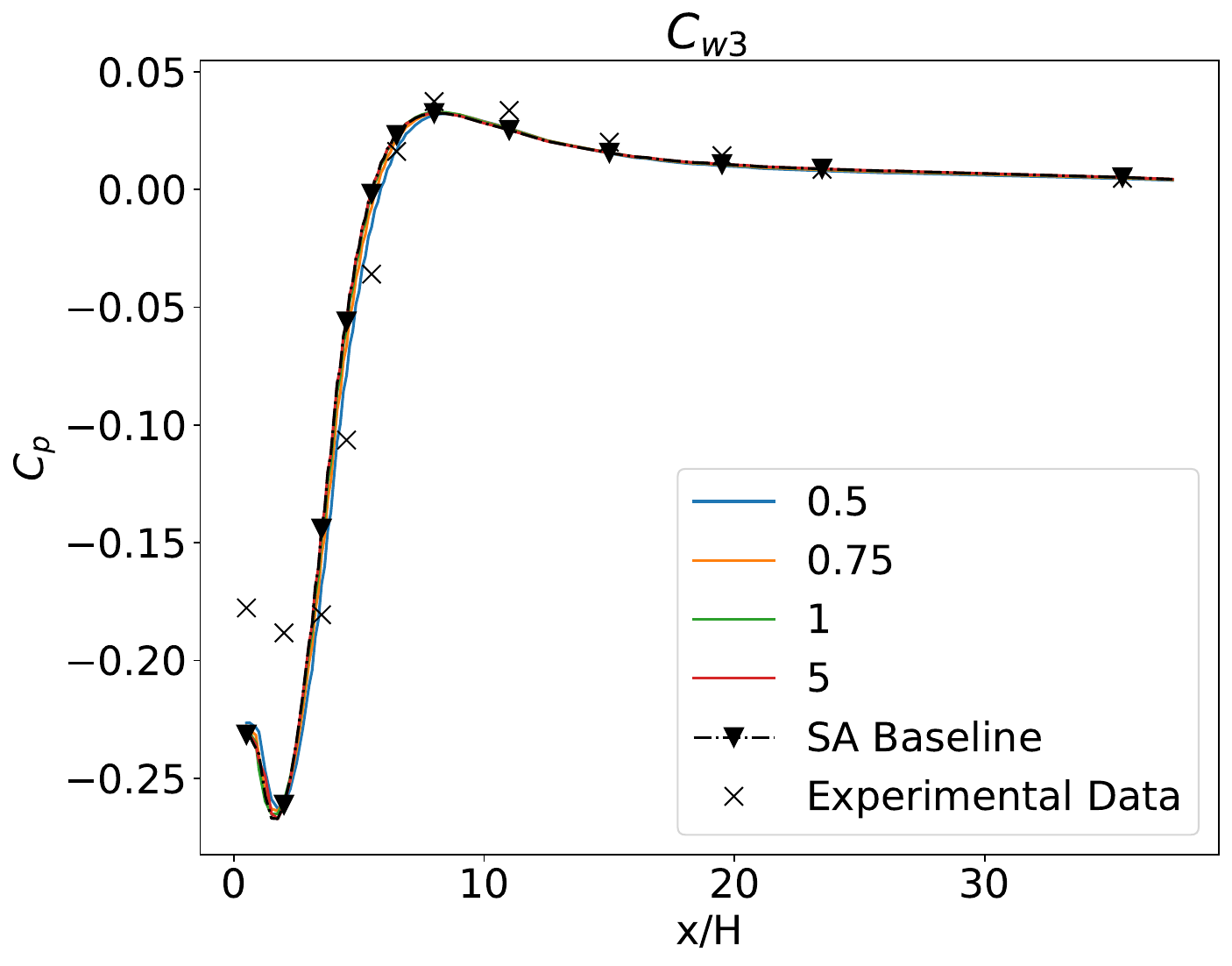}}\\
\end{center}
\caption{Parametric analysis for $C_f$ and $C_p$  along the bottom wall of the BFS by varying $\sigma$ (a,b), $C_{w2}$ (c,d), $C_{w3}$ (e,f). For reference, the solutions with original SA coefficients along with experimental data \cite{nasa} are also given. The title of each subplot is the SA coefficient that was varied for analysis.}
\label{sens_1}
\end{figure}

\begin{figure}{h!}
\begin{center}
\subfloat[]{\includegraphics[height=5.5cm]{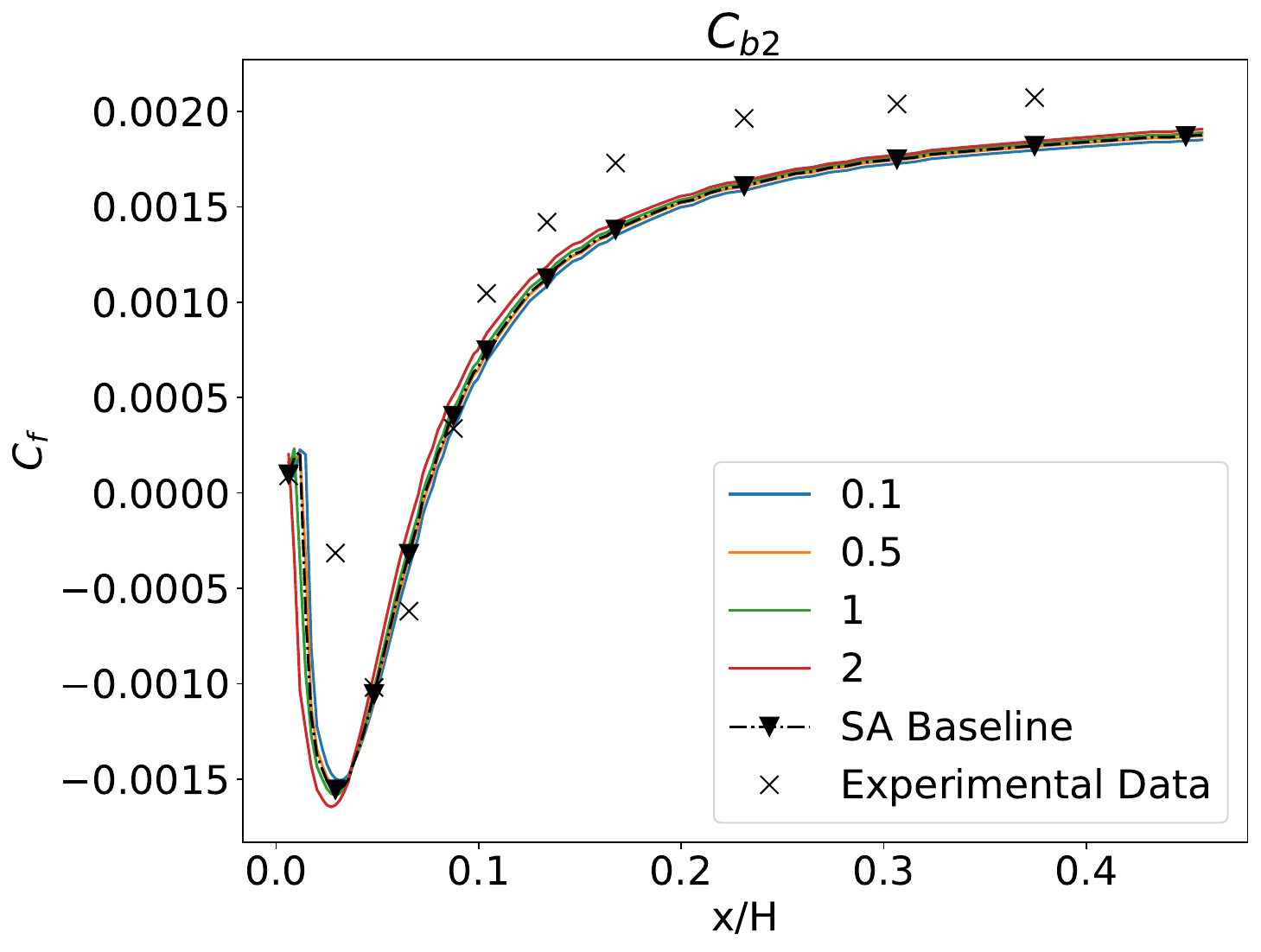}}
\subfloat[]{\includegraphics[height=5.5cm]{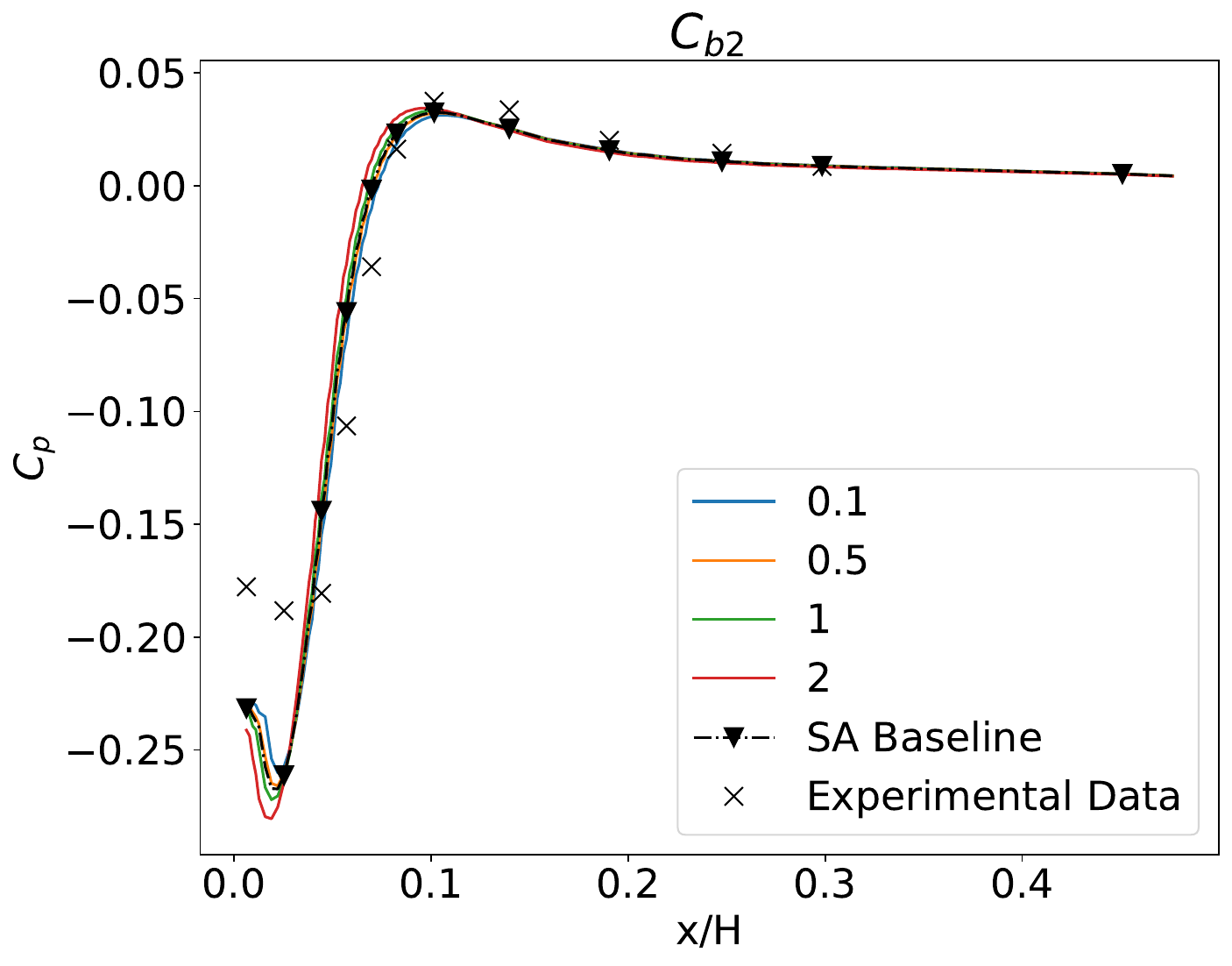}}\\
\subfloat[]{\includegraphics[height=5.5cm]{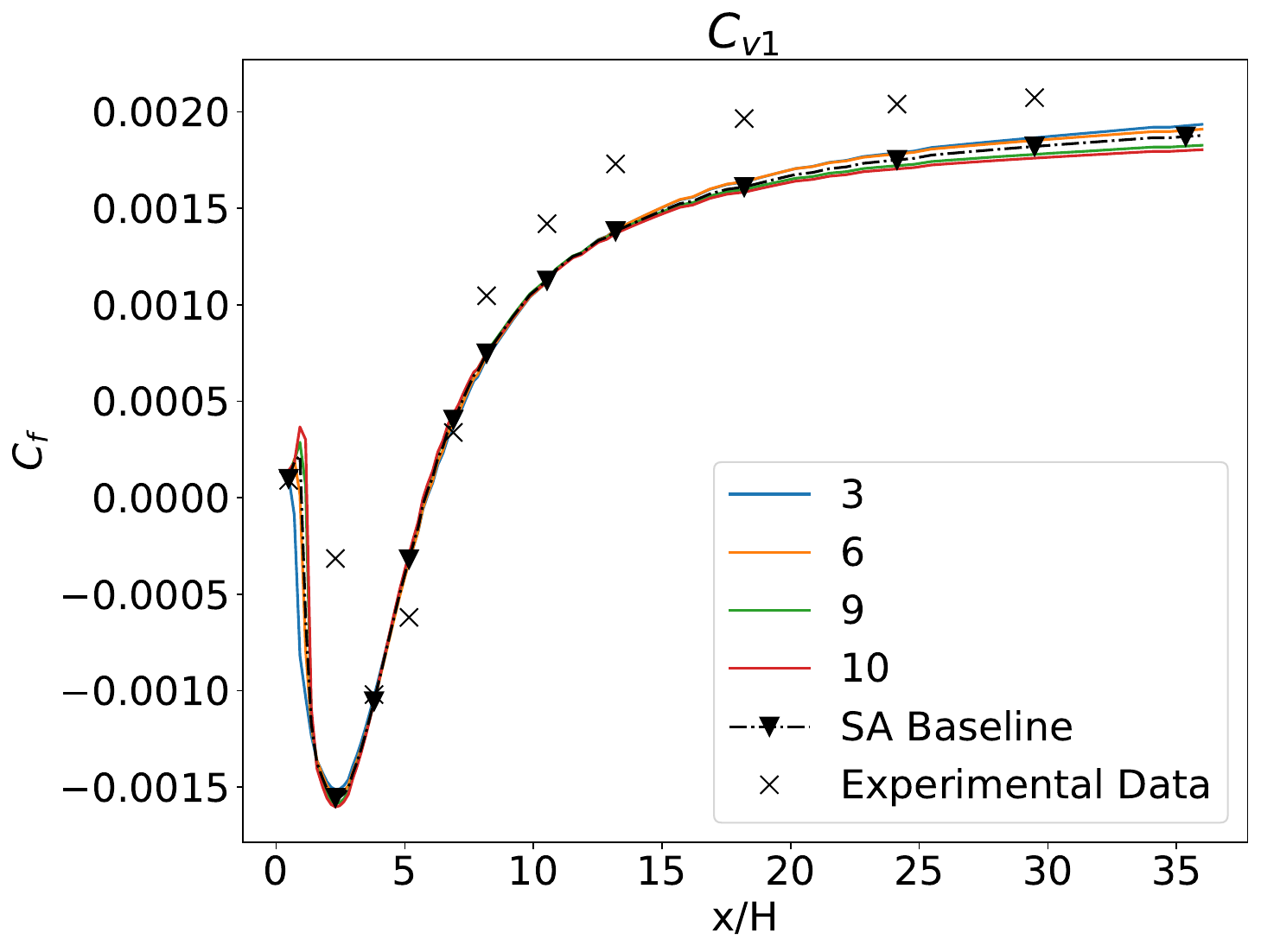}}
\subfloat[]{\includegraphics[height=5.5cm]{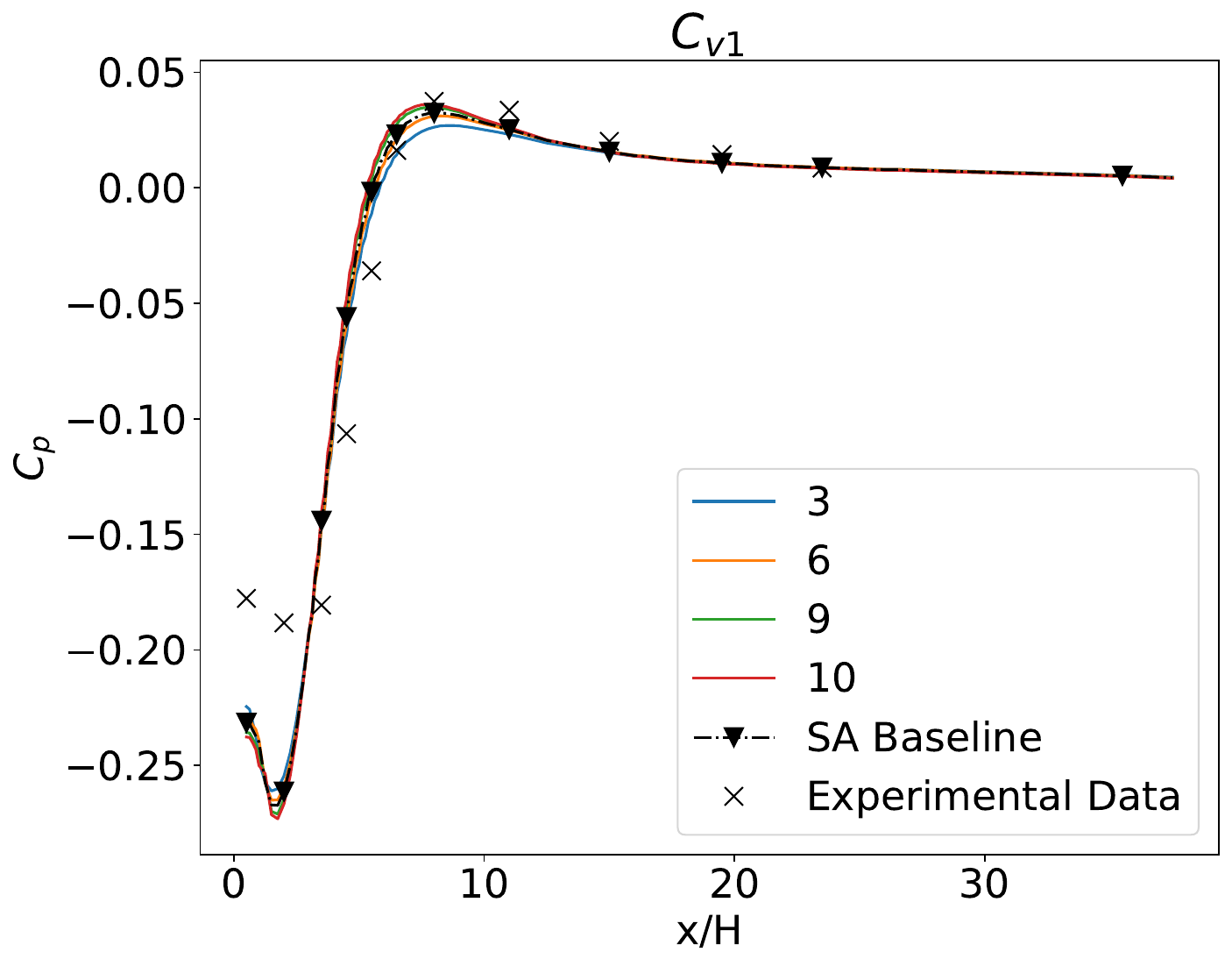}}\\
\subfloat[]{\includegraphics[height=5.5cm]{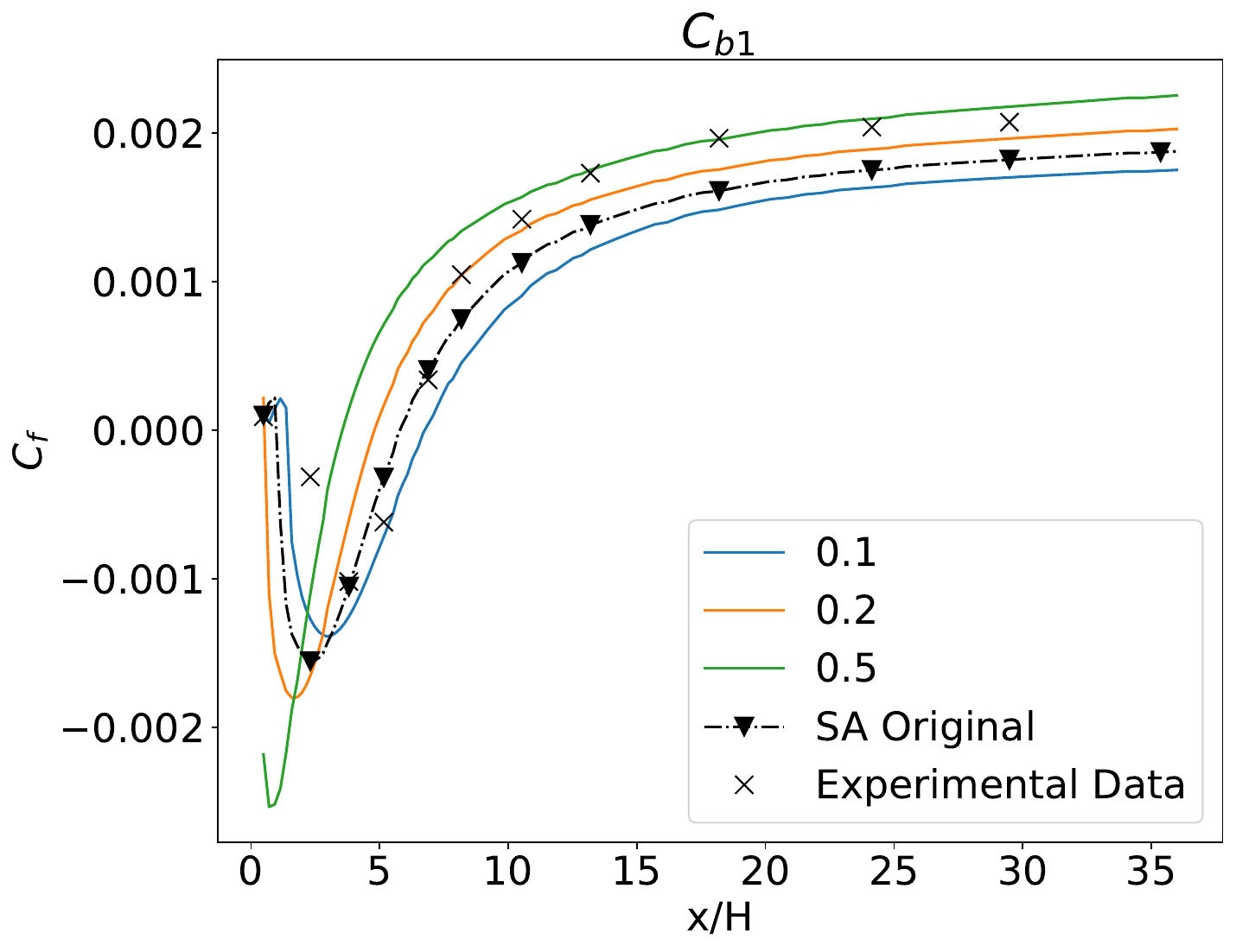}}
\subfloat[]{\includegraphics[height=5.5cm]{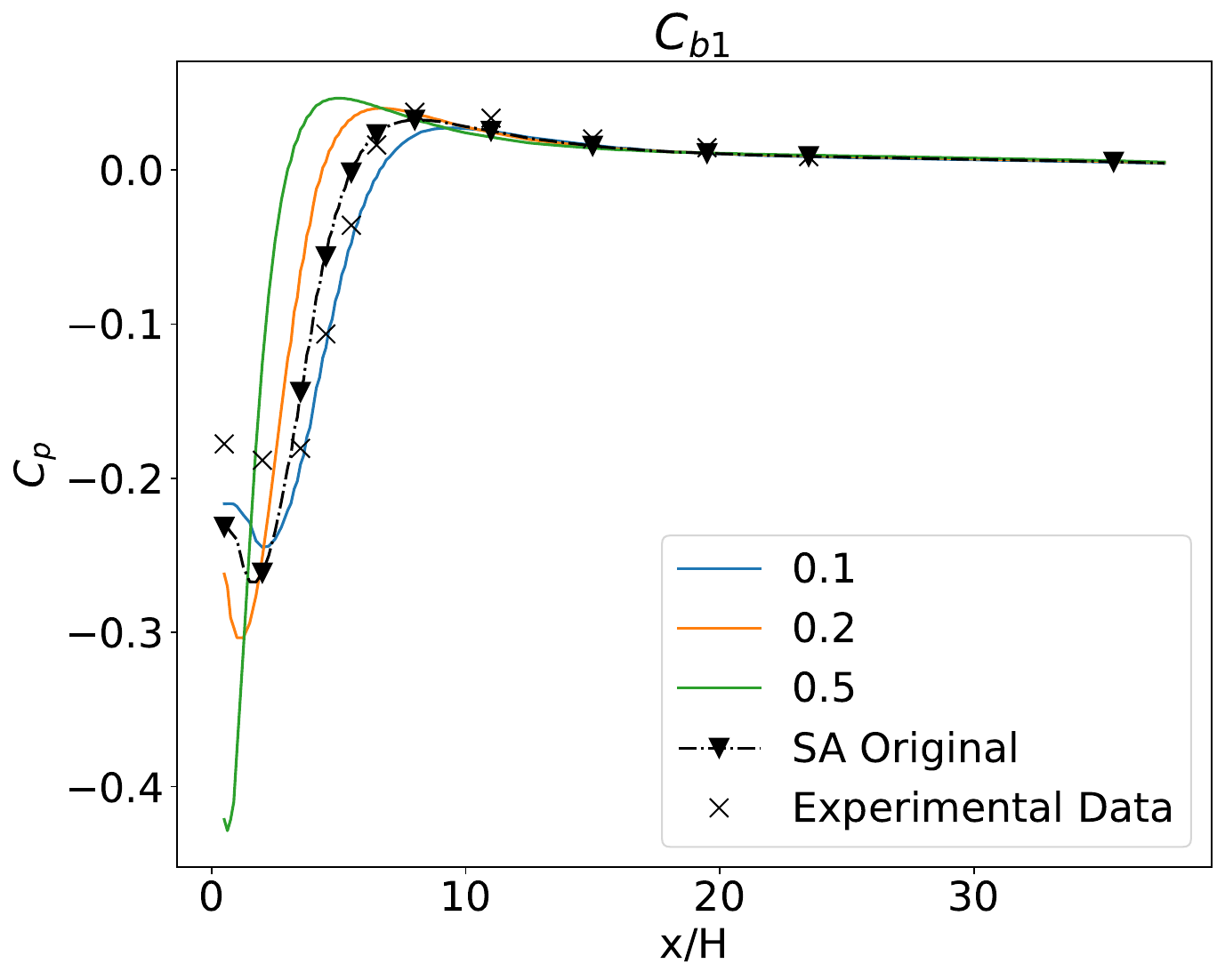}}\\
\end{center}
\caption{Parametric analysis by varying $C_{b2}$ (a,b), $C_{v1}$ (c,d), $C_{b1}$ (e,f). For reference, the solutions with original SA coefficients along with experimental data \cite{nasa} are also given.}
\label{sens_2}
\end{figure}

In this section, a parametric analysis by varying the SA coefficients is performed. The underlying objective is to enhance our comprehension of how alterations in coefficient values influence the QOIs, such as $C_f$ and $C_p$, in the context of separation phenomena. This analysis significantly contributes to the selection of coefficients, along with their respective bounding values, that collectively constitute the formulation of the matrix $X$. Figures \ref{sens_1} and \ref{sens_2} outline the results of the analysis. 
\newline Figure \ref{sens_1}a and b present the variations in $C_f$ and $C_p$ with different $\sigma$ values. The parameter $\sigma$ influences the diffusivity within the SA equation. Notably, observable changes in both $C_f$ and $C_p$ occur in response to alterations in $\sigma$ values. Consequently, the choice of $\sigma$ assumes importance as a target for optimization in EnKF.
\newline The range $\sigma \in [0.3, 2]$ yields results in the vicinity of experimental data. Values below 0.3, such as $\sigma = 0.1$, deviate substantially from the experimental data, while $\sigma = 5$ displays minimal variance compared to the results at 2. This lack of variation at $\sigma > 2$ does not incentivize an expansion of the sample space. Hence, the interval [0.3, 2] emerges as the preferred range for $\sigma$ sampling in the $X$ matrix.
\newline Figure \ref{sens_1} c-f, show the variation of $C_f$ and $C_p$ w.r.t $C_{w2}$ and $C_{w3}$. These two coefficients collectively contribute to the parameter $f_w$ (as depicted in eq. \ref{fw_1}). This parameter plays a pivotal role in governing the destruction of eddy viscosity.
\newline The selection of $C_{w2}$ and $C_{w3}$ for $X$ matrix stems from their direct influence on $f_w$. Notably, $f_w$ affects the SA's predictions for wall-bounded nonequilibrium flows \cite{spalart1992one,spalart2023old,bin2023data}. Particularly, Bin et al. \cite{bin2023data}, outline the impact of $f_w$ on the behavior of skin friction coefficient ($C_f$) in the recovery zone.
\newline The interval [0.75, 1.75], is chosen for sampling $C_{w2}$. This range is chosen because for different regions the values of $C_f$ and $C_p$ improve based on the values $C_{w2}$. In addition, $C_p $ and $C_f$, follow mutually opposite trends for increase or decrease of the $C_{w2}$. It was observed that for the values in the range [0.75, 1.75], the QOI takes the values that are closer to the experimental data for the entire region.  
\newline Similarly, the interval [1,2] is decided for the  $C_{w3}$. A slight improvement in the recovery zone for $C_f$ is observed at $C_{w3} =1 $, which is also selected as the lower bound of the sample space owing to deteriorated accuracy at $C_{w3} < 1$ at $x/H > 25$, i.e., 0.75 and 0.5. For the upper limit, the value of 2 (SA baseline) is selected owing to the non-observable difference between the results at the values of 2 and 5. Additionally, the upper bound is motivated by the original value of $C_{w3} =2 $ in the SA model.
\newline Figures \ref{sens_2}a and b depict the results of a parametric analysis involving the variation of $C_{b2}$. The observed variations in both $C_f$ and $C_p$ due to changes in $C_{b2}$ are not substantial. Consequently, $C_{b2}$ is not deemed influential enough to warrant selection for the optimization process.
\newline Similarly, Figures \ref{sens_2}c and d illustrate the impact of variations in $C_{v1}$ on $C_f$ and $C_p$. Although alterations in $C_{v1}$ have a limited effect on both coefficients, discernible variations emerge in the far downstream region for $C_f$ and the recovery region for $C_p$. As a result of these observations, $C_{v1}$ is chosen for the optimization process.
\newline For the optimization of $C_{v1}$, the selected range of variation is [6, 9]. This range is informed by the improved performance of $C_f$ at higher values of $C_{v1}$. Hence, the sample space is slightly biased towards values higher than baseline $C_{v1} = 7.1$.
\newline Figure \ref{sens_2} e and f shows significant influence of varying $C_{b1}$ on $C_f$ and $C_p$. Therefore, $C_{b1}$ is also selected for the optimization. However, for the purpose of this study the $C_{b1}$ is further parameterized as a function of $r$, where $ r \equiv \nu_t/Sk^2d^2$.
\section{Parametrization of $C_{b1}$} \label{senstivity2} From Figure \ref{sens_2}e, it becomes evident that when $C_{b1} = 0.1$, the behavior of skin friction coefficient ($C_f$) in the re-circulation region aligns more closely with the experimental data. Conversely, with $C_{b1} = 0.2$, better performance is observed in the recovery zone. While this region-specific impact is observable for other coefficients as well, it is particularly pronounced in the case of $C_{b1}$. Hence, $C_{b1}$ is parameterized further. This extension allows $C_{b1}$ to adopt different values contingent on the flow and domain characteristics. Drawing inspiration from eq. \ref{fw_1}, which parameterizes $f_w$ in terms of $r$, it was deemed appropriate to express $C_{b1}$ as a function of $r$ as well. Though the rest of the coefficients can also be parameterized further, for our study only $C_{b1}$ is parameterized.
\newline For the sake of simplicity, our study parameterized $C_b1$ as a linear function of $r$ as shown in eq. \ref{cb_1_eqn2}. The upper bound $C_{b1}$ is limited to its default value of 0.1355 using a $min$ function. This upper bound has the effect of plateauing on the $C_{b1}$ function which is similar to the $f_w$. This will help to ensure the consistency between the production and destruction term of the SA model, along with preserving the behavior of the model in equilibrium flows. As in section \ref{ensemble_matrix}, $max(,c*r+d)$, term spike the value for $C_{b1}$ as $r$ approached zero, this is important for maintaining the behavior of the model in free shear flows. The simplistic nature of the eq. \ref{cb_1_eqn2} and the improvements due to it are encouraging. However, the equation used here is not claimed to be optimum for the parameterization and mildly violates the soft constraint proposed by Spalart et al. \cite{spalart2023old} for the data-driven studies, i.e. not using $min$ or $max$ functions. Hence, a more focused study can be conducted to explore equations that are more consistent with the SA model. Another approach could be to replace the eq. \ref{cb_1_eqn2} by a NN.
\begin{equation}
C_{b1} = min\left( max\left(r*b,c*r + d\right), 0.1355 \right)
\label{cb_1_eqn2}
\end{equation}
\begin{figure}{h!}
\begin{center}
\subfloat[]{\includegraphics[height=6 cm]{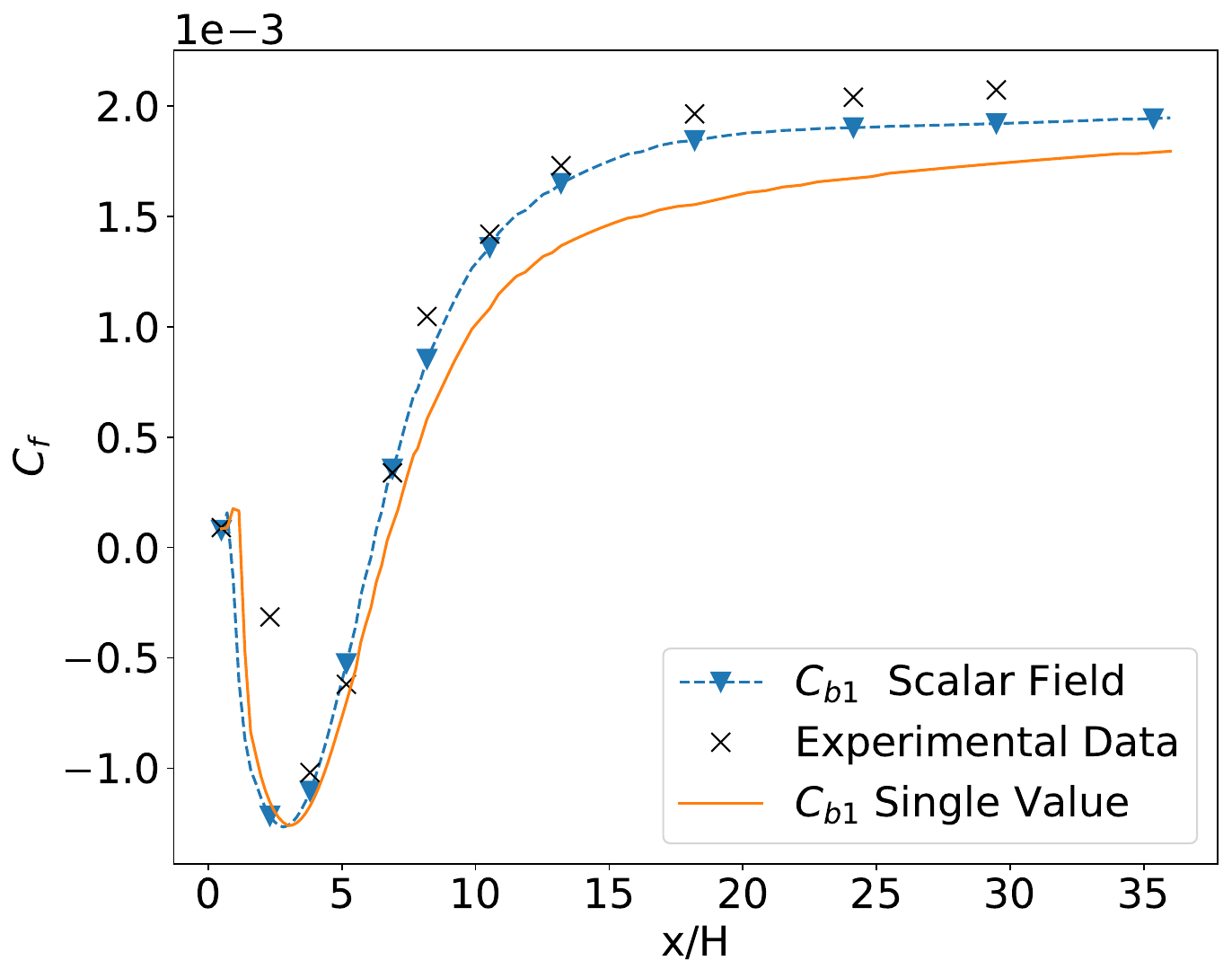}}
\subfloat[]{\includegraphics[height=6 cm]{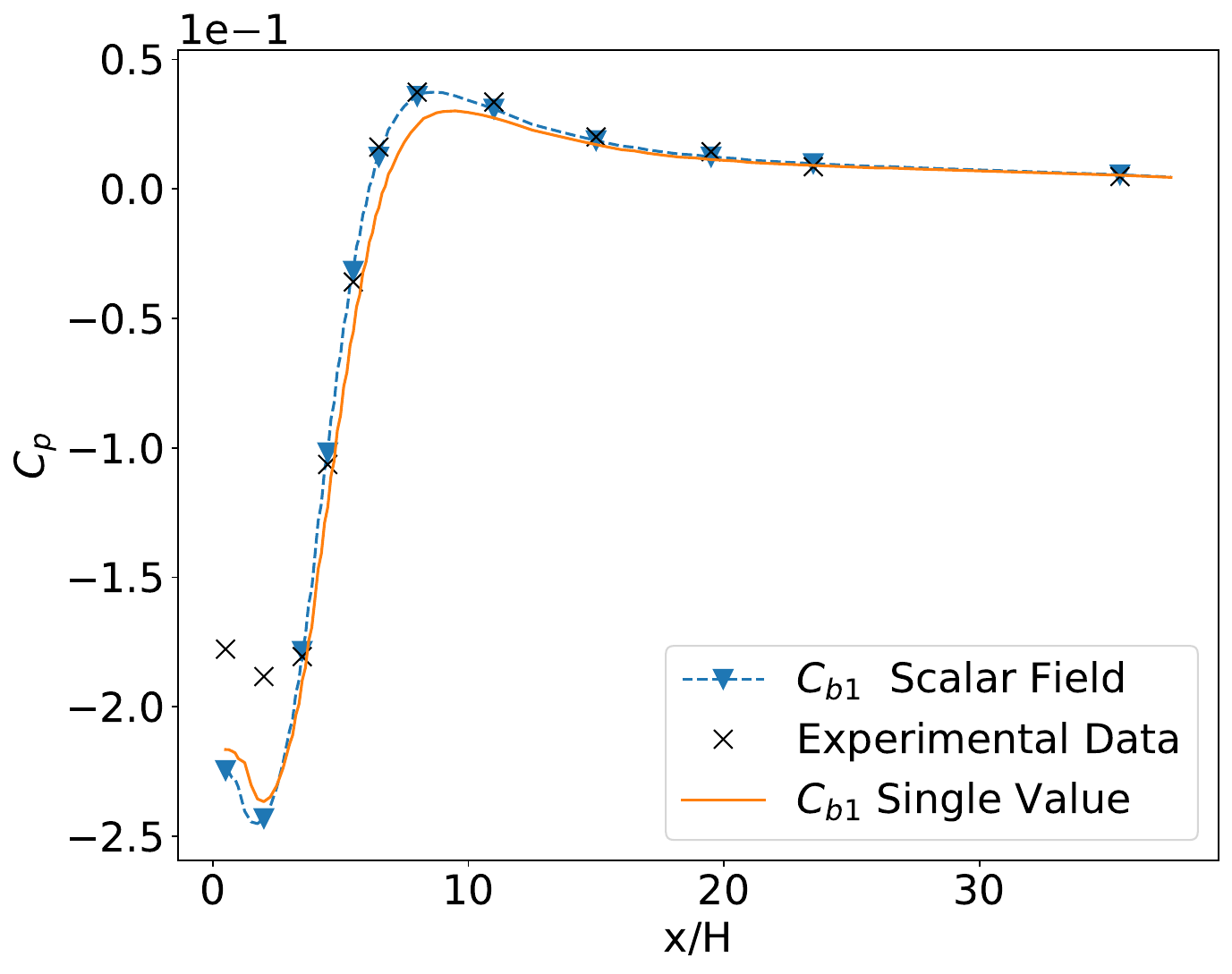}}\\
\end{center}
\caption{Comparison of EnKF calibration using $C_{b1}$ as a function of $r$ with $C_{b1}$ as constant scalar within the BFS domain. a.) $C_f$ and b.) $C_p$ along the bottom wall for both calibrations.}
\label{sens_3}
\end{figure}
\newline Figure \ref{sens_3} compares the EnKF calibration for both scenarios i.e. where the $C_{b1}$ is: \textbf{i.} Scalar field varying w.r.t $r$ and \textbf{ii.} Scalar field with a constant value throughout the domain. Notably, the rest of the coefficients in the ensemble are the same. Using $C_{b1}$ as a function of $r$ yields results closer to the experimental values and hence was the preferred approach for the current study.
\section{Backward facing step (BFS): Mesh 2-4} \label{meshes} 
\begin{figure}\centering
\subfloat[Mesh 2]{\label{Mesh 2}\includegraphics[height=5.5cm]{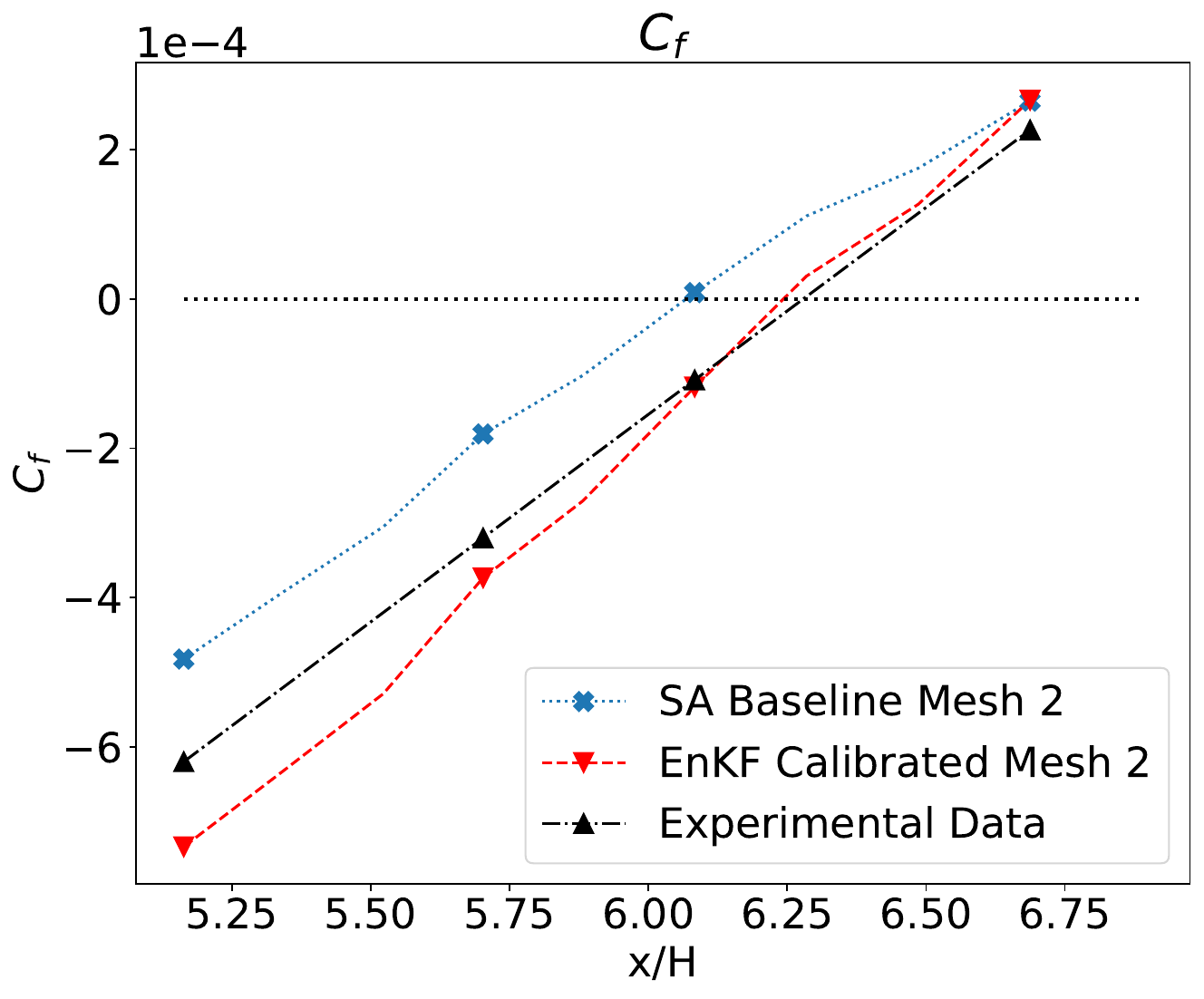}}\hfill
\subfloat[Mesh 3]{\label{Mesh 3}\includegraphics[height=5.5cm]{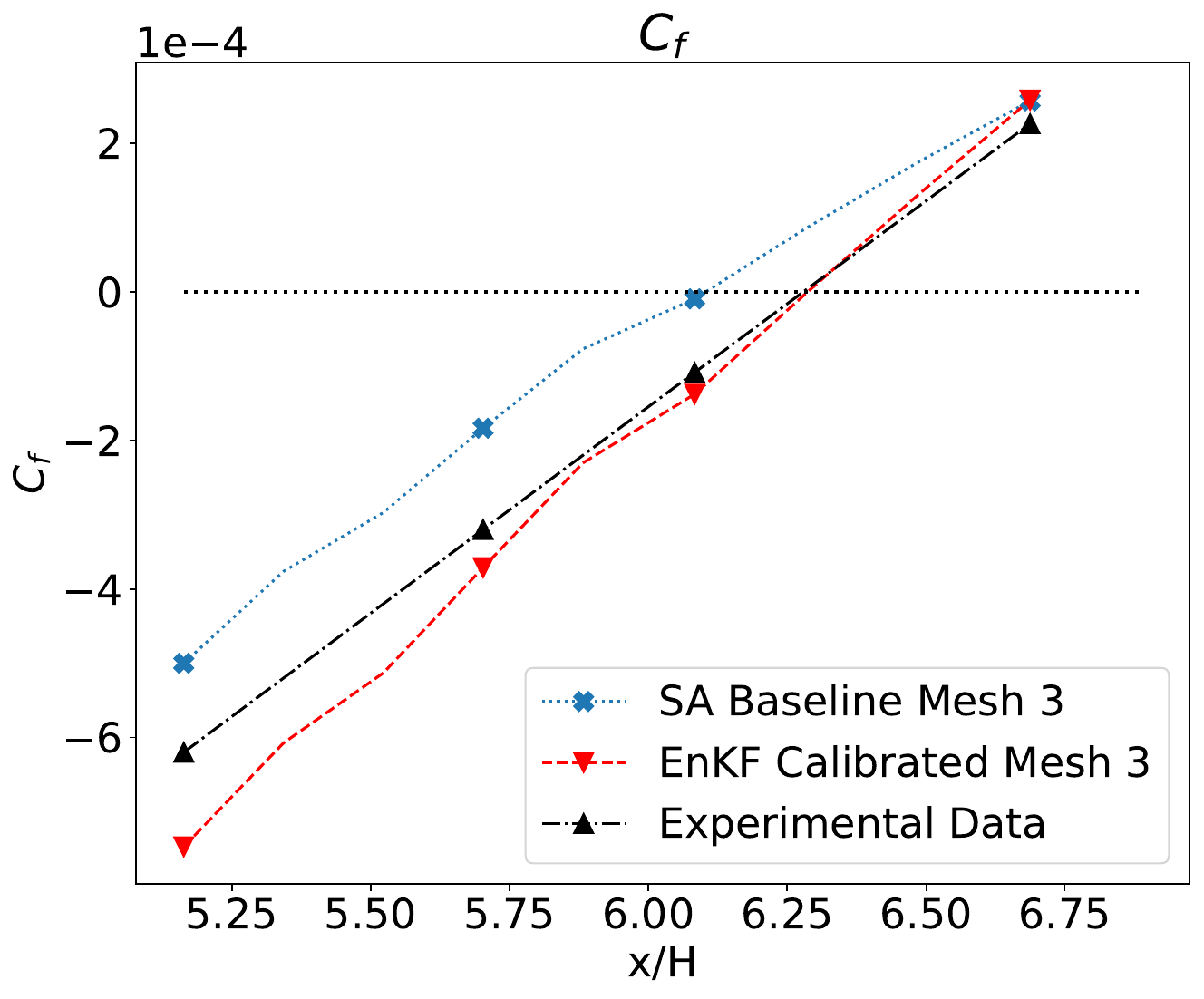}}\par 
\subfloat[Mesh 4]{\label{Mesh 4}\includegraphics[height=5.5cm]{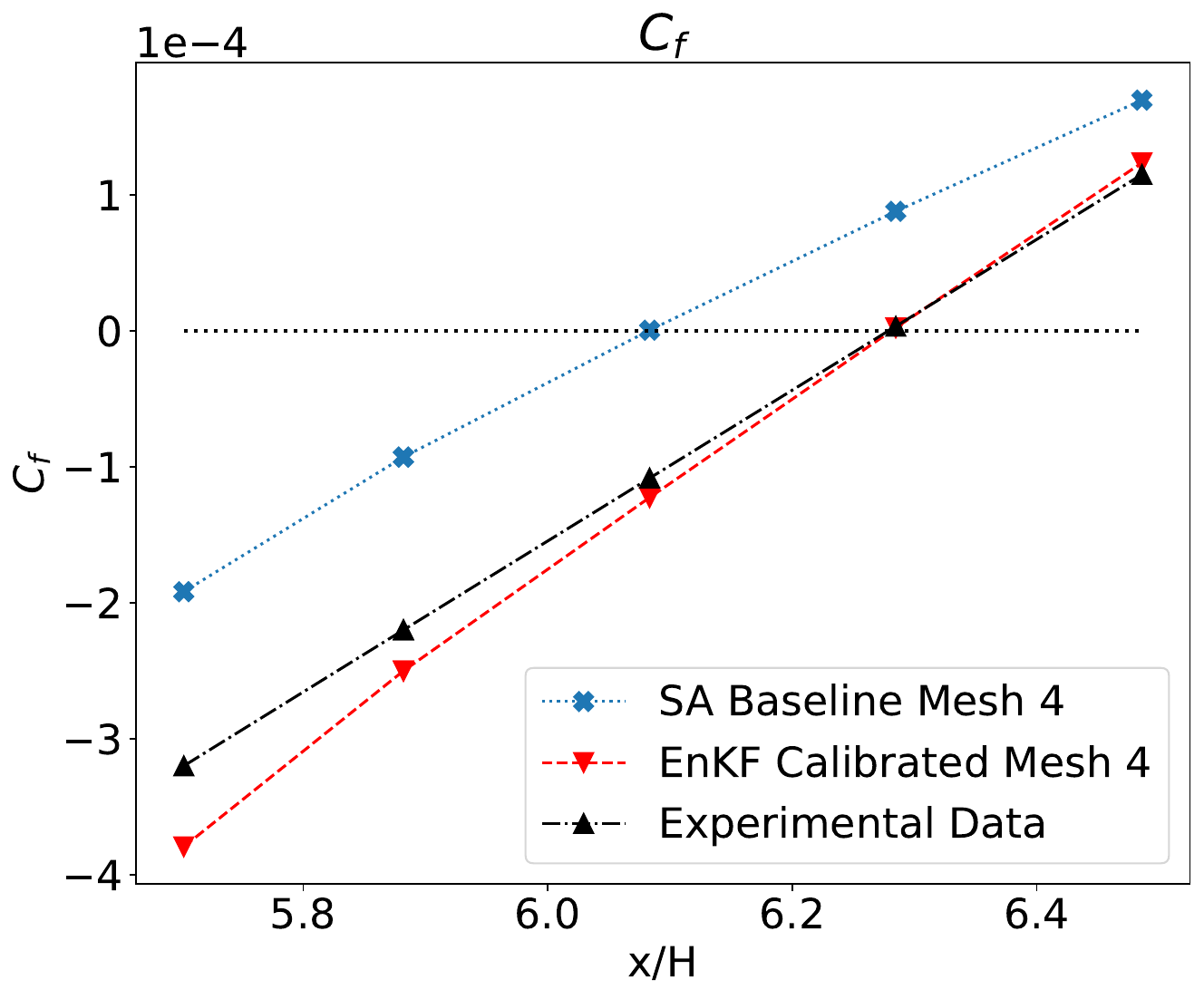}}
\caption{$C_f$ vs. $x/H$ showing the improved reattachment location ($C_f = 0$) for the calibrated model across meshes 2-4.}
\label{reattache_mesh}
\end{figure}

\begin{figure}[h!]\centering
\subfloat[Mesh 2]{\label{Mesh 2}\includegraphics[height=5.5cm]{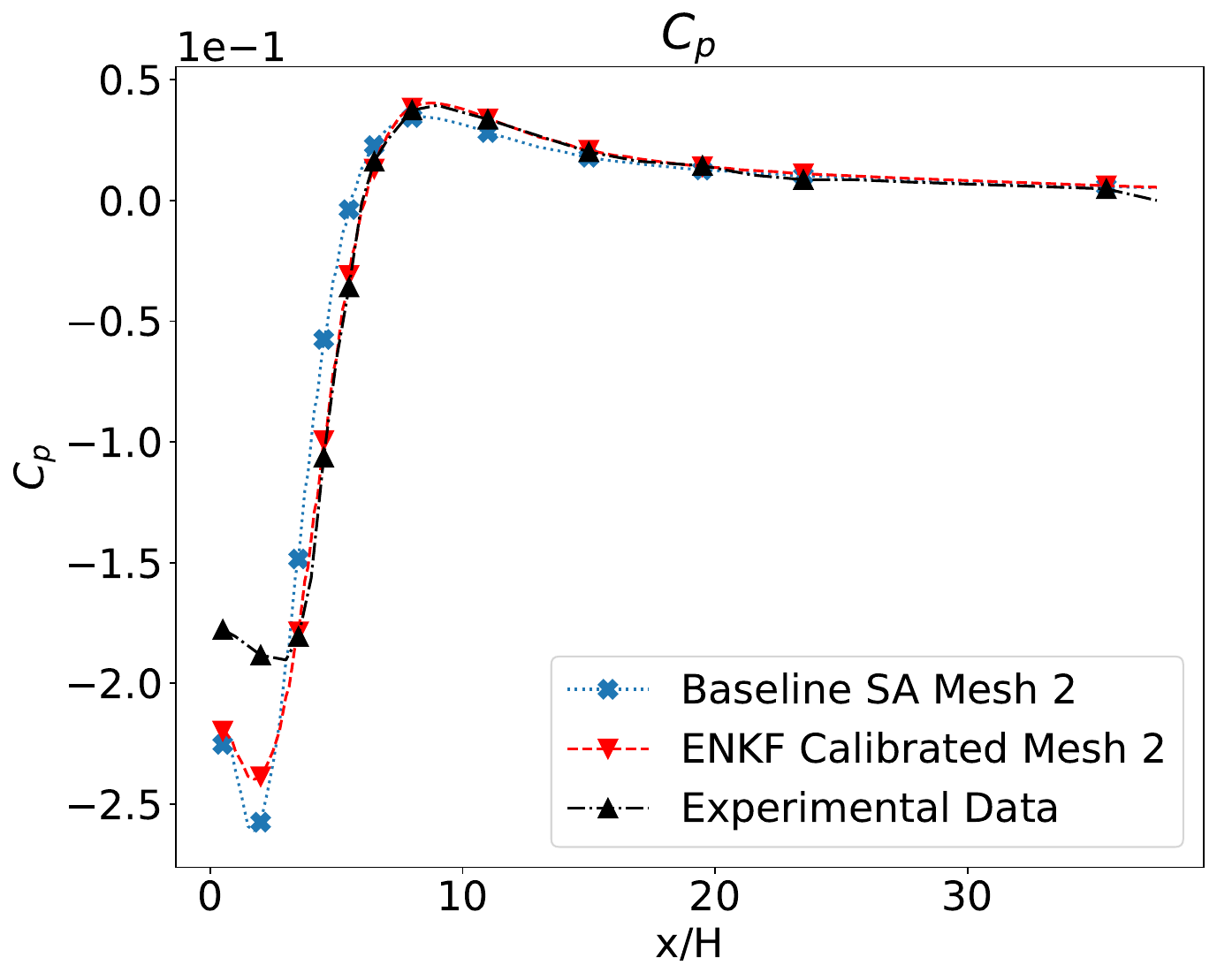}}\hfill
\subfloat[Mesh 3]{\label{Mesh 3}\includegraphics[height=5.5cm]{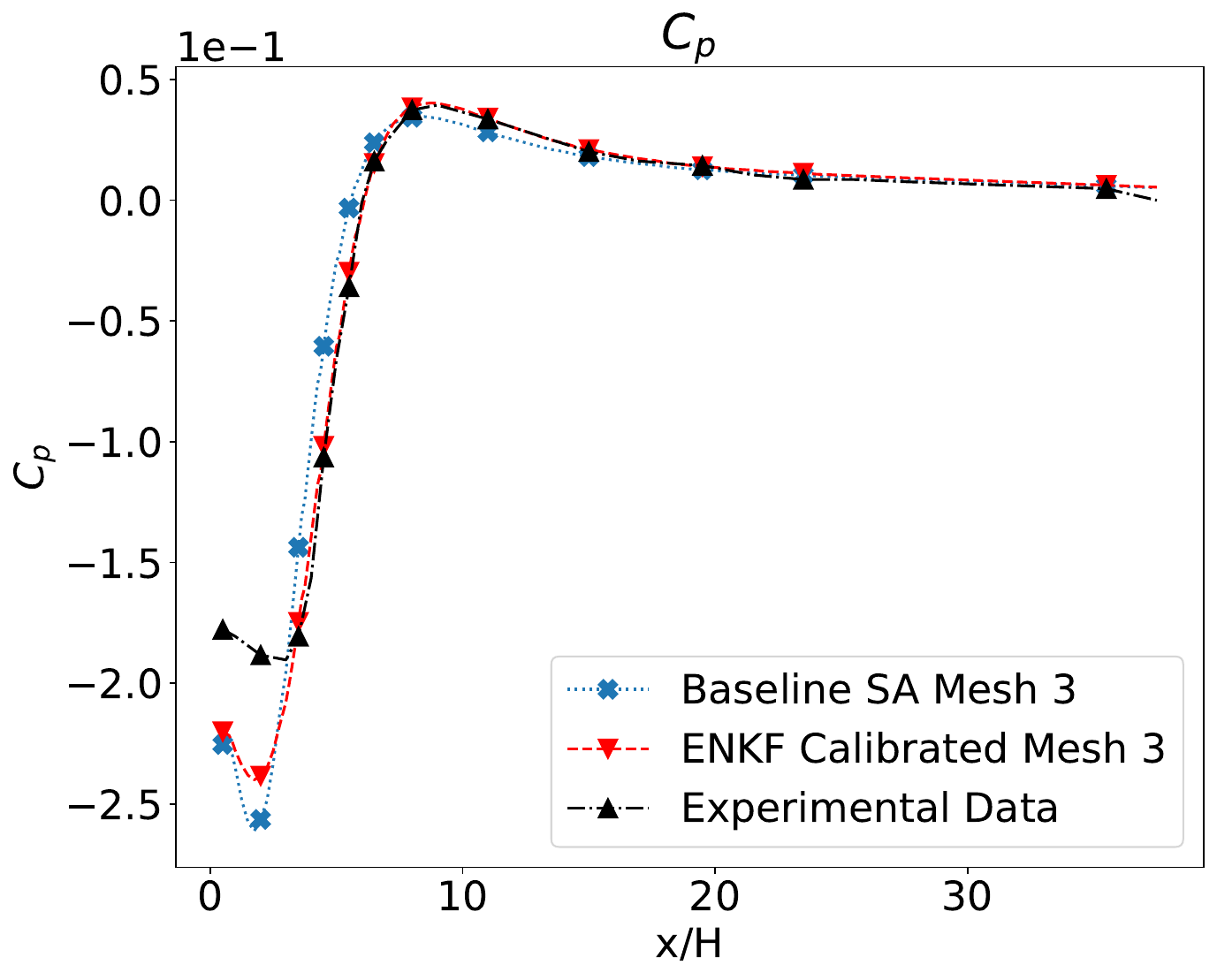}}\par 
\subfloat[Mesh 4]{\label{Mesh 4}\includegraphics[height=5.5cm]{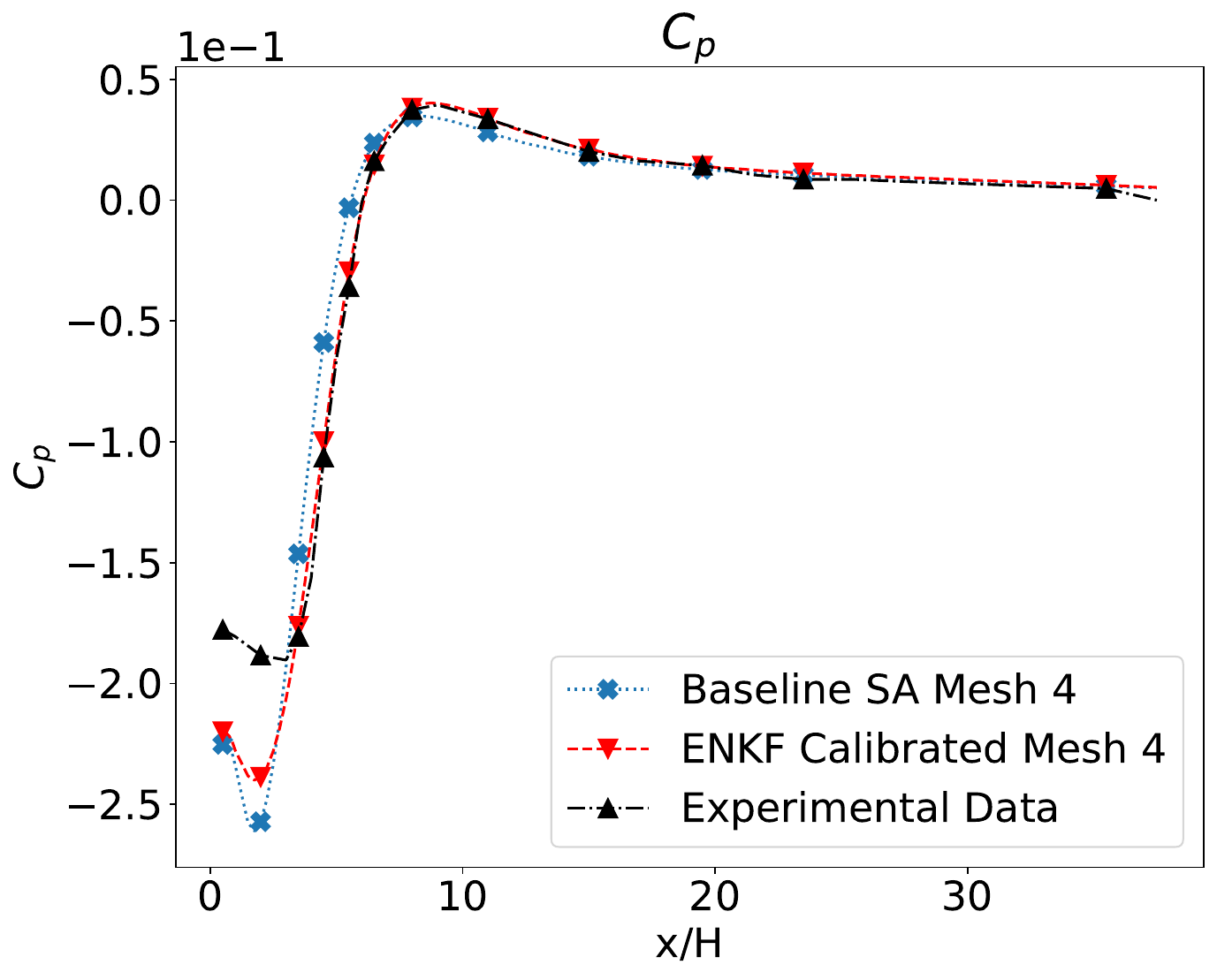}}
\caption{$C_p$ vs. $x/H$ across meshes 2-4.}
\label{Cp_mesh}
\end{figure}

In this section, the previously calibrated coefficients, are tested for meshes 2-4 (Section \ref{flow_over_bfs}), approximately 43,000, 47,000, and 53,000 cells, respectively, are outlined for BFS. Figure \ref{reattache_mesh} shows the reattachment location of the calibrated model compared to the baseline for each mesh, along with the experimental data. It is evident that the reattachment location is significantly improved for each mesh compared to the baseline SA model.   
\newline Figure \ref{Cp_mesh} presents the $C_p$ curves for each mesh. The calibrated model demonstrates better agreement with the experimental data across all meshes.
\newline For $C_f$, figure \ref{Cf_mesh}, the calibrated model shows agreement with the experimental data in the separation bubble and reattachment zone.
\begin{figure}[h!]\centering
\subfloat[Mesh 2]{\label{Mesh 2}\includegraphics[height=5.5cm]{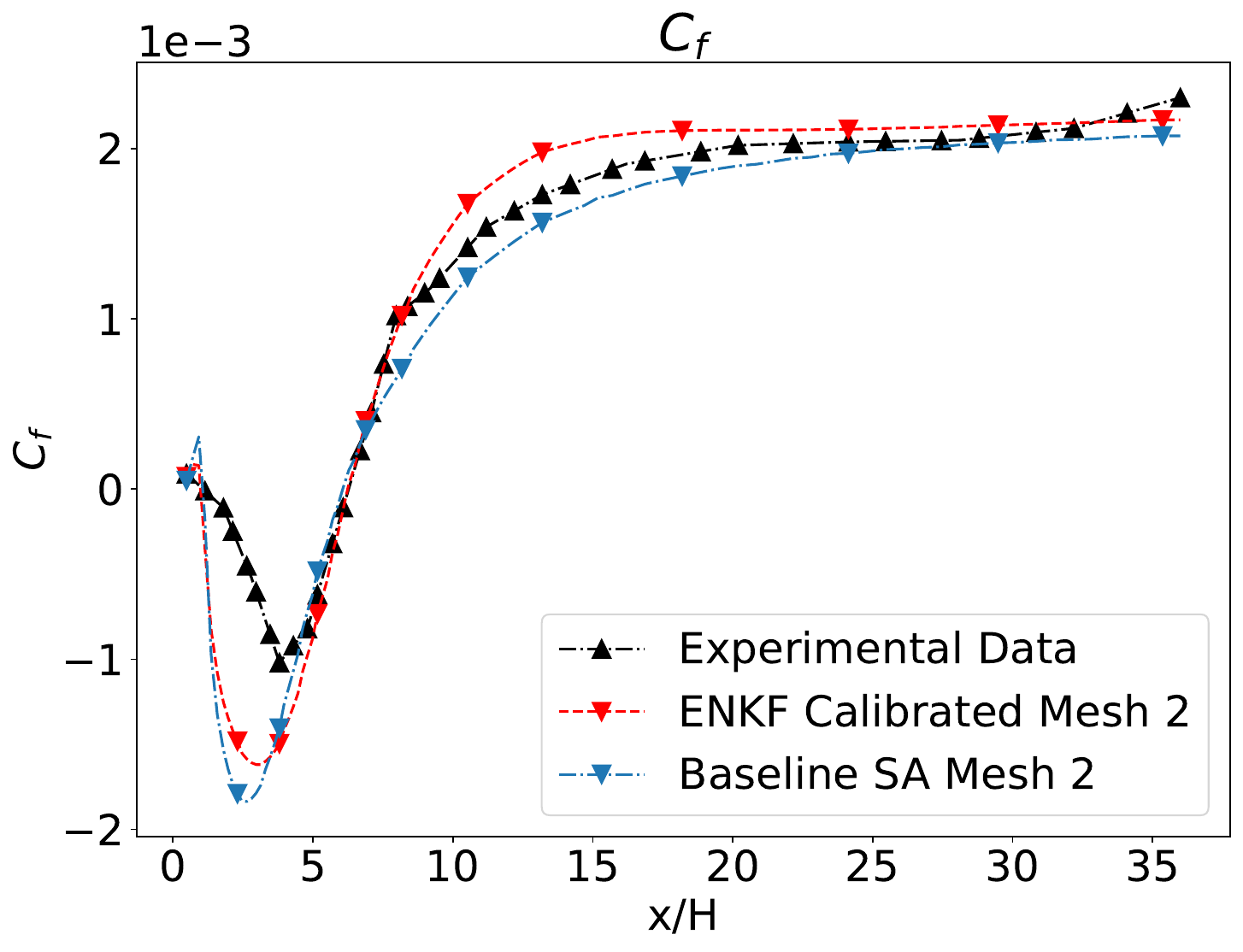}}\hfill
\subfloat[Mesh 3]{\label{Mesh 3}\includegraphics[height=5.5cm]{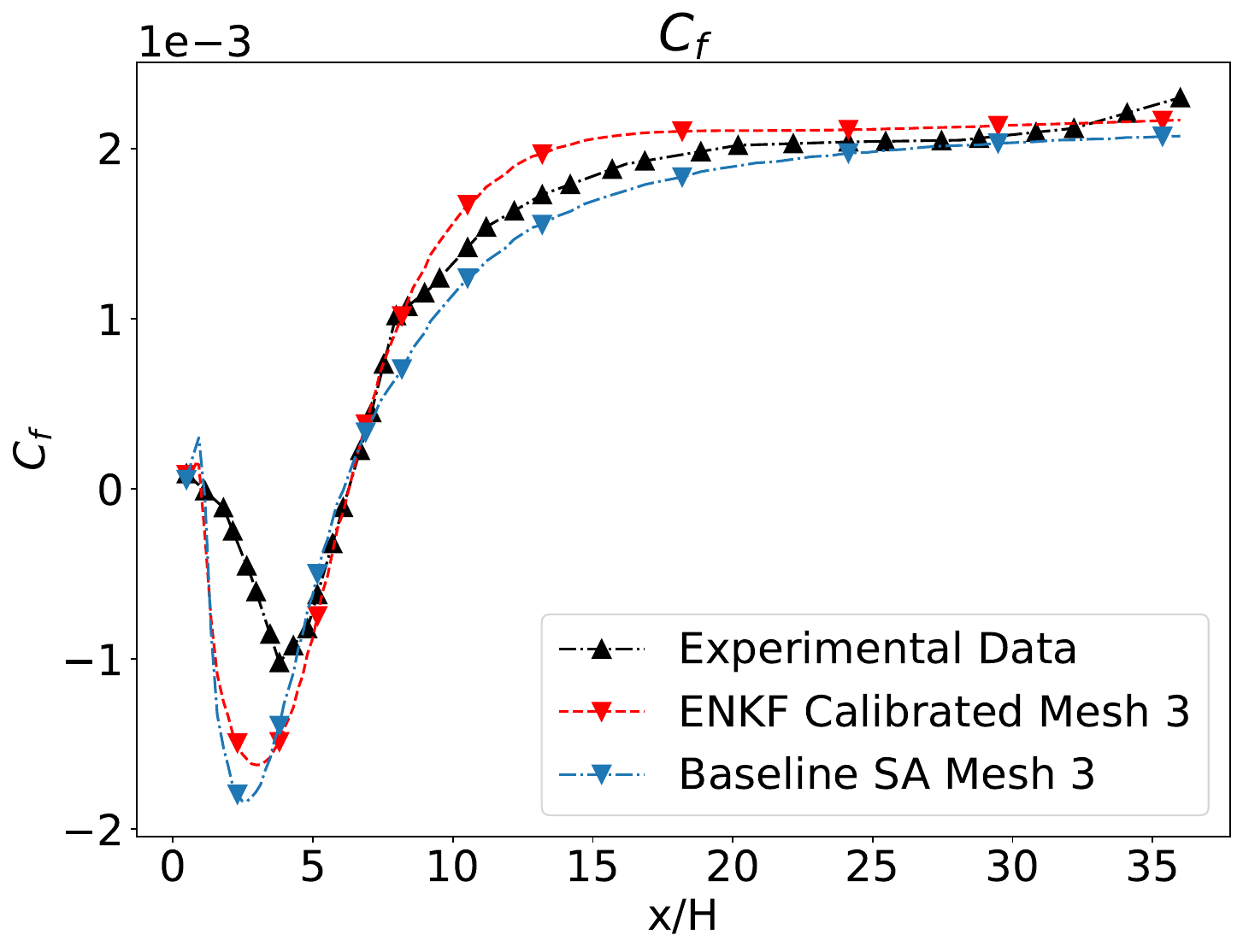}}\par 
\subfloat[Mesh 4]{\label{Mesh 4}\includegraphics[height=5.5cm]{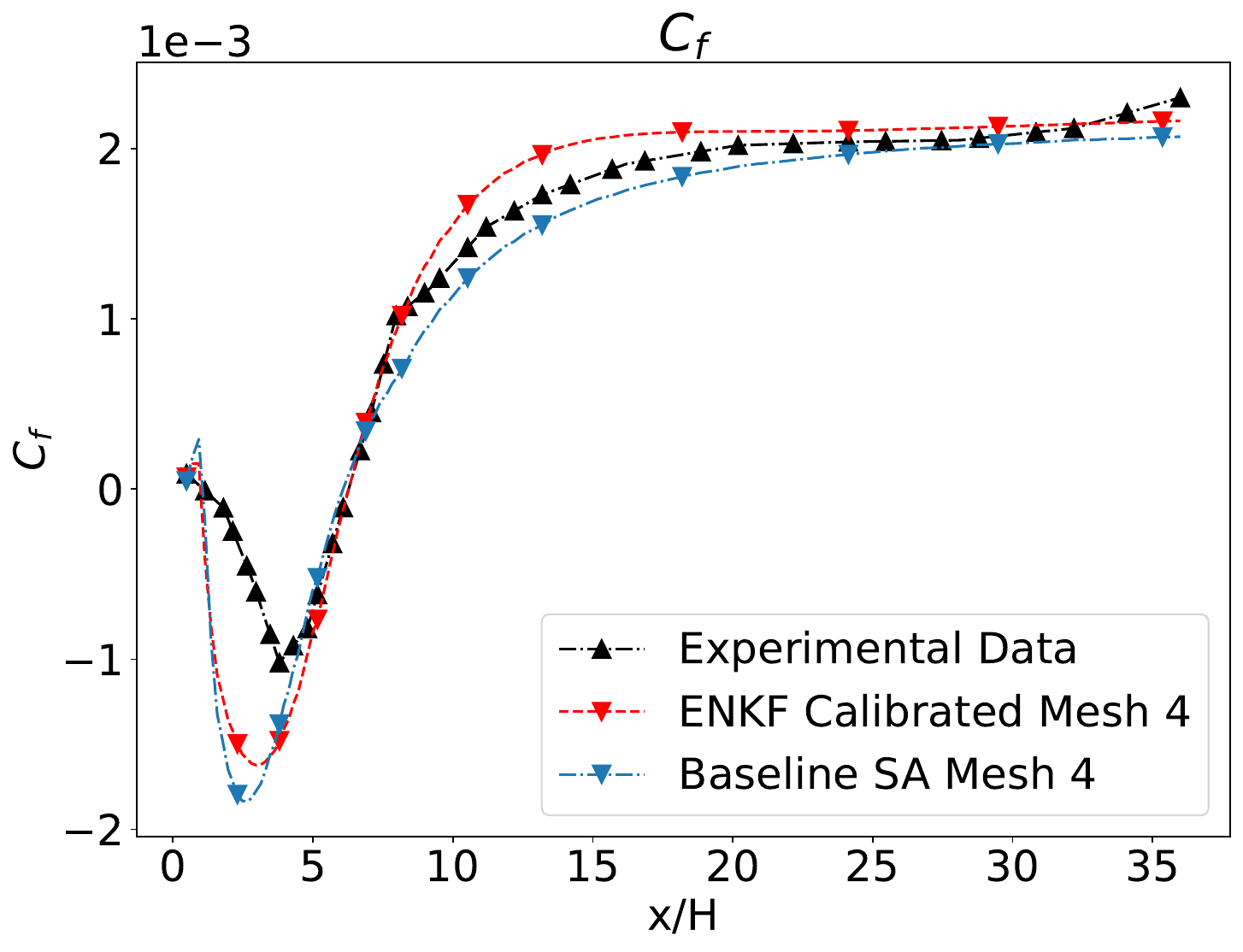}}
\caption{$C_f$ vs. $x/H$ across meshes 2-4.}
\label{Cf_mesh}
\end{figure}
\newline For each mesh, figure \ref{Cf_mesh} shows deviation of $C_f$ from experimental data. It can be seen that the calibrated model have a better agreement with the experimental data in the separation bubble and reattachment zone.
\begin{figure}[h!]\centering
\subfloat[Mesh 2]{\label{Mesh 2}\includegraphics[height=5.5cm]{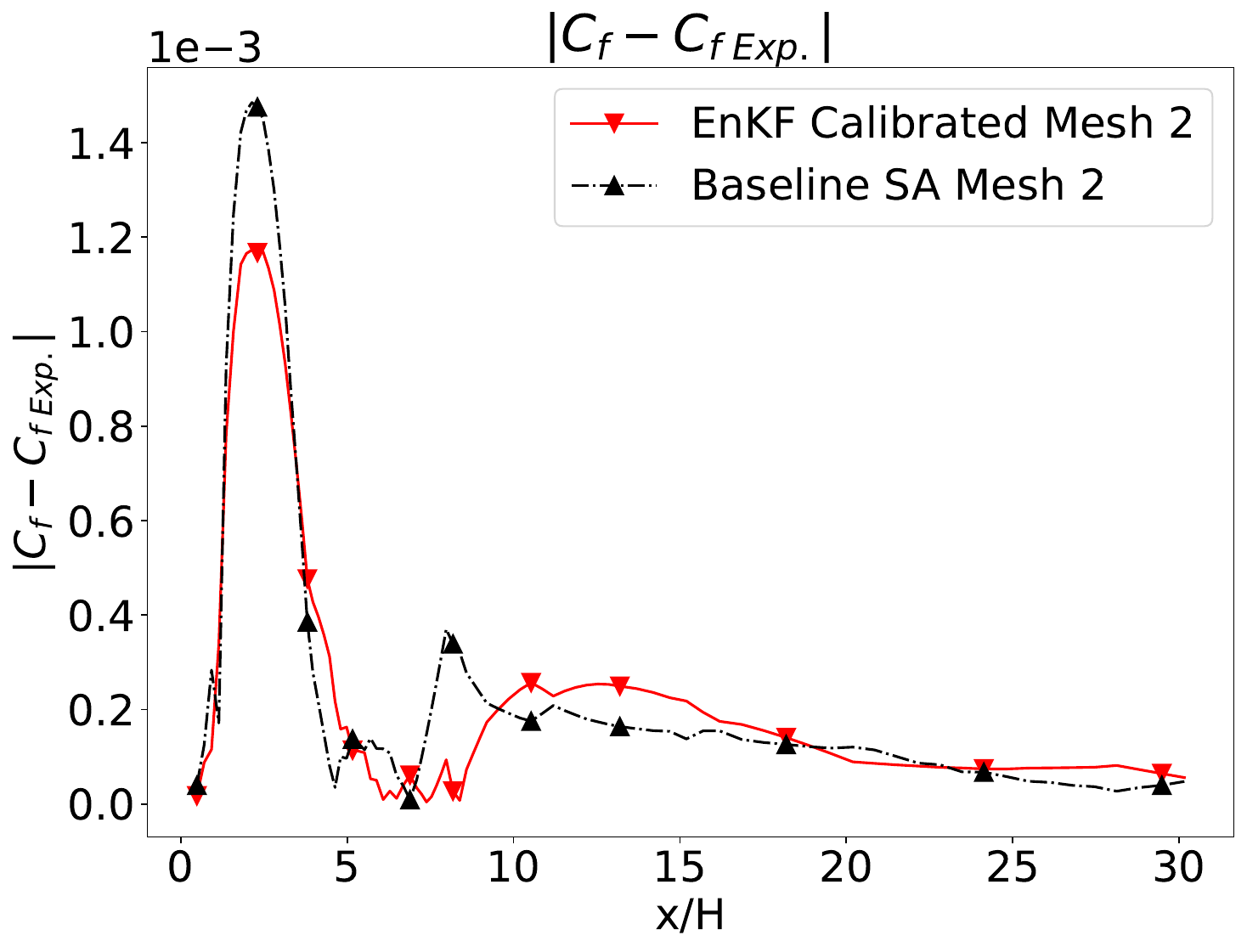}}\hfill
\subfloat[Mesh 3]{\label{Mesh 3}\includegraphics[height=5.5cm]{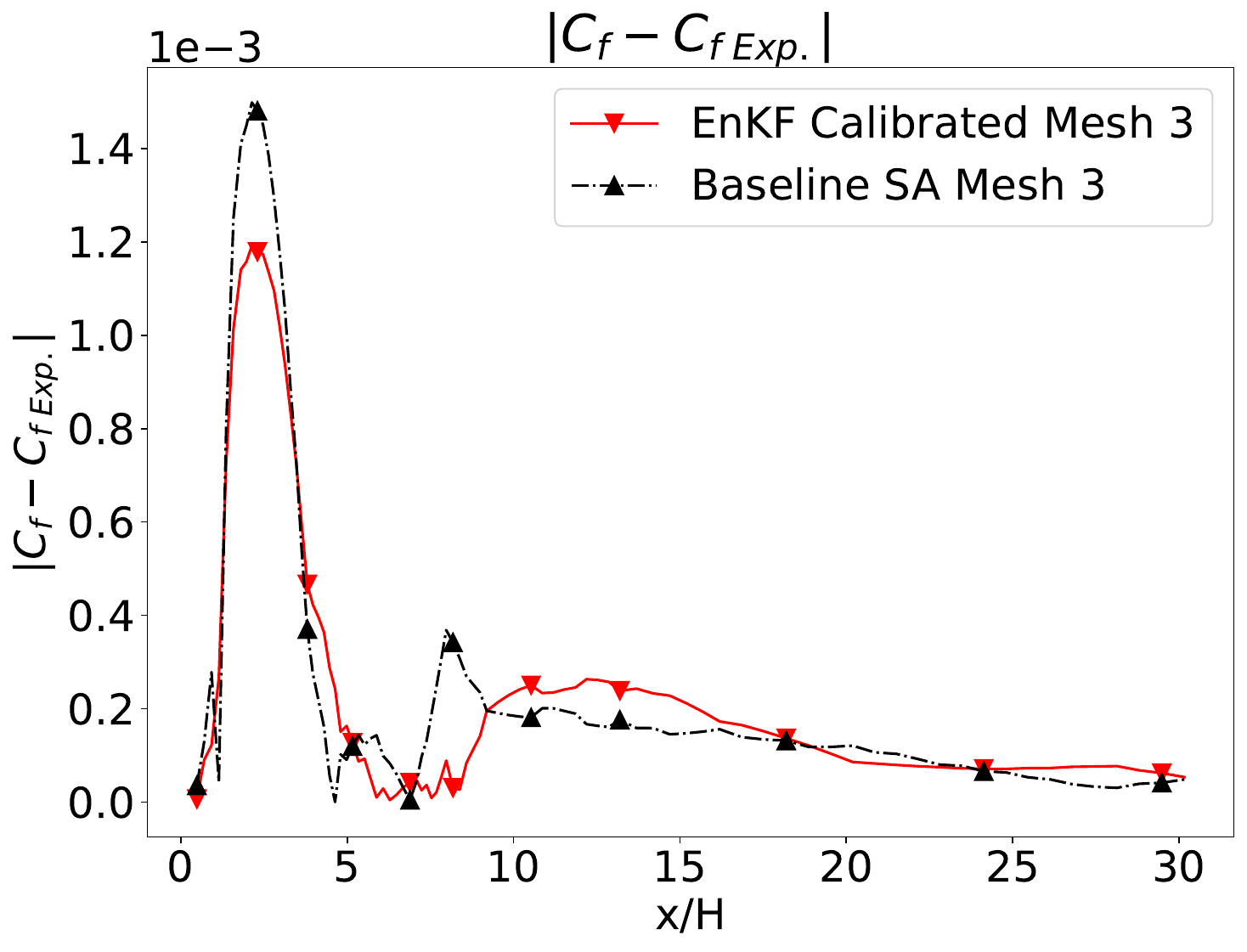}}\par 
\subfloat[Mesh 4]{\label{Mesh 4}\includegraphics[height=5.5cm]{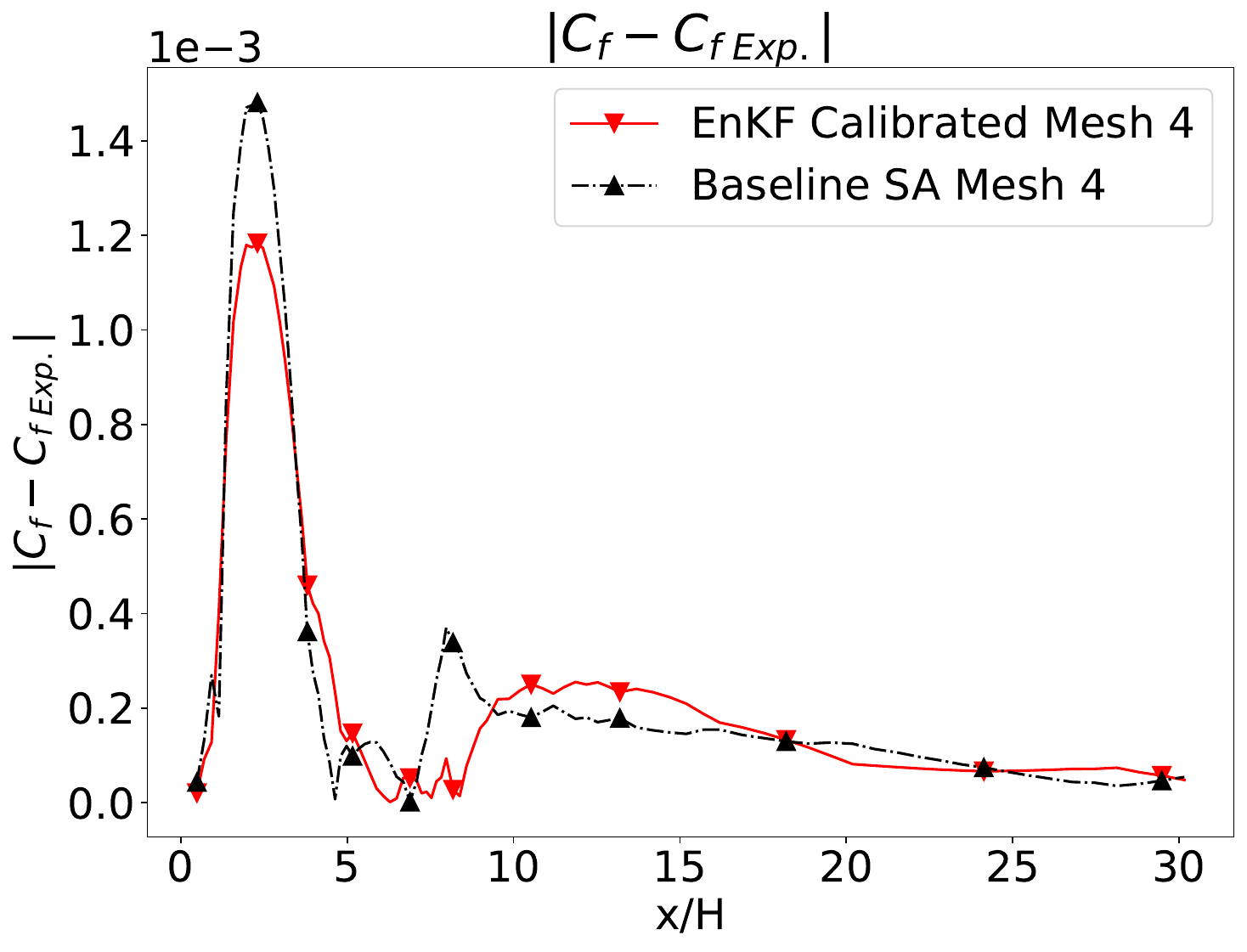}}
\caption{$C_f$ vs. $x/H$ across meshes 2-4.}
\label{Cf_mesh}
\end{figure}
\clearpage 
\bibliography{apssamp}

\end{document}